\documentclass[apj]{emulateapj}
\usepackage{graphicx} 
\usepackage{appendix}

\newcommand{\nii}{[N\thinspace{II}]}
\newcommand{\oii}{[O\thinspace{II}]}
\newcommand{\oiii}{[O\thinspace{III}]}

\slugcomment{Accepted for publication in The Astrophysical Journal }

\begin{document}
\title{An Atlas of $\lowercase{z}=5.7$ and $\lowercase{z}=6.5$ Ly$\alpha$ Emitters
\altaffilmark{*}\altaffilmark{\textdagger}
}
\author{
E.~M.~Hu,$\!$\altaffilmark{1} \email{hu@ifa.hawaii.edu}
L.~L.~Cowie,$\!$\altaffilmark{1} \email{cowie@ifa.hawaii.edu}
A.~J.~Barger,$\!$\altaffilmark{1,2,3} \email{barger@astro.wisc.edu}
P.~Capak,$\!$\altaffilmark{4} \email{capak@astro.caltech.edu}
Y.~Kakazu,$\!$\altaffilmark{5} \email{kakazu@astro.caltech.edu}
L.~Trouille$\!$\altaffilmark{2} \email{trouille@astro.wisc.edu}
}

\altaffiltext{*}{Based in part on data collected at the Subaru Telescope,
which is operated by the National Astronomical Observatory of Japan.}
\altaffiltext{\textdagger}{Based in part on data obtained at the W.~M.~Keck 
Observatory, which is operated as a scientific partnership among the 
California Institute of Technology, the University of California, and 
NASA and was made possible by the generous financial support of the 
W.~M.~Keck Foundation.}
\altaffiltext{1}{Institute for Astronomy, University of Hawaii,
2680 Woodlawn Drive, Honolulu, HI, 96822}
\altaffiltext{2}{Department of Astronomy, University of Wisconsin-Madison, 
475 North Charter Street, Madison, WI 53706}
\altaffiltext{3}{Department of Physics and Astronomy,
University of Hawaii, 2505 Correa Road, Honolulu, HI 96822}
\altaffiltext{4}{Spitzer Science Center, 314-6 Caltech, 1201 E. California
Blvd., Pasadena, CA, 91125}
\altaffiltext{5}{Department of Astronomy, 249-17 Caltech, 1201 E. California
Blvd., Pasadena, CA, 91125}

\shorttitle{High-Redshift Ly$\alpha$ Emitters}
\shortauthors{Hu et al.}

\begin{abstract}
We present an atlas of 88 $z\sim5.7$ and 30 $z\sim6.5$ Ly$\alpha$
emitters obtained from a wide-field narrowband survey.
We combined deep narrowband imaging 
in 120~\AA\ bandpass filters centered at
8150~\AA\ and 9140~\AA\ with deep $BVRIz$ broadband imaging
to select high-redshift galaxy candidates over an area of 4180~arcmin$^2$.
The goal was to obtain a uniform selection of comparable
depth over the 7 targeted fields in the two filters. 
For the GOODS-N region of the HDF-N field, we also selected candidates 
using a 120~\AA\ filter centered at 9210~\AA.
We made spectroscopic observations with Keck DEIMOS of nearly 
all the candidates to obtain the final sample of Ly$\alpha$ emitters. 
At the 3.3~\AA\ resolution of the DEIMOS observations the asymmetric
profile for Ly$\alpha$ emission 
can be clearly seen in the spectra of nearly all the galaxies.
We show that the spectral profiles are surprisingly similar for many of the
galaxies and that the composite spectral profiles are nearly identical at
$z=5.7$ and $z=6.5$. 
We analyze the distributions of line widths and
Ly$\alpha$ equivalent widths and find that
the lines are marginally narrower at the higher
redshift, with median values of $0.77$~\AA\ at $z=6.5$
and $0.92$~\AA\ at $z=5.7$.
The line widths have a dependence on the Ly$\alpha$ luminosity of 
the form $\sim L_{\alpha}^{0.3}$.
We compare the surface densities and
the luminosity functions at the two redshifts and find that there
is a multiplicative factor of 2 decrease in the number density of
bright Ly$\alpha$ emitters from $z=5.7$ to $z=6.5$, while the 
characteristic luminosity is unchanged. 
\end{abstract}

\section{Introduction}
\label{secintro}

\begin{deluxetable*}{lcccccc}
\tablecolumns{7}
\tablecaption{Narrowband Survey Fields}
\tablewidth{0pt}
\tablehead{
\colhead{Field}  &  \colhead{R.A.} & \colhead{Decl.} &
\colhead{($l^{\rm{II}},b^{\rm{II}}$)} & \colhead{E$_{B-V}$\tablenotemark{a}} &
\colhead{NB816}  & \colhead{NB912} \\[0.5ex]
& \colhead{(J2000)} & \colhead{(J2000)} &    &   &   \colhead{(hrs)} & \colhead{(hrs)}}
\startdata
\noalign{\vskip-2pt}
SSA22 &  22:17:57.00 & +00:14:54.5 &  (63.1,$-44.1$) &0.07 & 3.8  & 8.4\\
SSA22\_new &  22:18:24.67 & +00:36:53.4 & (63.6,$-43.9$) & 0.06 & 4.8  & 9.0\\
A370 & 02:39:53.00 & $-$01:34:35.0 & (173.0,$-53.6$) & 0.03 & 4.6 & 7.9\\
A370\_new &  02:41:16.27 & $-$01:34:25.1 & (173.4,$-53.3$) & 0.03 & 4.6 & 8.0\\
HDF-N   &  12:36:49.57 & +62:12:54.0 & (125.9,$+54.8$) & 0.01 & 9.7 & 8.3 \\
HDF-N\_new   &  12:40:26.40 & +62:21:45.0 & (125.1,$+54.7$) & 0.01 & 7.1 & 10.0 \\
SSA17 & 17:06:36.22 & +43:55:39.5 & (69.1,$+36.8$) & 0.02 & 5.1 & 4.5 \\
\enddata
\tablenotetext{a\ }{Estimated using \url{http://irsa.ipac.caltech.edu/applications/DUST/} 
based on Schlegel, Davis, \& Finkbeiner (1998)}
 \label{tbl-1}
\end{deluxetable*}

Ever since the first discovery of Ly$\alpha$ emitters above redshift
$z\sim6$ (Hu et al.\ 2002), the goal has been to identify substantial 
samples of high-redshift Ly$\alpha$ emitters in order 
to study such diverse topics as the formation of galaxies
in the early universe, early structure formation, reionization, and the
interactions of galaxies with the intergalactic medium (IGM).
The advent of deep, wide-field, narrowband surveys 
(e.g., Hu et al.\ 2004; hereafter, H04;
Wang et al.\ 2005; Shimasaku et al.\ 2006; Kashikawa et al.\ 2006;
Ouchi et al.\ 2008) has recently made it possible to obtain substantial 
numbers of galaxies at $z\sim5.7$ and $z\sim6.5$, where gaps in the night 
sky emission permit deep studies, for addressing these topics.
However, the straightforward interpretation of these survey results 
has been plagued by statistical and cosmic variance
from field to field (e.g., Hu \& Cowie 2006) and by the absence of
extensive spectroscopic follow-up of the primarily photometrically
selected samples.  As we shall show in the present work, photometric
samples can have a substantial degree of contamination, and spectroscopy
is necessary to remove these interlopers. 
With high-resolution spectroscopic follow-ups of the candidates, it is
possible both to confirm that the emission is due to redshifted 
Ly$\alpha$ and to estimate the statistics of interlopers.

Here we present the results of a wide-field, narrowband survey with
highly complete spectroscopic follow-up. We used
the $34'\times 27'$ field-of-view SuprimeCam mosaic CCD
camera (Miyazaki et al.\ 2002) on the Subaru 8.2~m telescope to observe
seven different target fields to search for Ly$\alpha$ emitters at
$z\sim5.7$ (NB816) and $z\sim6.5$ (NB912, NB921).
In Section~\ref{secnarrow} we summarize the narrowband and
continuum imaging together with the candidate selection.
The data obtained in these narrowbands are comparable
in depth, and we have analyzed them with uniform selection
and processing criteria. We have spectroscopically observed nearly 
all the photometrically selected objects with the 
DEep Imaging Multi-Object Spectrograph (DEIMOS; Faber et al.\ 2003)
on the Keck~II 10~m telescope, which we also describe in
Section~\ref{secnarrow}.

The resulting photometric and spectroscopic samples form a large, 
consistent data set with which to tackle the cosmological problems.
We provide spectra for 88 $z\sim5.7$ and 30 $z\sim6.5$ Ly$\alpha$ 
emitting galaxies. This forms the largest sample of confirmed
high-redshift galaxies in the very distant universe. We give the 
catalogs, thumbnail images, and spectra in the Appendix.

In Section~\ref{secdisc} we discuss the relative
properties of the samples at $z=5.7$ and $z=6.5$.
The emission-line properties provided by the spectra are key 
diagnostics of the intergalactic gas properties at these redshifts. 
They may be used to probe the neutral fraction of the
surrounding intergalactic medium (IGM; e.g., Gnedin \& Prada\ 2004;
Santos 2004; Zheng et al.\ 2010) and thus to study the ionization state 
of the IGM and its redshift evolution at early times.  Theoretical
models make predictions of the impact of the neutral portion of the 
IGM on the line profile of Ly$\alpha$ emission and also upon the redshift
evolution of the Ly$\alpha$ luminosity function (e.g., Barkana \&
Loeb 2001; Mesinger \& Furlanetto 2004; Dijkstra et al.\ 2007;
Dijkstra \& Wyithe 2010). In Section~\ref{secdisc} we analyze
the line width distributions at $z=5.7$ and $z=6.5$, showing
that there are only small changes in the properties of the
lines over this range and that the line properties are remarkably
similar for most of the objects. We find that there is a dependence
of line width on Ly$\alpha$ luminosity, with the more luminous
objects being broader. We show that the luminosity function (LF)
drops by a factor of two at $z=6.5$ relative to that at $z=5.7$,
and the $z=5.7$ value is a factor of four lower than
that at $z=3.1$. However, the characteristic luminosity is unchanged.
The results are for the observed luminosities, and the fall-off 
will be less for the intrinsic LFs when
the effects of intergalactic scattering are allowed for.  

In a subsequent paper we will combine the optical data 
with longer wavelength observations from the {\em Hubble Space 
Telescope (HST)\/} and discuss the UV continuum properties of 
the high-redshift sample and the Ly$\alpha$ escape fraction.

We assume $\Omega_M=0.3$, $\Omega_\Lambda=0.7$, and
$H_0=70$~km~s$^{-1}$~Mpc$^{-1}$ throughout.
All magnitudes are given in the AB magnitude system
(Oke et al.\ 1983, 1990),
where an AB magnitude is defined by
$m_{AB}=-2.5\log f_\nu - 48.60$.
Here $f_\nu$ is the flux of the source in units of
erg~cm$^{-2}$~s$^{-1}$~Hz$^{-1}$.

\section{Narrowband Selection}
\label{secnarrow}

\subsection{Observed Fields}
\label{secfield}

\begin{figure*}
\includegraphics[width=2.9in,angle=90,scale=0.9]{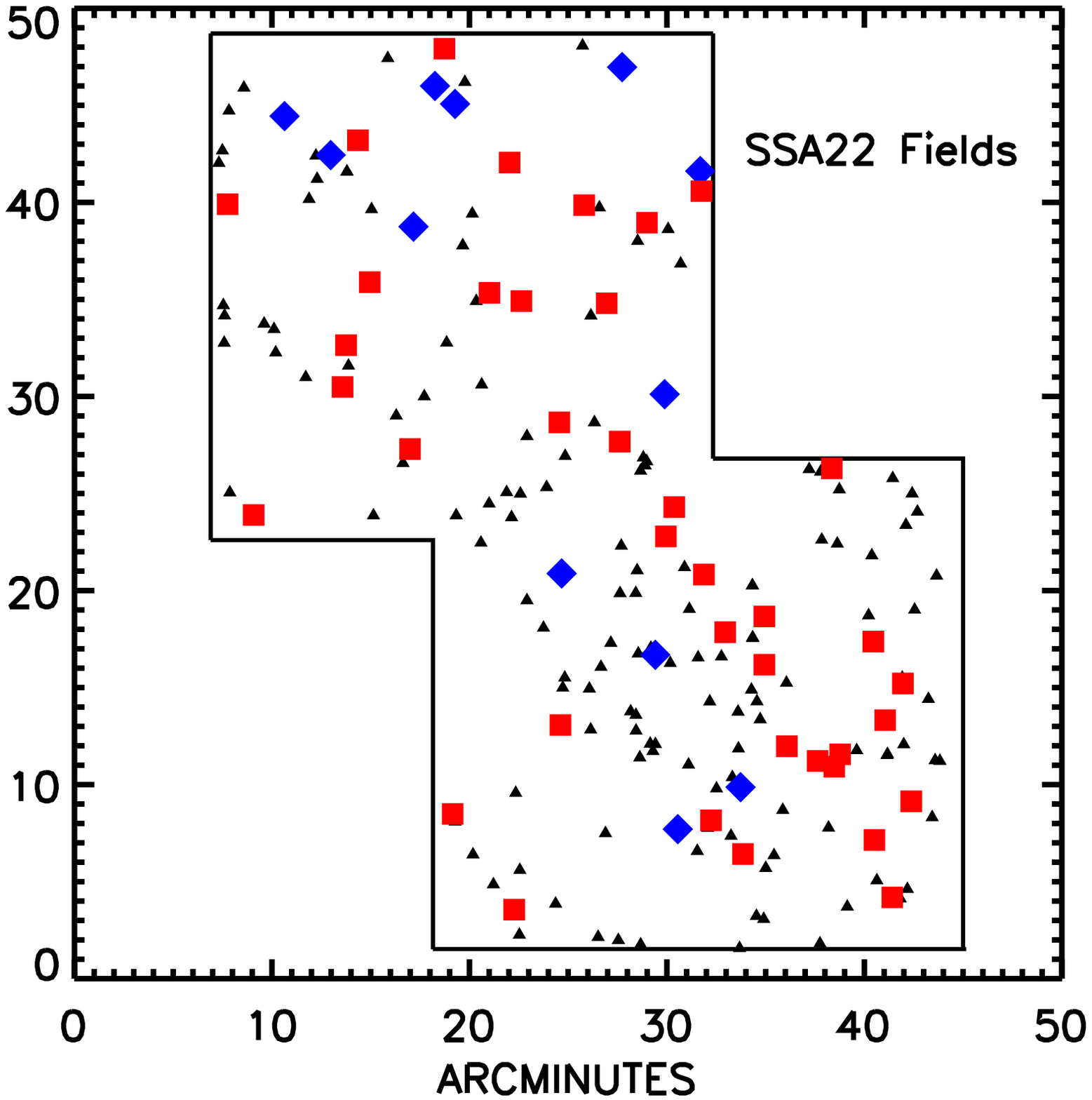}
\includegraphics[width=2.9in,angle=90,scale=0.9]{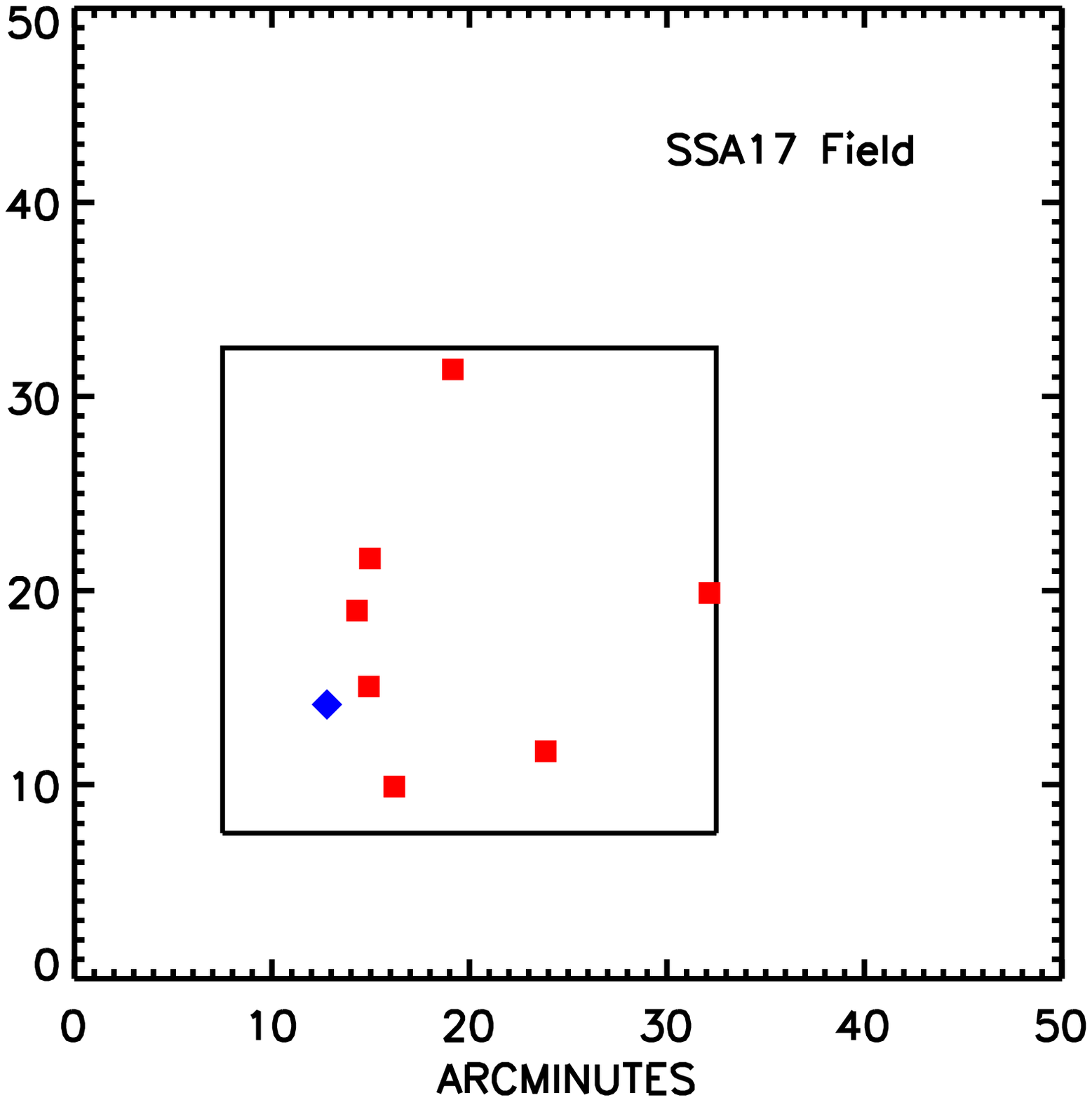}
\includegraphics[width=2.9in,angle=90,scale=0.9]{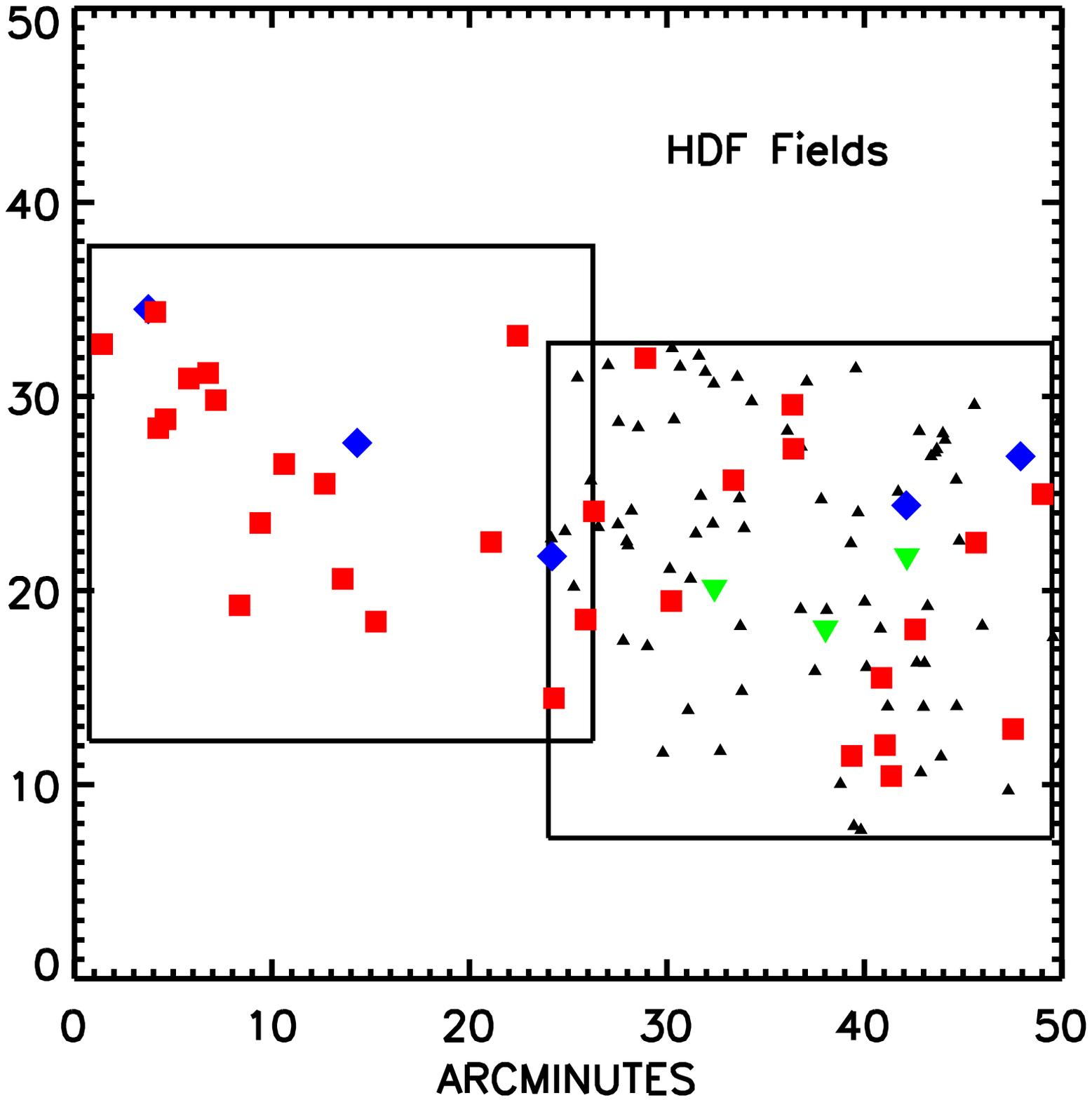}
\includegraphics[width=2.9in,angle=90,scale=0.9]{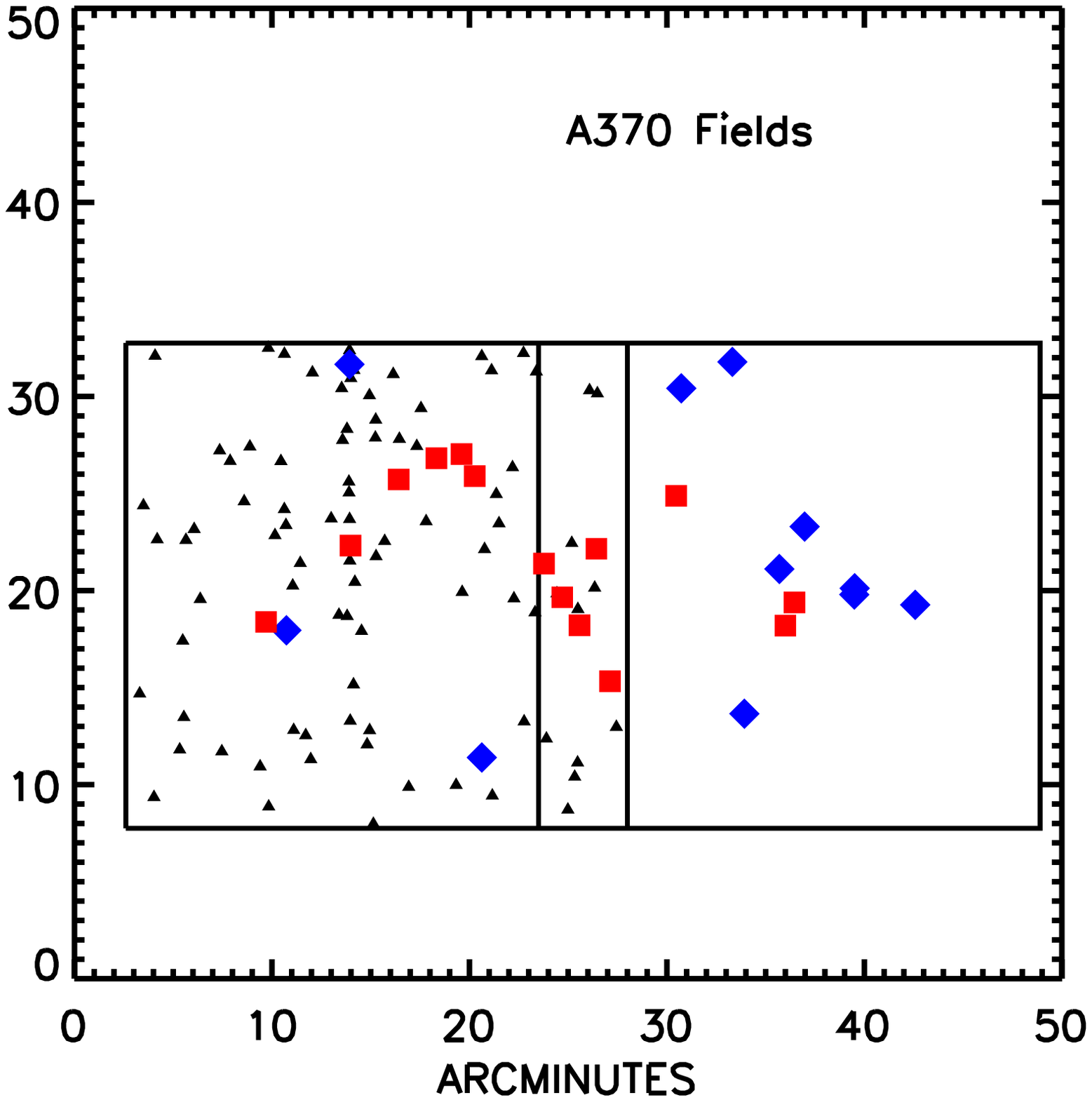}
\caption{The geometric configurations of the seven SuprimeCam
fields. Six of the fields are grouped into pairs with small
amounts of overlap, while one is isolated (SSA17). 
The solid lines outline the observed areas. The red squares 
show the detected Ly$\alpha$ emitters at $z\sim5.7$, the blue 
diamonds show those at $z\sim6.5$, and the green downward pointing 
triangles show sources found using the 9210~\AA\ filter (GOODS-N only). 
Four of the fields were used in the Kakazu et al.\ (2007) survey of 
$z<1.6$ ultra-strong emission-line galaxies. These
objects are denoted by black small triangles.
\label{fieldout}
}
\end{figure*}

The present survey covers seven SuprimeCam fields. We summarize these
in Table~\ref{tbl-1}, where we give the name of the field in column~1, 
the J2000 right ascension and declination of the field centers in 
columns~2 and 3, the galactic longtitude and latitude in
column~4, the galactic extinction to the field
in column~5, and the exposure times in hours through
the NB816 and NB912 filters in columns~6 and 7.
The field geometry is shown in Figure~\ref{fieldout}. 
Six of the fields are grouped into three neighboring and slightly
overlapping pairs to allow a study of clustering. The seventh
field (SSA17) is isolated. The very central region
of the A370 field is lensed by the foreground massive cluster A370 at
$z=0.37$.

\begin{figure}
\includegraphics[width=2.9in,angle=90,scale=0.9]{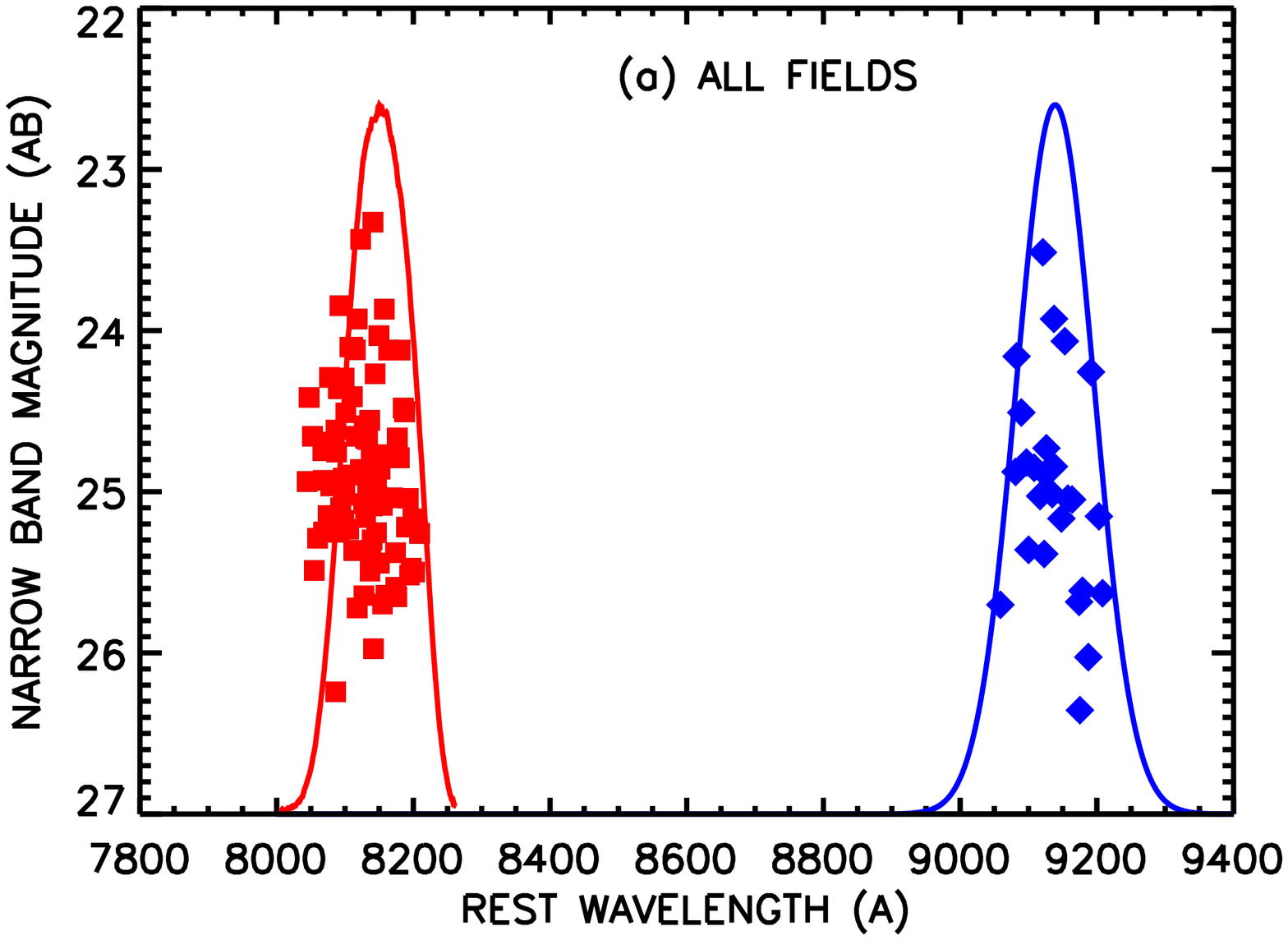}
\includegraphics[width=2.9in,angle=90,scale=0.9]{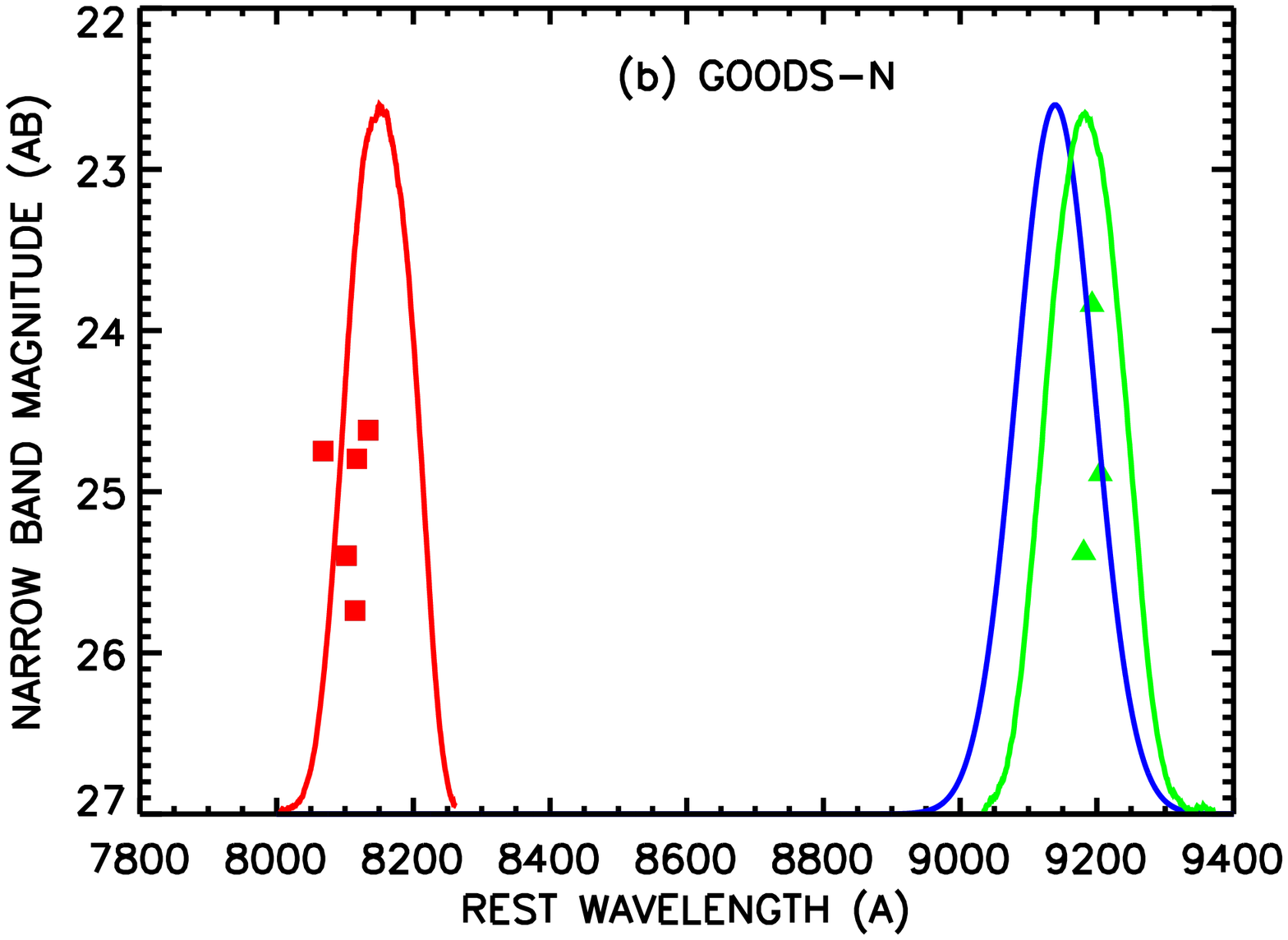}
\caption{(a) The transmission profiles versus rest-frame
wavelength for the two primary filters (NB816/8150~\AA\
in red and NB912/9140~\AA\ in blue). We also show the wavelength positions
of all the spectroscopically confirmed $z\sim5.7$ Ly$\alpha$ emitters
(red squares) and $z\sim6.5$ Ly$\alpha$ emitters (blue diamonds) in the 
fields. (b) The transmission profiles from (a) plus the transmission 
profile of the 9210~\AA\ filter (green) used in the GOODS-N region 
of the HDF-N field. 
We also show the wavelength positions of all the spectroscopically 
confirmed $z\sim6.5$ Ly$\alpha$ emitters found with this filter 
(green triangles; 3 objects), along with the spectroscopically 
confirmed $z\sim5.7$ Ly$\alpha$ emitters found with the NB816 filter 
(red squares; 5 objects) in this field. No $z\sim6.5$ Ly$\alpha$ 
emitters were found in the GOODS-N with the NB912 filter.
\label{filtshape}
}
\end{figure}

We obtained the narrowband images of each field with SuprimeCam
under photometric or near-photometric conditions.  
We observed each field with a 120~\AA\ (FWHM) filter (NB816) 
centered at a nominal wavelength of 8150~\AA\ corresponding to 
a $z\sim5.7$ selection and a 120~\AA\ (FWHM) filter (NB912) 
centered at a nominal wavelength of 9140~\AA\ corresponding to 
a $z\sim6.5$ selection.  Both lie in regions of low sky background 
between the OH bands. (The nominal specifications for the Subaru 
filters may be found at
\url{www.naoj.org/Observing/Instruments/SCam/sensitivity.html}
and are described in Ajiki et al.\ 2003.)\ We show the location and 
shape of both filter profiles in Figure~\ref{filtshape}(a).
We also show the wavelength positions of all the spectroscopically
confirmed $z\sim5.7$ Ly$\alpha$ emitters (red squares) and
$z\sim6.5$ Ly$\alpha$ emitters (blue diamonds) in all seven fields.
The Gaussian shape of the SuprimeCam filter profiles
may be compared with the more square profile of the 
108~\AA\ filter centered at 8185~\AA\ that has been
used in the Low-Resolution Imaging Spectrograph (LRIS; Oke et al.\ 1995) 
parallel beam on Keck~I (Hu et al.\ 1999). The complex shape
of the SuprimeCam filter profiles requires careful treatments 
of the conversion of narrowband magnitudes to line fluxes and
of the determination of the accessible volume
(e.g., Gronwall et al.\ 2007).
We carry this out in Section~\ref{secdisc}. 

The NB816 filter is well centered on the
Cousins $I$-band filter, which we use as the reference
continuum bandpass for the $z\sim5.7$ selection.
The NB912 filter is well centered on the $z$-band, which
we use as the reference continuum bandpass for the $z\sim6.5$ selection.
For the Great Observatories Origins Deep Survey-North 
(GOODS-N; Giavalisco et al.\ 2004) region of the Hubble Deep Field-North
(HDF-N), we also selected objects using a 120~\AA\ filter 
centered at 9210~\AA. 
We show this filter profile (green) in Figure~\ref{filtshape}(b), 
along with the other two filter profiles from Figure~\ref{filtshape}(a).
We show the wavelength positions of all the spectroscopically
confirmed $z\sim 6.5$ Ly$\alpha$ emitters found with this filter
(green triangles; 3 objects), along with the positions of the 
spectroscopically confirmed $z\sim 5.7$ Ly$\alpha$ emitters 
found with the NB816 filter in this field (red squares; 5 objects). 
No $z\sim6.5$ Ly$\alpha$ emitters were found in the GOODS-N with the 
NB912 filter.

Table~\ref{tbl-1} summarizes the observations made in the primary bands,
including giving the total photometric exposure time in hours for each 
of the filters in each of the fields.  We obtained $\sim4-5$ hour 
exposures for the NB816 filter and $\sim8-10$ hour exposures for the 
NB912 filter. The longer exposures in the NB912 filter partially compensate for 
the lower camera throughput at this wavelength, so the observations 
provide comparable depth exposures in the two bands. However, the
NB912 images are still slightly shallower than those in the NB816 band. 
We took the data as a sequence of dithered 
background-limited exposures with alternate sequences rotated by
90~deg. We always obtained the corresponding continuum exposures
in the same observing run as the narrowband
exposures to avoid falsely identifying transients---such as 
high-redshift supernovae or Kuiper belt objects---as Ly$\alpha$
candidates. Capak et al.\ (2004) gives a detailed description of the 
full reduction procedure that we used to process the images.
We calibrated the SuprimeCam data using photometric and
spectrophotometric standard stars (Turnshek et al.\ 1990; Oke et al.\ 1990)
and faint Landolt standard stars (Landolt et al.\ 1992). 
We obtained an astrometric solution using stars from the USNO survey.
The final narrowband selected samples were drawn from the more uniformly 
covered central $25'\times 25'$ region of each of the fields. Allowing 
for overlaps, the combined area of the 7 fields
in the survey is 4168~arcmin$^2$. We note that the present calibration
of the SSA22 field is about 0.2~mag fainter than that given in H04. 
This is typical of the errors introduced
in the calibrations and in the choice of method for measuring
the magnitudes. Thus, we adopt it as an estimate of the
systematic error in the absolute measurements. 
The relative magnitudes are extremely well
determined in the field compared to the absolute calibration,
so the relative counts and LFs in the NB912 and 
NB816 bands are insensitive to this issue.  However, it does 
affect the normalization of the LF.
The FWHM seeing on the final reduced images ranges
from $0.5''$ to $1''$.
The typical limiting magnitudes of the images 
($5 \sigma$ for the corrected $3''$ diamteer aperture mags, 
see Section~\ref{secselect}) 
expressed as AB magnitudes are 26.9 ($B$), 26.8 ($V$), 26.6 ($R$), 
25.6 ($I$), 25.4 ($z$), 25.3 (NB816), and 25.2 (NB912), though the 
exact values vary slightly depending on the exposure time and the 
observing conditions.

\subsection{Photometric Candidate Selection}
\label{secselect}

We used the SExtractor package (Bertin \& Arnouts 1996) to generate
catalogs of objects in each of the fields.  We measured all of the 
magnitudes in $3''$ diameter apertures and applied average aperture
corrections to obtain total magnitudes. The typical
correction is just under 0.2 mags. (Hereafter, we refer to these
as corrected $3''$ diameter aperture magnitudes.) Throughout the
paper a negative sign in front of a magnitude means that the
flux in the aperture was negative. The numerical value of the
magnitude then corresponds to the absolute value of the flux.
We follow this procedure so that the tables may be used to
properly average fluxes including negative values.

\begin{figure*}
\centerline{
\includegraphics[width=5.0in,angle=0,scale=0.9]{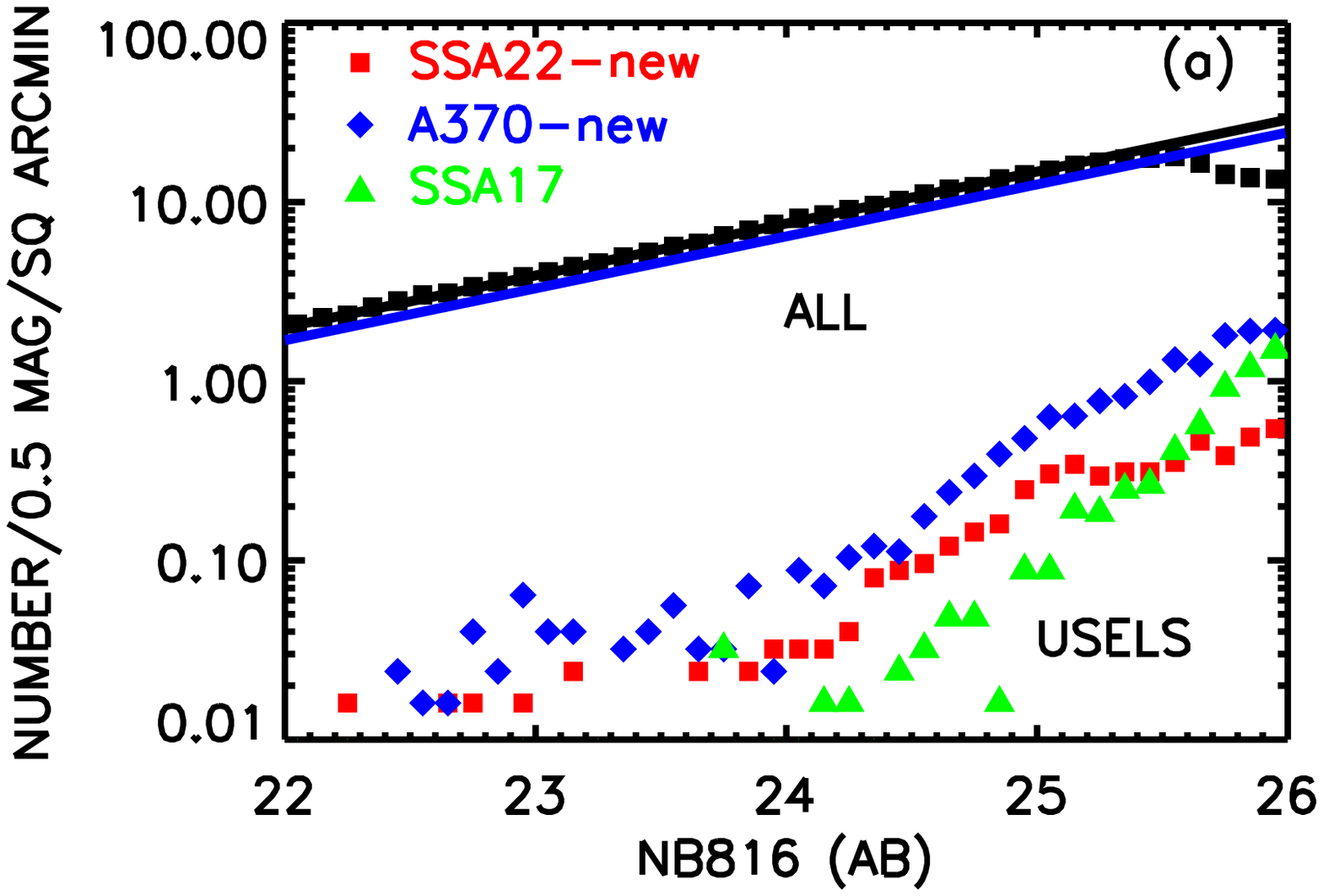}
}
\centerline{
\includegraphics[width=5.0in,angle=0,scale=0.9]{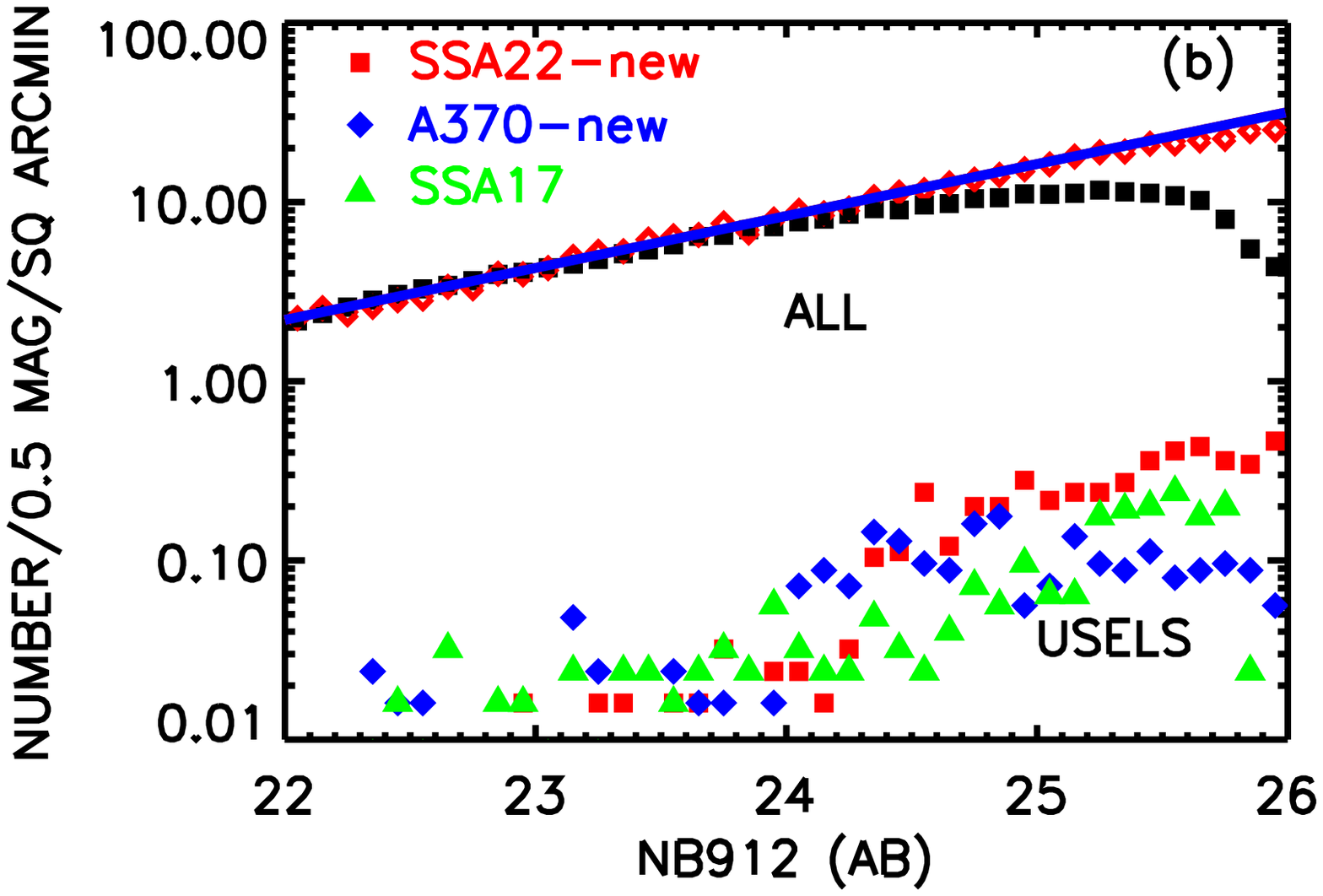}
}
\caption{Measured number counts of narrowband selected objects 
in the (a) NB816 and (b) NB912 bands (black squares). A power-law fit 
to the NB816 number counts [black line in (a)] may be compared with a
power-law fit to the incompleteness corrected number 
counts measured in this band by Ouchi et al.\ (2008) [blue line in (a)].
The NB912 number counts may be compared with the expected incompleteness
corrected counts in this band determined from the {\em HST\/} ACS
GOODS F850LP selected number counts [red open diamonds in (b)], which 
we fitted with a power law and then renormalized to match the bright-end 
counts in the NB912 band [blue line in (b)]. For three of the fields we 
also show the number density of USEL candidates (see figure legends).
\label{ncounts}
}
\end{figure*}

In Figure~\ref{ncounts} we show our measured number counts in the 
(a) NB816 and (b) NB912 bands (black squares).  We have plotted 
the $1\sigma$ error bars, but they are generally smaller than
the symbol size. These counts are averaged over all fields, with 
the exception of the A370 field, where lensing effects from the massive 
cluster may be important. In (a) we show a power-law fit to the NB816
counts (black line), which we compare with a power-law fit 
to the incompleteness corrected NB816 counts of Ouchi et al.\ (2008;
their Figure~11) (blue line). The shapes of the counts are in
extremely good agreement, but the normalization is slightly
higher in our counts. This corresponds to our magnitudes being 
about 0.2~mag brighter, on average, than those of Ouchi et al.\ (2008).
This again suggests that 0.2~mag is the level of uncertainty arising 
from the calibrations and from the magnitude measurements. All of 
our fields give consistent counts, and all show the same offset.
By comparing the actual counts with the power-law fit, we
see that our sample is highly complete to magnitudes 
just above NB816~$=25.5$.

We do not have a previous comparison for the NB912 sample,
so in Figure~\ref{ncounts}(b) we compare with the F850LP selected 
number counts from the {\em HST\/} ACS observations of
the GOODS fields (Giavalisco et al.\ 2004) (red open diamonds).
The F850LP filter has a very similar color response to the
$z$ filter, which we use as the continuum for the NB912 filter.
Thus, it should provide a good approximation to the NB912 counts.
Indeed, the {\em HST\/} counts agree closely with the NB912 counts.
We used these much deeper counts to determine the form of the 
power-law fit, and then we renormalized the fit to match the 
bright-end counts in the NB912 band. We show this fit as the 
blue line in Figure~\ref{ncounts}(b).
We see that the NB912 counts become progressively incomplete
above NB912~$\sim25$. We use the ratio of this power-law fit 
to the observed counts to calculate an incompleteness correction
as a function of the NB912 magnitude. For the NB816 band we 
compute the incompleteness correction using the power-law fit to
Ouchi et al.\ (2008)'s incompleteness corrected counts. 
At NB912~$=25.5$ the correction is a multiplicative factor of 2.3, 
while at NB816~$=26.0$ the correction is 2.0.

For each field we formed a sample of galaxies with narrowband
magnitudes $N_{\rm AB}<25.5$ satisfying either the criterion 
$(I-N)_{\rm AB}>0.8$ in the NB816 band or
$(z-N)_{\rm AB}>0.9$ in the NB912 band. Following 
Kakazu et al.\ (2007), we refer to these objects as 
ultra-strong emission-line galaxies or USELs.
Our narrowband excess selection criteria are slightly lower 
than those used by Ouchi et al.\ (2008) [$(i'-$NB816)$>1.2$]
or by Taniguchi et al.\ (2005) [$(z-$NB921)$>1$]. These
groups wanted to choose more securely genuine emission-line galaxies 
in their photometric samples, while we seek 
completeness and rely on the subsequent spectroscopic observations 
to eliminate false objects where the narrowband excess is not 
produced by an emission line lying in the filter bandpass.

\begin{figure*}
\centerline{
\includegraphics[width=4.0in,angle=0,scale=0.9]{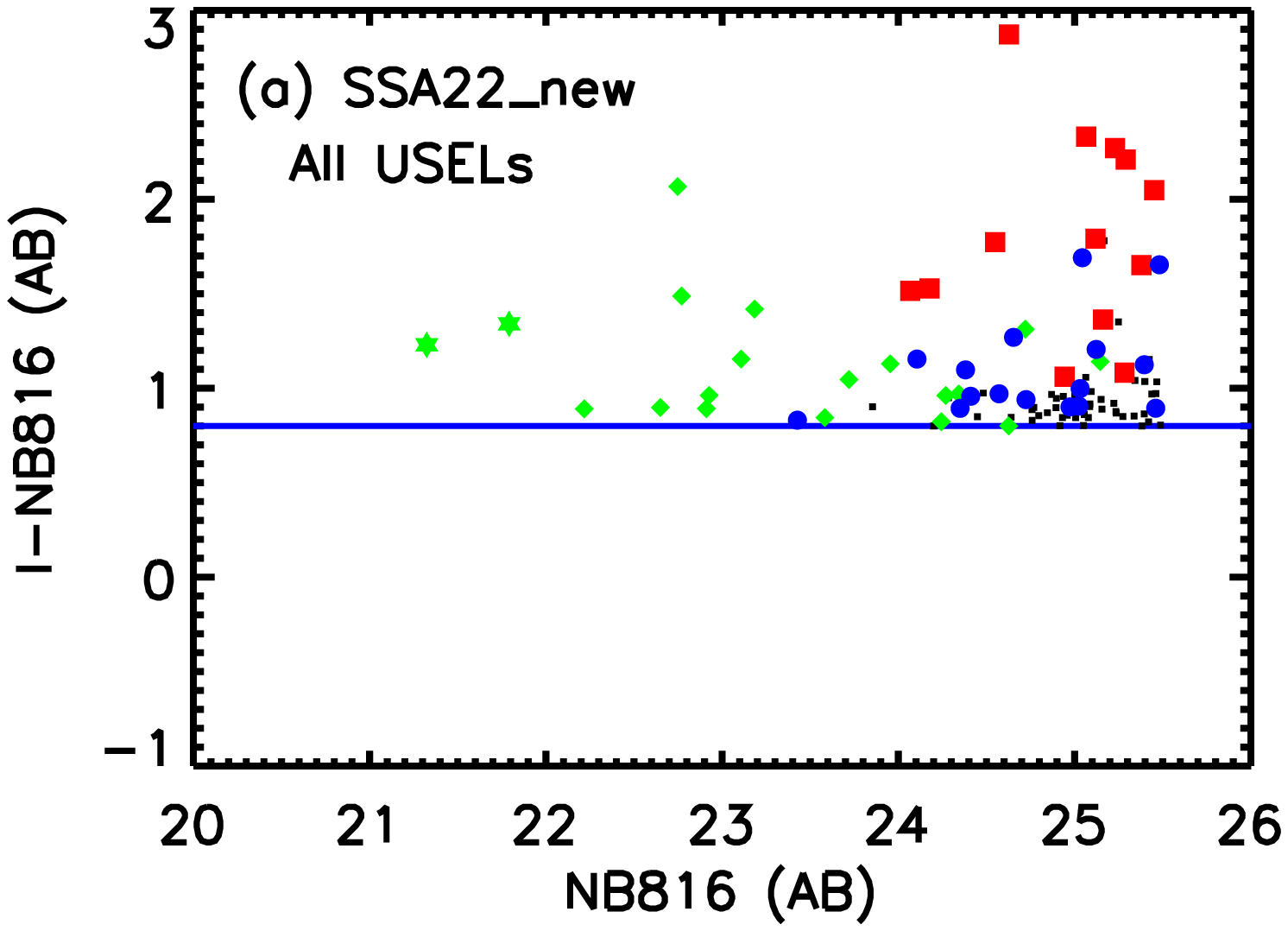}
\includegraphics[width=4.0in,angle=0,scale=0.9]{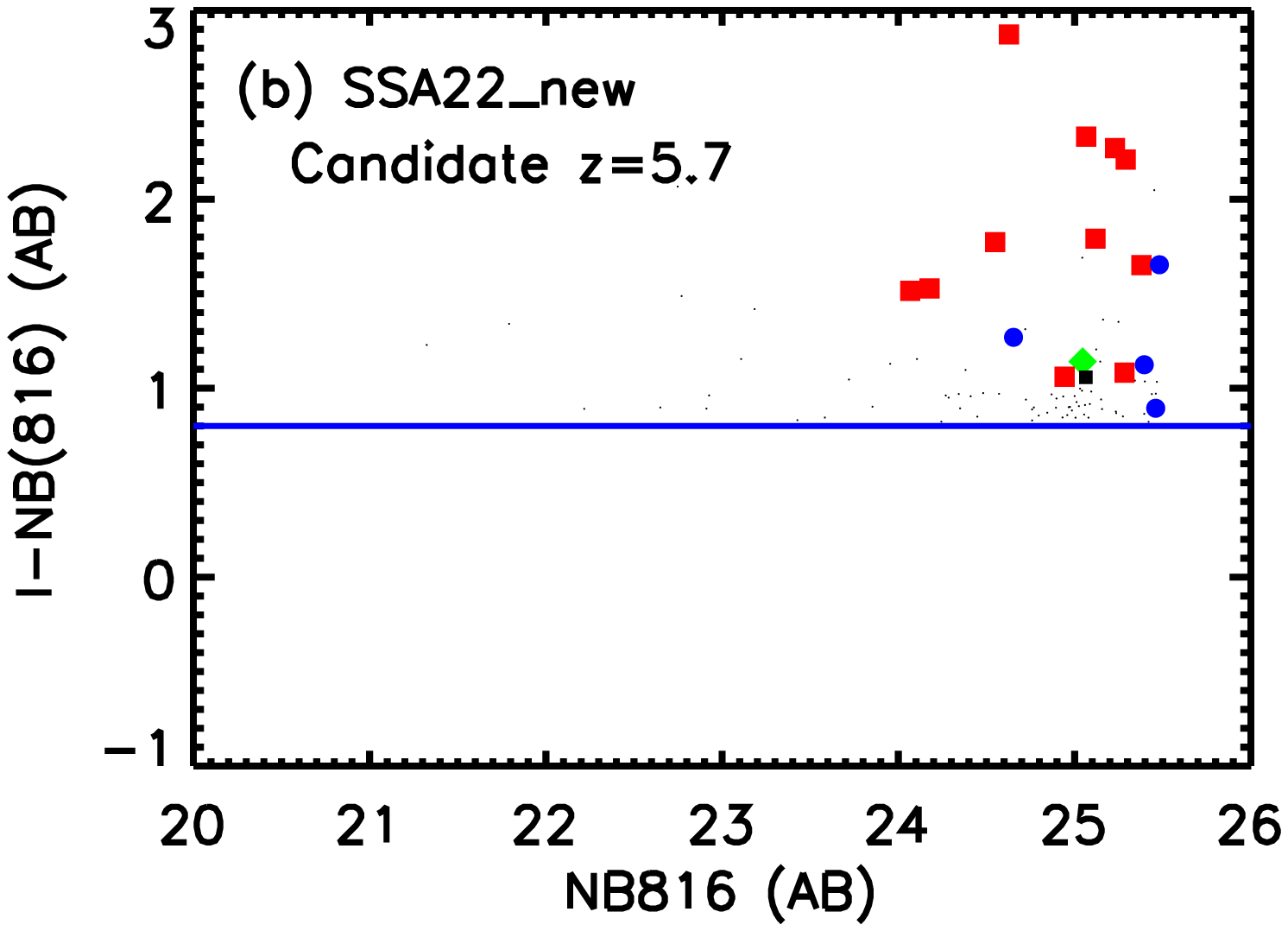}
}
\centerline{
\includegraphics[width=4.0in,angle=0,scale=0.9]{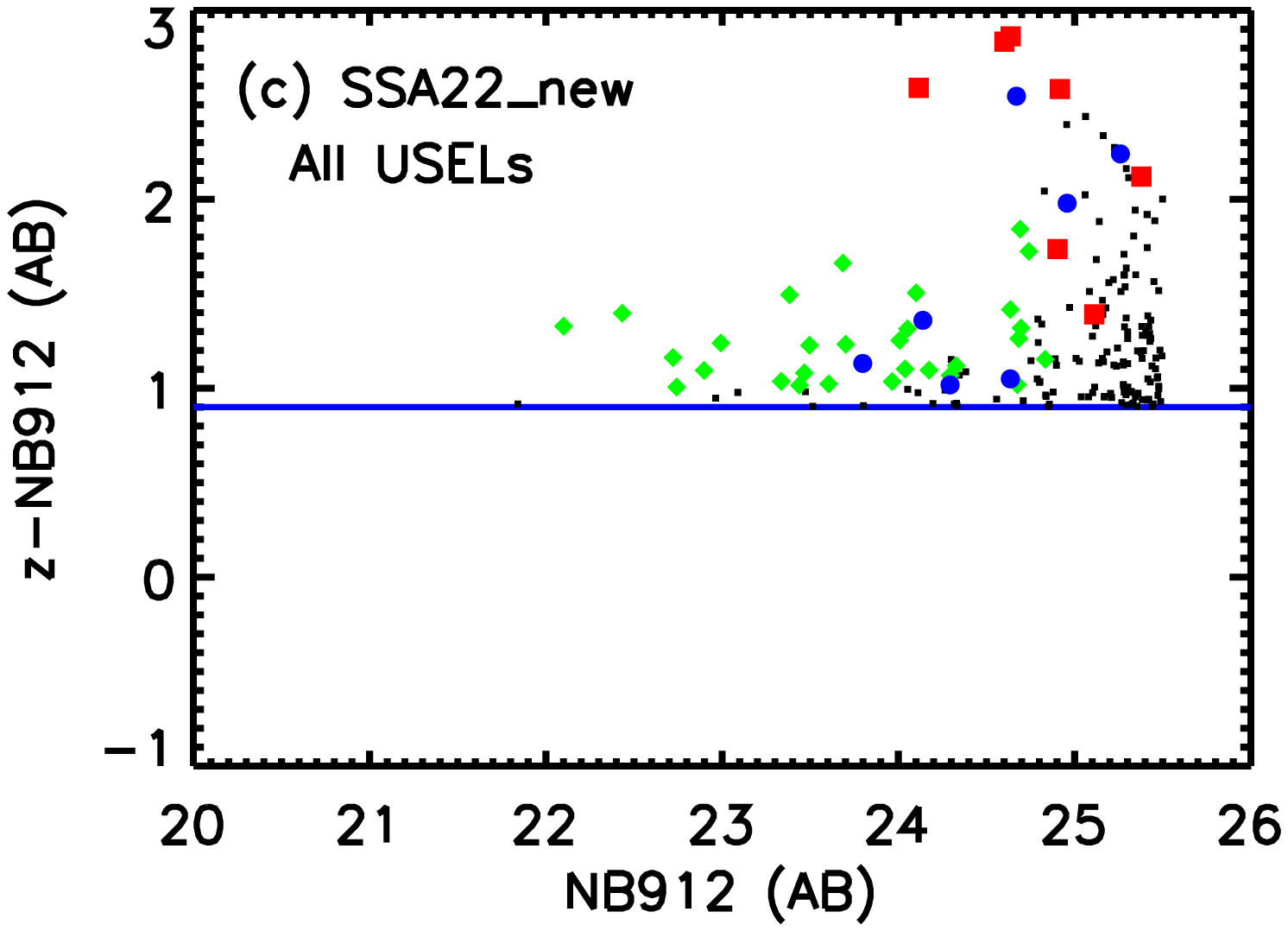}
\includegraphics[width=4.0in,angle=0,scale=0.9]{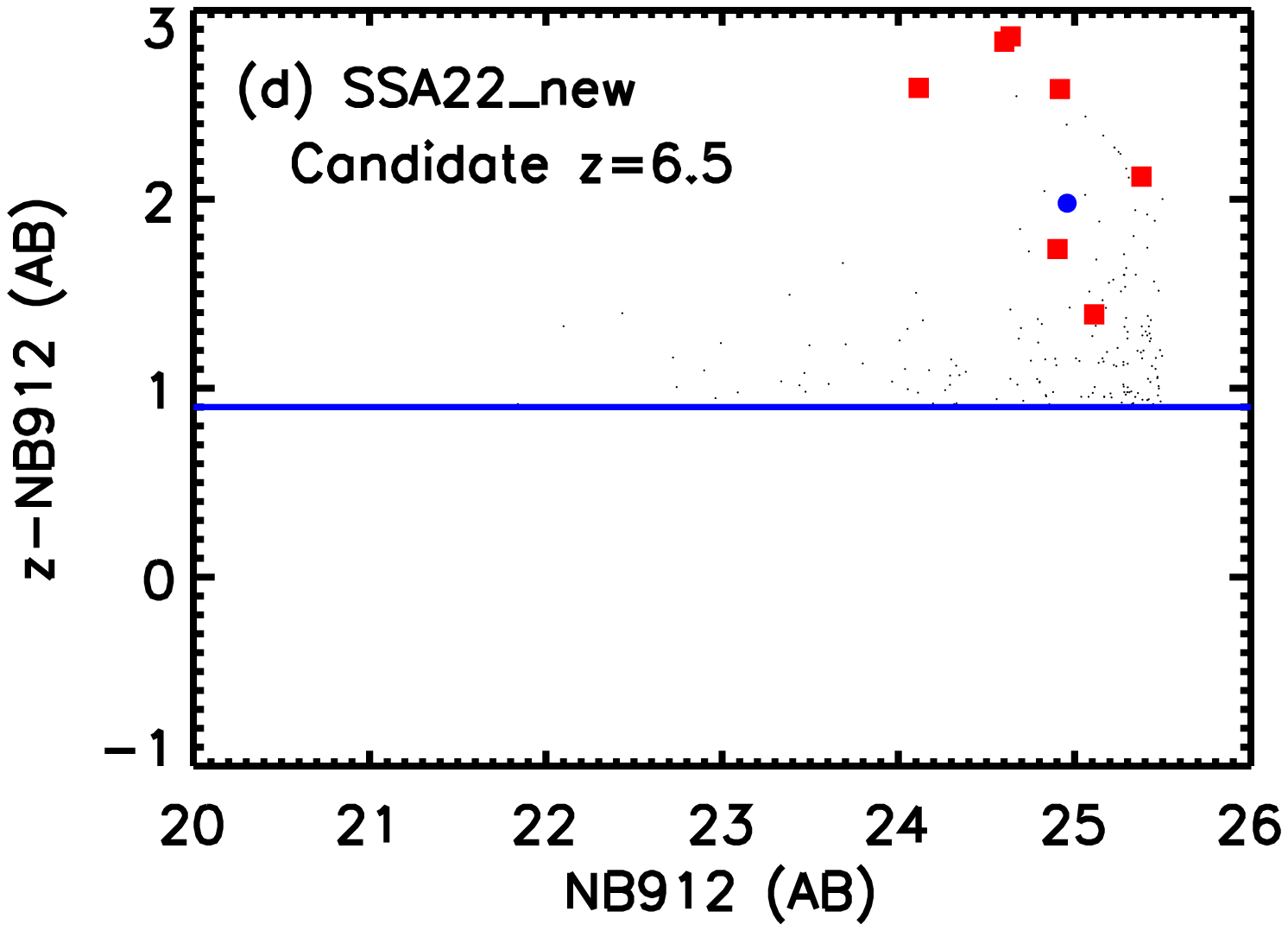}
}
\caption{Emission-line excess objects found
using (a) the NB816 filter in the SSA22$\_$new
field and (c) the NB912 filter in the SSA22$\_$new field
vs. $N_{\rm AB}$ magnitude. The small symbols show the entire sample
of $N_{\rm AB}<25.5$ galaxies in the field. The blue horizontal line
shows the narrowband excess selection of (a) $(I-N)_{\rm AB}>0.8$
for NB816 and (c) $(z-N)_{\rm AB}>0.9$ for NB912.
Objects satisfying these criteria are shown with large symbols:
green diamonds---spectroscopically classified low-redshift emitters ($z<1.6$);
green stars---spectroscopically classified stars; red squares---spectroscopically
confirmed $z\sim5.7$ emitters or $z\sim6.5$ emitters; 
blue circles---spectroscopically observed but unidentified objects;
black squares---objects with no spectroscopic measurements.
(b) and (d) are similar to (a) and (c) but the large symbols are restricted
to either objects with a strong break between the $R$- and $z$-bands
($I$- and $z$-bands) for NB816 (NB912) or objects which are undetected 
at the $2\sigma$ level in the $R$-band ($I$-band) for NB816 (NB912)
and which are not detected in the $B$- or $V$- (or $R$-) bands.
[Because of space constraints only points satisfying the narrowband
excess selection are shown in this version of the paper]
\label{line_excess}
}
\end{figure*}

We visually inspected each USEL candidate to
eliminate artifacts, and we also visually searched the images
for USELs that might have been missed in the initial catalog 
because they were blended with neighboring objects. 
In Figure~\ref{line_excess} we show examples of the final USEL 
selection. In (a) we show $(I-N)_{\rm AB}$ 
versus $N_{\rm AB}$ for the NB816 
selected objects in the SSA22$\_$new field, 
and in (c) we show $(z-N)_{\rm AB}$ versus $N_{\rm AB}$
for the NB912 selected objects in the SSA22$\_$new field.
The small symbols show the entire sample of $N_{\rm AB}<25.5$
galaxies in the field.  The blue horizontal line shows
the narrowband color selection.  Objects satisfying these
criteria are shown with large symbols.
As can be seen from the figures, this sample will 
include a number of objects at the faint end that have scattered 
into the selection region. Thus, as mentioned above, with this approach 
we rely on the subsequent spectroscopic observations to eliminate 
any spurious objects. 
For the NB816 selected sample in the SSA22$\_$new field
we find 101 objects with $(I-N)_{\rm AB}>0.8$ over the 
$25'\times 25'$ field, which is about 0.3\% of the 36816
objects included in the initial $N_{\rm AB}<25.5$ sample.
For the NB912 selected sample in this field we find 183 USEL
candidates, which
is about 0.5\% of the 38120 objects.  We show the number density
of USEL candidates versus narrowband magnitude for three of the individual 
fields in Figure~\ref{ncounts}. All the fields have similar
number densities of USEL candidates, which rise rapidly to fainter magnitudes.

We next eliminated USELs whose continuum colors rule them
out as candidate $z\sim5.7$ or $z\sim6.5$ galaxies. 
For $z=5.7$ we restricted either to objects with $(R-z)_{\rm AB}>1.5$ 
or to objects which were not detected at the $2\sigma$
level in the $R$-band. We also required that the objects not
be detected above the $2\sigma$ level in the $B$- and $V$-bands.
For $z\sim6.5$ we restricted either to objects with $(I-z)_{\rm AB}>1.5$
or to objects which were not detected at the $2\sigma$ level in the
$I$-band. We also required that the objects not be detected above the 
$2\sigma$ level in the $B$-, $V$- and $R$-bands. The results from 
these restrictions on the USEL sample are shown in 
Figures~\ref{line_excess}(b) and (d) with the large symbols. 
We call these objects $z=5.7$ and $z=6.5$
candidate Ly$\alpha$ emitters, and we made an effort to obtain
spectra of all of them. Typically, there are
$\sim20$ candidates in each field at $z\sim5.7$
and $\sim10$ at $z\sim6.5$. In four of
the fields (A370\_new, HDF-N, SSA22, and SSA22\_new; see
Figure~\ref{fieldout})
we obtained spectra of all of the USELs with $N_{\rm AB}<24$
and many of the USELs with $N_{\rm AB}=24-25$,
regardless of their colors. These results are described in 
Kakazu et al.\ (2007) and Hu et al.\ (2009).

As can be seen from Figures~\ref{line_excess}(b) and 
\ref{line_excess}(d),
the number of candidates is not particularly sensitive
to the choice of narrowband excess. If we had used the Ouchi
et al.\ (2008) cut of 1.2 in Figure~\ref{line_excess}(b), then
it would have eliminated 6 of the 17 candidate $z=5.7$ galaxies.
Five of the eliminated candidates have been spectroscopically observed:
two are spurious, one is a low-redshift emitter, 
and two are genuine $z=5.7$ emitters. Thus, using the higher
cut slightly improves 
the accuracy of the selection but at the expense of losing some 
$z=5.7$ emitters. If we had used the Taniguchi et al.\ (2005) cut 
of 1.0 in Figure~\ref{line_excess}(d), then it would not have 
changed the candidate selection at all.

\subsection{Spectroscopic Confirmation}
\label{secspec}

 Spectroscopic followup of the candidates is essential
to rule out contaminants in the photometric selection.
These can include lower redshift
emission line galaxies, red galaxies or stars, transients
and artifacts in the data. The spectroscy also provides
precise redshifts which allow us to make an accurate determination
of the line fluxes as well as providing us with the
shapes and widths of the Ly$\alpha$ lines.

We used the DEIMOS spectrograph on the Keck~II 10~m 
telescope for our spectroscopic follow-up of the candidate 
$z\sim 5.7$ and $z\sim 6.5$ Ly$\alpha$ emitters. For most of the 
observations we used the G830~$\ell$/mm grating blazed at 
8640~\AA\ with $1''$ wide slitlets because
of its high resolution and excellent red sensitivity. 
In a small number of cases we 
used the slightly lower resolution G600~$\ell$/mm grating.
Both configurations have sufficient resolution 
(e.g., 3.3~\AA\ with G830~$\ell$/mm) 
to distinguish the $z\sim1.19$ \oii\ doublet structure from the 
profile of redshifted Ly$\alpha$ emission (see Figure~2 of 
H04).  Redshifted \oiii\
emitters ($z\sim0.62$) show up frequently as emission-line objects in
the narrowband and can easily be identified by the doublet signature.
The observed wavelength coverage is $\sim 3840$~\AA\ 
with a typical
range of $\sim5900-9700$~\AA. It generally encompasses redshifted
H$\beta$ and \oiii\ lines in cases where the detected emission line
might be H$\alpha$ at $z\sim 0.24$. This is particularly useful for 
dealing with the problematic instance of extragalactic \ion{H}{2} regions 
with strongly suppressed \nii. The G830~$\ell$/mm grating used with 
the OG550~$\ell$/mm blocker gives a throughput
greater than 20\% for most of this range and $\sim28\%$ at 8150~\AA.

We made the observations during a number of runs in the $2005-2010$ period.
Just over 90\% of the candidates were observed in each of the bands.
We apply a spectroscopic incompleteness correction to allow for
the missing fraction in the analysis of the LFs in Section~\ref{secdisc}.
Exposure times ranged from one to six hours for each object.

We filled the DEIMOS masks with color-selected and magnitude-selected 
samples, which will be described elsewhere.
All of the spectra (emission-line objects and field objects) 
were spectroscopically classified without reference to either their 
narrowband strengths or their color
properties to avoid any subjective biasing of the line interpretation.
We classified each candidate object as either a confirmed
high-redshift emitter, a low-redshift emitter, a red star,
or a false emission-line object (i.e., when the observed narrowband 
excess does not appear to correspond to an emission line). 
For each high-redshift emitter the redshift is measured at the peak of 
the emission line. Slightly more than half of the $z=5.7$ candidate 
Ly$\alpha$ emitters were spectroscopically confirmed. The fraction of 
confirmation for the $z=6.5$ candidate Ly$\alpha$ emitters is higher, 
but there is still a substantial degree of contamination.

\subsection{GOODS-N}
\label{secgood}

We summarize the results for objects in the GOODS-N 
in Tables~\ref{goods816} (NB816) and \ref{goods921} (NB921).  
The GOODS-N comprises about a quarter of the HDF-N field and has the
advantage of having extremely deep broadband images from the
{\em HST\/} ACS imaging (Giavalisco et al.\ 2004).  In addition,
Ajiki et al.\ (2006) carried out an independent analysis using
the NB816 data obtained in the present program in combination
with the ACS data to identify candidate $z=5.7$ Ly$\alpha$ emitters
in the field. Since the Ajiki et al.\ analysis is based on their 
reductions of the SuprimeCam images and uses different calibrations 
and magnitude measurements, it provides an excellent check on the 
present work.

We identified 12 candidate $z=5.7$ Ly$\alpha$ emitters in the 
NB816 band in the GOODS-N.  We summarize their properties
in Table~\ref{goods816}, where we give the object number in column~1, 
the R.A.(J2000) and Decl.(J2000) in decimal degrees in columns~2 and 3, 
the NB816 magnitude in column~4, the $I-$NB816 color in column~5, 
the SExtractor auto magnitudes for the four ACS bandpasses taken from 
the ACS catalogs in columns~6-9, and the measured redshift in column~10. 
While the selection of the present
candidates was made solely using the ground-based observations,
all of the objects are contained in the ACS catalogs, and all
have colors in the ACS observations that are consistent
with their being $z\sim5.7$ galaxies.

All of the objects were spectroscopically observed
with exposures ranging from 1 to 6 hours.  Where an
object was spectroscopically confirmed as a $z\sim5.7$ emitter, 
we give the redshift of the source in Table~\ref{goods816}.  
Of the 12 candidates, we identified 
6 as Ly$\alpha$ emitters. Some of the remaining
cases may be genuine emitters that we have failed to
identify with the spectra.  However, for many of the unidentified
objects, we have obtained multiple repeated spectra and have
failed to find an emission line. One object appears
to be a portion of a larger galaxy, where the photometry may be 
contaminated.  Some of the remaining objects could be red stars 
or objects at high redshift whose continuum break is just below 
the narrowband filter, simulating an emission-line object.
Other cases may simply be spurious detections.
This is representative of the fields in general. 

Ajiki et al.\ (2006) used a more restricted area of the GOODS-N
and found 10 candidate $z\sim5.7$ Ly$\alpha$ emitters. 
Eight of these overlap with the present sample. Their narrowband
magnitudes show an average of $-0.1$~mag offsets and a spread
of up to 0.3~mag relative to the present values, probably
reflecting the different apertures used. We spectroscopically
observed one of the two objects from the Ajiki et al.\ 
sample that did not overlap with ours, but we did not confirm it 
as a $z=5.7$ Ly$\alpha$ emitter. 
All of the confirmed $z=5.7$ Ly$\alpha$ emitters in the area used 
by Ajiki et al.\ are common to the two samples. In both samples
we have confirmed spectroscopically half of the candidate
emitters.

We did not find any candidate $z\sim6.5$ Ly$\alpha$ emitters
in the NB912 band in the GOODS-N.  However, a deep 12.7~hr exposure 
that we obtained with a 120~\AA\ narrowband filter at 9210~\AA\ 
yielded four candidates.  We summarize their properties
in Table~\ref{goods921}. We spectroscopically observed three of 
the four candidates and confirmed all of them as $z\sim6.5$ Ly$\alpha$ 
emitters.  Their redshifts place them at wavelengths where the NB912
filter is becoming insensitive (see Figure~\ref{filtshape}),
so they are not picked out in the NB912 observations.
One of the objects is extremely luminous (HC123725+621227)
and has a very high S/N spectrum (see
Figure~\ref{figA6:z6_spectra_plus}). This object is very faint
in the continuum, and it is not detected in the GOODS-N ACS 
catalogs. In contrast to the NB816 sample, where all of the 
objects are detected in the ACS F850LP filter, only two of the 
four objects are present in the ACS F850LP catalog.  This 
reflects the fading in the magnitude produced by the continuum 
break at the Ly$\alpha$ line, which lies near the middle of 
the F850LP bandpass.

\begin{deluxetable*}{cccccccccc}
\renewcommand\baselinestretch{1.0}
\tablewidth{0pt}
\tablecaption{GOODS-N: NB816 Selected $z\sim 5.7$ Ly$\alpha$ Emitter Candidates}
\scriptsize
\tablehead{Number & R.A. & Decl. & $N_{\rm AB}$ & $(I-N)_{\rm AB}$ & F850LP & F775W & F606W & F435W &  Redshift  \\  & (J2000) & (J2000) &  & & &  &  &  &  \\ (1) & (2) & (3) & (4)  & (5) & (6) & (7) & (8) & (9) & (10)}
\startdata
       1  &  189.215240  &  62.32683  &  $ 24.51$  &  $ 1.049$  &  $ 26.89$  &
$ 27.66$  &  $ 29.16$  &  \nodata  &    5.675  \cr
       2  &  189.216751  &  62.36460  &  $ 24.60$  &  $ 2.032$  &  $ 26.37$  &
$ 27.21$  &  \nodata  &  $ 28.56$  &    5.689  \cr
       3  &  189.056106  &  62.12994  &  $ 24.68$  &  $ 1.133$  &  $ 25.99$  &
$ 26.61$  &  $ 28.35$  &  $ 28.49$  &    5.635  \cr
       4  &  189.254120  &  62.35397  &  $ 24.86$  &  $0.8972$  &  $ 26.32$  &
$ 27.59$  &  $ 29.22$  &  $ 34.04$  &  \nodata  \cr
       5  &  189.399750  &  62.23944  &  $ 24.88$  &  $ 1.642$  &  $ 26.06$  &
$ 26.56$  &  $ 28.90$  &  $ 27.67$  &  \nodata  \cr
       6  &  189.324677  &  62.29974  &  $ 25.06$  &  $ 2.342$  &  $ 26.52$  &
$ 27.30$  &  $ 28.45$  &  \nodata  &    5.663  \cr
       7  &  189.033004  &  62.14394  &  $ 25.08$  &  $ 1.853$  &  $ 26.36$  &
$ 26.57$  &  $ 29.44$  &  \nodata  &    5.640  \cr
       8  &  189.456543  &  62.22942  &  $ 25.18$  &  $0.8409$  &  $ 24.56$  &
$ 25.93$  &  $ 29.52$  &  \nodata  &  \nodata  \cr
       9  &  189.342285  &  62.26277  &  $ 25.43$  &  $ 1.387$  &  $ 25.55$  &
$ 26.70$  &  $ 31.33$  &  \nodata  &  \nodata  \cr
      10  &  189.045471  &  62.17144  &  $ 25.49$  &  $ 2.116$  &  $ 26.54$  &
$ 27.13$  &  \nodata  &  $ 30.56$  &    5.673  \cr
      11  &  189.366013  &  62.19613  &  $ 25.51$  &  $ 2.557$  &  $ 26.72$  &
$ 27.26$  &  \nodata  &  \nodata  &  \nodata  \cr
      12  &  189.320419  &  62.23344  &  $ 25.67$  &  $ 1.265$  &  $ 25.60$  &
$ 26.47$  &  \nodata  &  $ 29.81$  &  \nodata  \cr
\enddata
\label{goods816}
\end{deluxetable*}

\begin{deluxetable*}{cccccccccc}
\renewcommand\baselinestretch{1.0}
\tablewidth{0pt}
\tablecaption{GOODS-N: NB921 Selected $z\sim 6.5$ Ly$\alpha$ Emitter Candidates}
\scriptsize
\tablehead{Number & R.A. & Decl. & $N_{\rm AB}$ & $(z-N)_{\rm AB}$ & F850LP & F775W & F606W & F435W &  Redshift  \\  & (J2000) & (J2000) &  & & &  &  &  &  \\ (1) & (2) & (3) & (4)  & (5) & (6) & (7) & (8) & (9) & (10)}
\startdata
       1  &  189.358170  &  62.20769  &  $ 23.68$  &  $ 9.472$  &  \nodata  &
\nodata  &  \nodata  &  \nodata  &    6.559  \cr
       2  &  189.356873  &  62.29541  &  $ 24.36$  &  $ 2.016$  &  $ 26.38$  &
\nodata  &  \nodata  &  \nodata  &  no obs  \cr
       3  &  189.093689  &  62.23458  &  $ 25.30$  &  $ 3.476$  &  $ 27.16$  &
$ 31.87$  &  \nodata  &  \nodata  &    6.560  \cr
       4  &  189.157135  &  62.17277  &  $ 25.86$  &  $ 1.092$  &  \nodata  &
\nodata  &  \nodata  &  \nodata  &    6.546  \cr
\enddata
\label{goods921}
\end{deluxetable*}


\subsection{Atlas of the $z=5.7$ and $z=6.5$ Ly$\alpha$ Emitters}
\label{secatlas}

Our final spectroscopically confirmed sample consists of
87 $z=5.7$ Ly$\alpha$ emitters found with the NB816 filter, 
27 $z=6.5$ Ly$\alpha$ emitters found with the NB912 filter, 
and 3 $z=6.5$ Ly$\alpha$ emitters found with the 9210~\AA\ filter 
in the GOODS-N only.  We include in these figures the small number 
of spectroscopically identified, high-redshift Ly$\alpha$ emitters 
that are slightly fainter than our $N_{\rm AB}<25.5$ selection limit 
but were contained in our spectroscopic observations.
We summarize all of the spectroscopically confirmed Ly$\alpha$
emitters in Tables~\ref{la_table}, \ref{high_la_tab}, 
and \ref{very_high_la.tab} in the Appendix, where we give the
object number in column~1, the object name in column~2, the 
R.A.(J2000) and Decl.(J2000) in decimal degrees in columns~3 
and 4, the narrowband magnitude in the selection filter in
column~5, the corresponding continuum magnitude in column~6, 
the redshift in column~7, the exposure time in hours in column~8, 
the quality flag (1=secure, 2=clear emission line but redshift 
may be more questionable, 3=weak emission line) in column~9, 
the FWHM of the line and its $1\sigma$ error in \AA\ in column~10,
and the logarithm of the luminosity in the line in 
erg~s$^{-1}$ in column~11. (Column 11 is not included for the
small sample in Table~\ref{very_high_la.tab}.) The calculation of the line
fluxes and luminosities is described in section 3.3.  We
show the finding charts for the $z\sim5.7$ Ly$\alpha$ emitters
in Figure~\ref{figA1:z5_images}, for the $z\sim6.5$ Ly$\alpha$ emitters 
in Figure~\ref{figA3:z6_images}, and for the 9210~\AA\ selected
$z\sim6.5$ Ly$\alpha$ emitters in Figure~\ref{figA5:z6_images}, 
and we show their corresponding spectra in Figures~\ref{figA2:z5_spectra}, 
\ref{figA4:z6_spectra}, and \ref{figA6:z6_spectra_plus}.

\section{Discussion}
\label{secdisc}

The escape of Ly$\alpha$ light from high-redshift galaxies
is determined by two processes: the escape
from the galaxy itself, and the subsequent propagation through
the neighboring IGM. Although the scattering process in the IGM is 
inherently conservative of the Ly$\alpha$ photons, both processes
involve the loss of light from the observed emitter. 

In the case of the escape from the galaxy, the well-known
random walk process---which causes the photons to diffuse in
frequency and finally allows them to leave---will extend the escape 
path and combine with any extinction in the galaxy to
destroy Ly$\alpha$ photons. However, the level of destruction is
dependent on the exact escape route, which depends in turn
on the structure of the interstellar medium 
(e.g., Neufeld 1991; Finkelstein et al.\ 2007).
The shape and width of the final Ly$\alpha$ spectrum will also depend
on the escape process, and observations and modeling of low-redshift 
galaxies (e.g., Kunth et al.\ 2003; Schaerer \& Verhamme 2008) suggest 
that there will be considerable variation in the output Ly$\alpha$ line.

The subsequent propagation of the Ly$\alpha$ line through the
IGM also reduces the strength of the emitter
and modifies its shape. The blue side of the Ly$\alpha$
line scatters on the neutral hydrogen in the IGM.
These photons will ultimately be rescattered to form
an extended halo of Ly$\alpha$ around the object 
(Loeb \& Rybicki 1999),
but in practice these halos are too faint to observe, and the
process may simply be viewed as a loss of light from the
emitting galaxy. The net effect is to truncate the shorter wavelength
light in the line and hence leave a red component, but the exact
effect and the fraction of the line which will finally be
observed depends on numerous modeling parameters, such
as the infall velocity of gas to the galaxy, the density
profile, the peculiar velocity of the galaxy, and whether
there is enhanced ionization around the galaxy from ionizing
photons from the galaxy or its neighbors (Haiman \& Cen 2005).
Zheng et al.\ (2010) have recently presented extensive modeling
within the context of a detailed numerical simulation that
produces redshifted Ly$\alpha$ lines with sharp blue cut-offs
that are very similar to those observed. In addition, they
show that there is a wide range of observed Ly$\alpha$ 
luminosities relative to the intrinsic luminosity. However, 
Zheng et al.\ use a very simplified galaxy Ly$\alpha$ profile, 
and the variation of these profiles may also play an important 
role in the process (e.g., Dijkstra \& Wyithe 2010). 
In particular, redder and wider galaxy Ly$\alpha$ profiles will 
be more likely to produce observable Ly$\alpha$ emission lines 
after the subsequent IGM propagation. We refer to this as 
pre-stretching before the shortening imposed by the IGM.

These processes will determine both the output shape of
the line and the distribution of line widths; the latter
may provide one of the strongest constraints on the modeling.
In Section~\ref{secwid} we show that the shapes of the lines
are remarkably invariant and that they span a fairly narrow range 
in width. We might also expect that there would be a progressive
reduction in the fraction of galaxies having strong Ly$\alpha$
as we move to higher neutral fractions in the IGM at higher
redshifts. We show in Section~\ref{secnum} that this is
not the case and that the number of Ly$\alpha$ emitters 
falls more slowly as we move from $z=5.7$ to $z=6.5$ 
than the number of UV-continuum selected galaxies does. 
However, there is some weak
evidence that the lines are becoming narrower and have
slightly smaller equivalent widths at $z=6.5$ than they
have at $z=5.7$.

\subsection{Spectral Shapes and the Distribution of Line Widths}
\label{secwid}

\begin{figure*}
\centerline{
\includegraphics[width=4.0in,angle=90,scale=0.95]{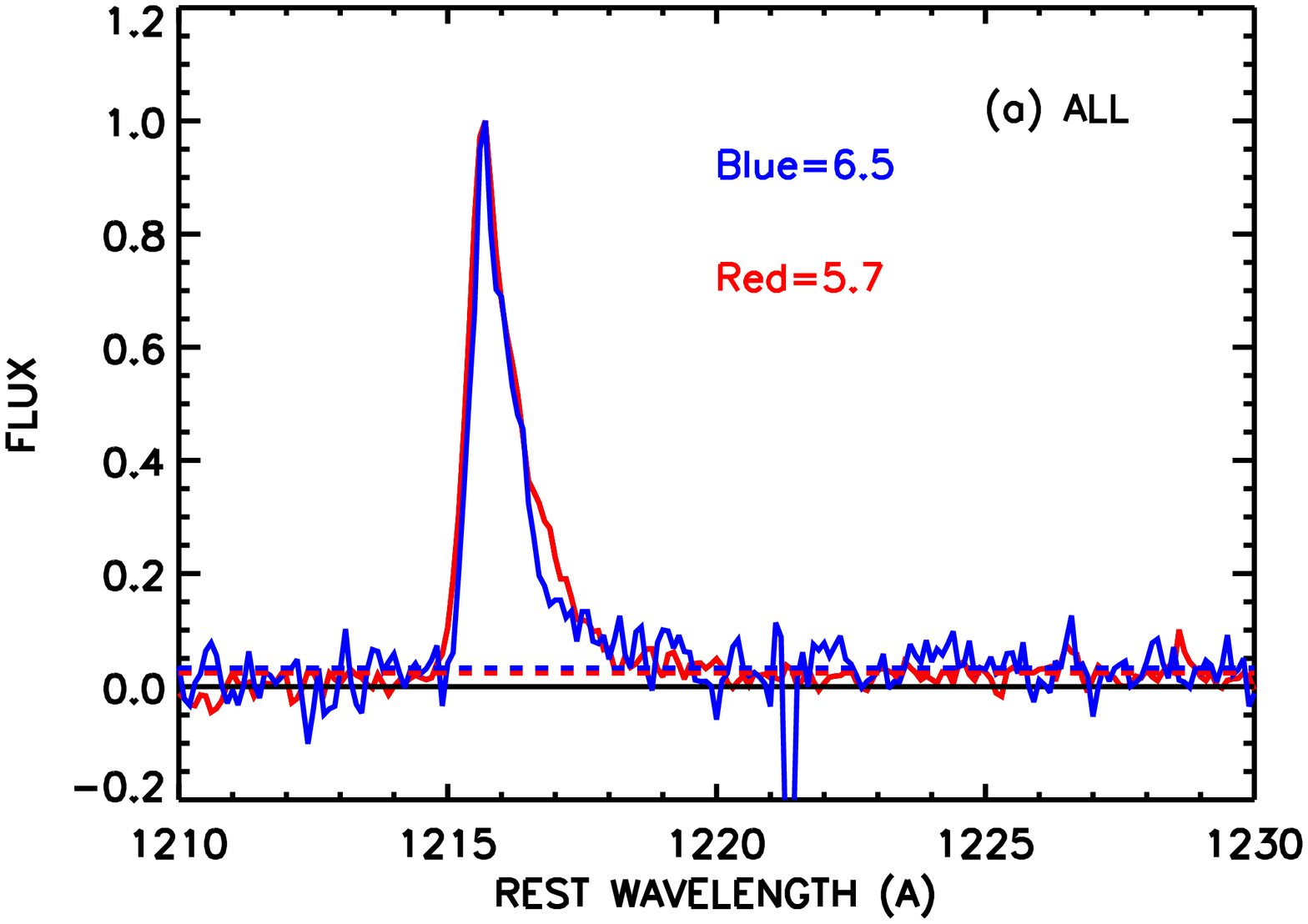}
}
\centerline{
\includegraphics[width=4.0in,angle=90,scale=0.95]{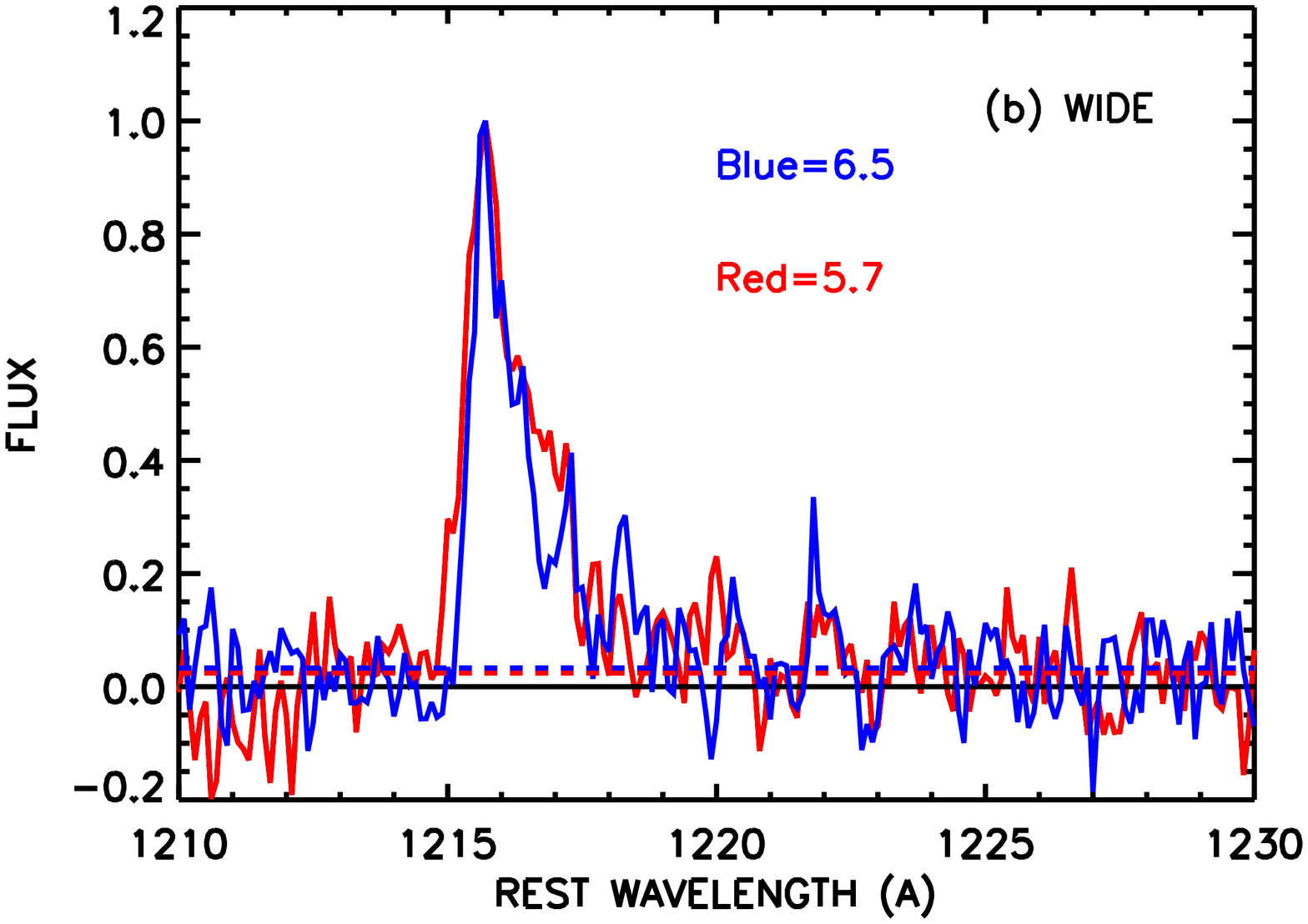}
}
\caption{(a) Comparison of the averaged spectra made from all the 
``quality one'' spectra in the NB816 ($z=5.7$; red line) and NB912 
($z=6.5$; blue line) samples. These were formed by normalizing each 
individual spectrum's Ly$\alpha$ peak to one and then averaging the 
normalized spectra. In each case we show the level of the continuum 
measured redward of the Ly$\alpha$ line with the dashed line of 
the same color. (b) Comparison of the averaged spectra made from only 
the FWHM$>1.6$~\AA\ objects in the two samples. These wider spectra 
have a more developed red elbow.
\label{compare_spectra}
}
\end{figure*}

The Ly$\alpha$ lines presented in this paper and in previous work are 
surprisingly uniform in their properties. In Figure~\ref{compare_spectra}(a) 
we compare the averaged spectra at $z=5.7$ and $z=6.5$. These were 
formed by normalizing each ``quality one'' individual spectrum to make the 
maximum value of the Ly$\alpha$ line be one and then averaging the spectra.
As can be seen, the line profiles at both redshifts are nearly identical.
They also have some broad general properties: a fairly sharp cut-off at the 
short wavelength side, a narrower peak, an elbow (by which
we mean the slight plateau at wavelengths redward of the
peak and at fluxes of about 0.3-0.5 of
the maximum and which is most clearly seen in 
the wider spectra), and then a trailing
long-wavelength edge. They may be compared with Figure~15 of Hu
et al.\ (2004) for the $z=5.7$ emitters and Figure~7 of Kashikawa
et al.\ (2006) for the $z=6.5$ emitters, though the lower resolution
spectra in Kashikawa et al.\ do not show the blue-side cut-off so clearly.

As noted in Section~\ref{secatlas},
in the Appendix we show figures of all the individual spectra. In
each case we overplot the averaged spectrum at the same redshift 
(red dashed line). The similarity of the individual spectra to the averaged 
spectrum is remarkable. However, there are some slight differences. For
example, when we separate out the wider spectra at the two redshifts using 
the directly measured FWHM from the individual spectra, we find that the 
wider spectra have a more developed long-wavelength elbow. 
This can be seen in Figure~\ref{compare_spectra}(b), where we have
made the averaged spectra at the two redshifts only from sources whose 
line widths are greater than $1.6$~\AA\ in the rest frame. We hereafter 
refer to these as wide averaged spectra.

\begin{figure*}
\centerline{
\includegraphics[width=2.7in,angle=90,scale=0.95]{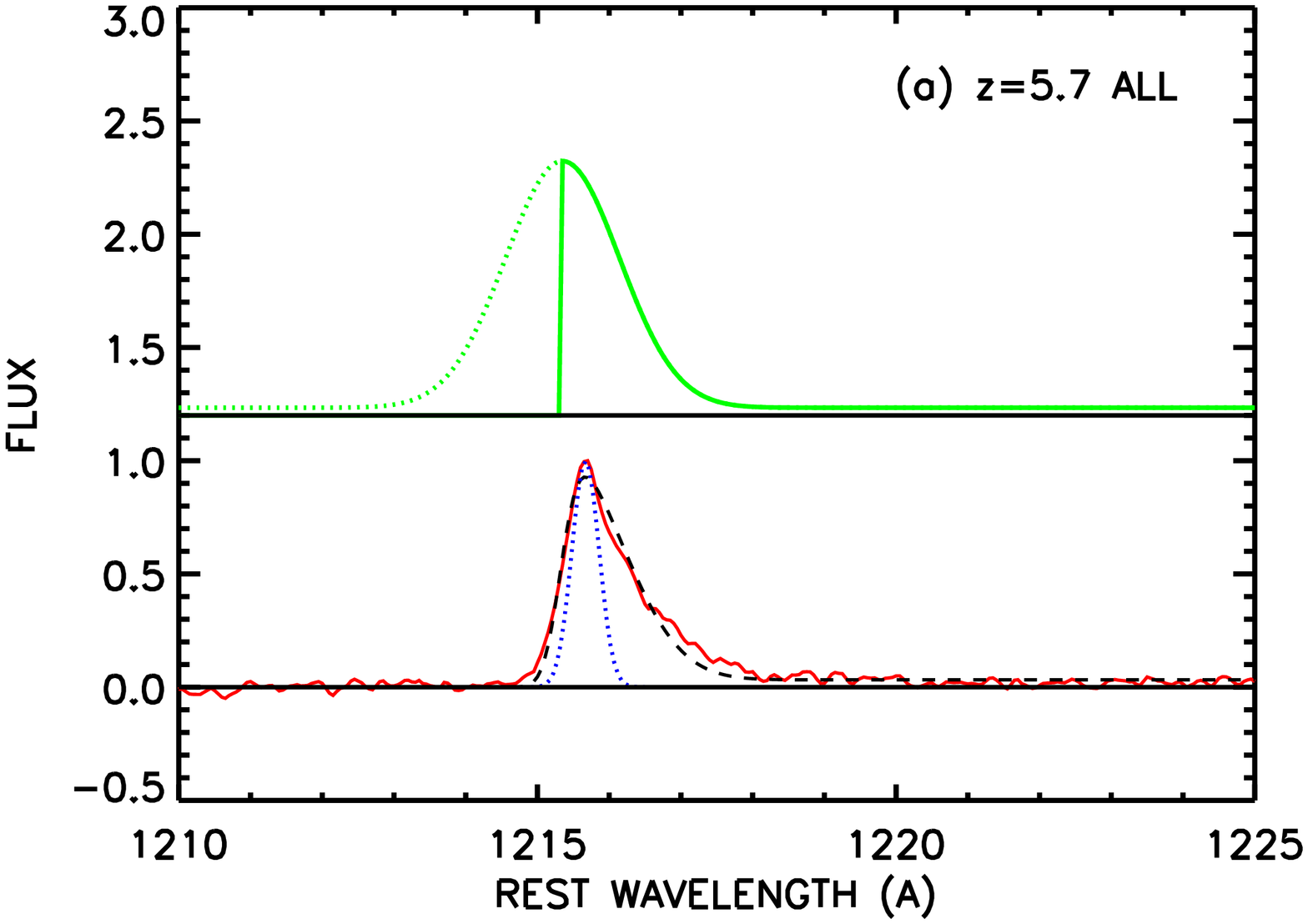}
\includegraphics[width=2.7in,angle=90,scale=0.95]{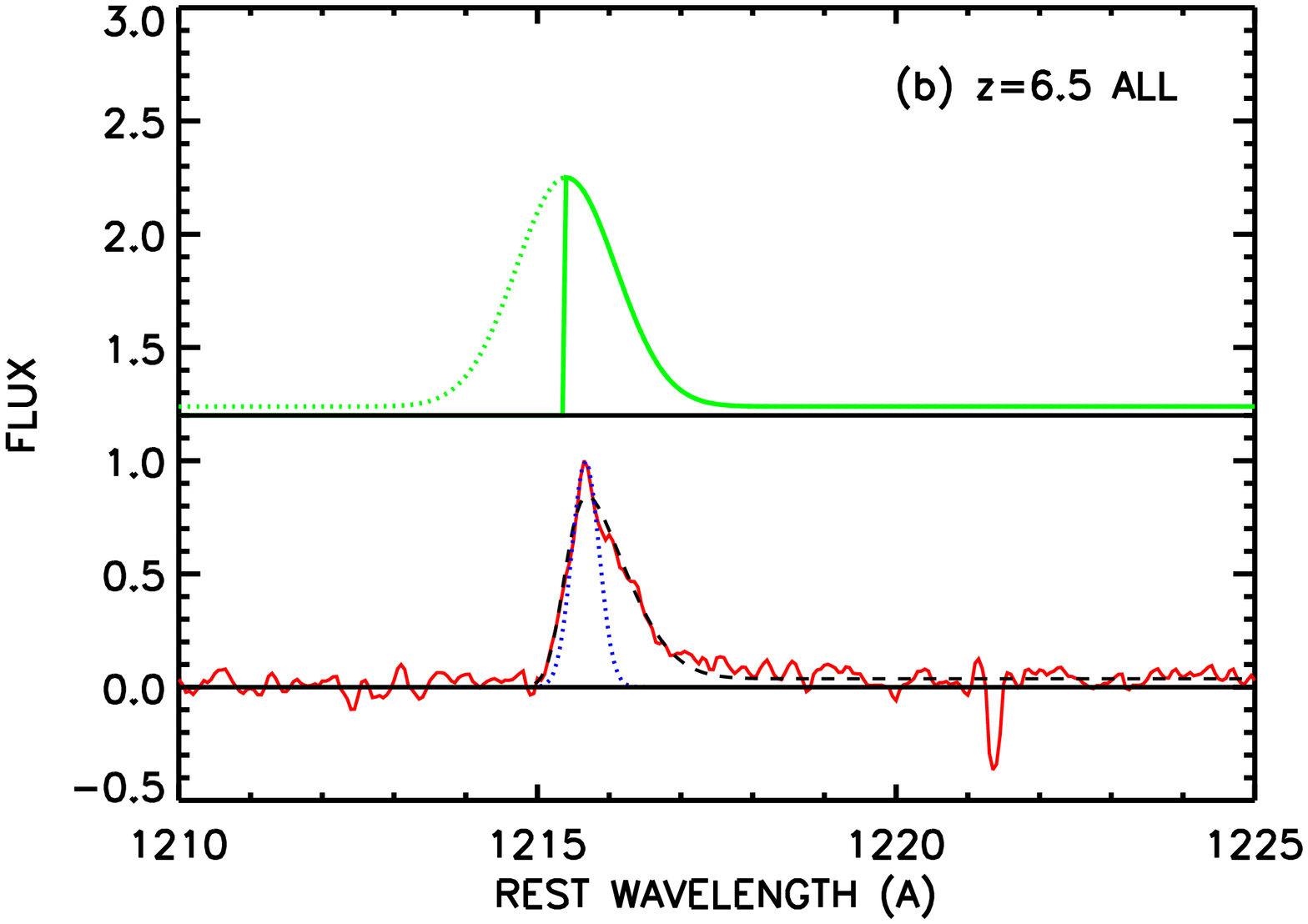}
}
\centerline{
\includegraphics[width=2.7in,angle=90,scale=0.95]{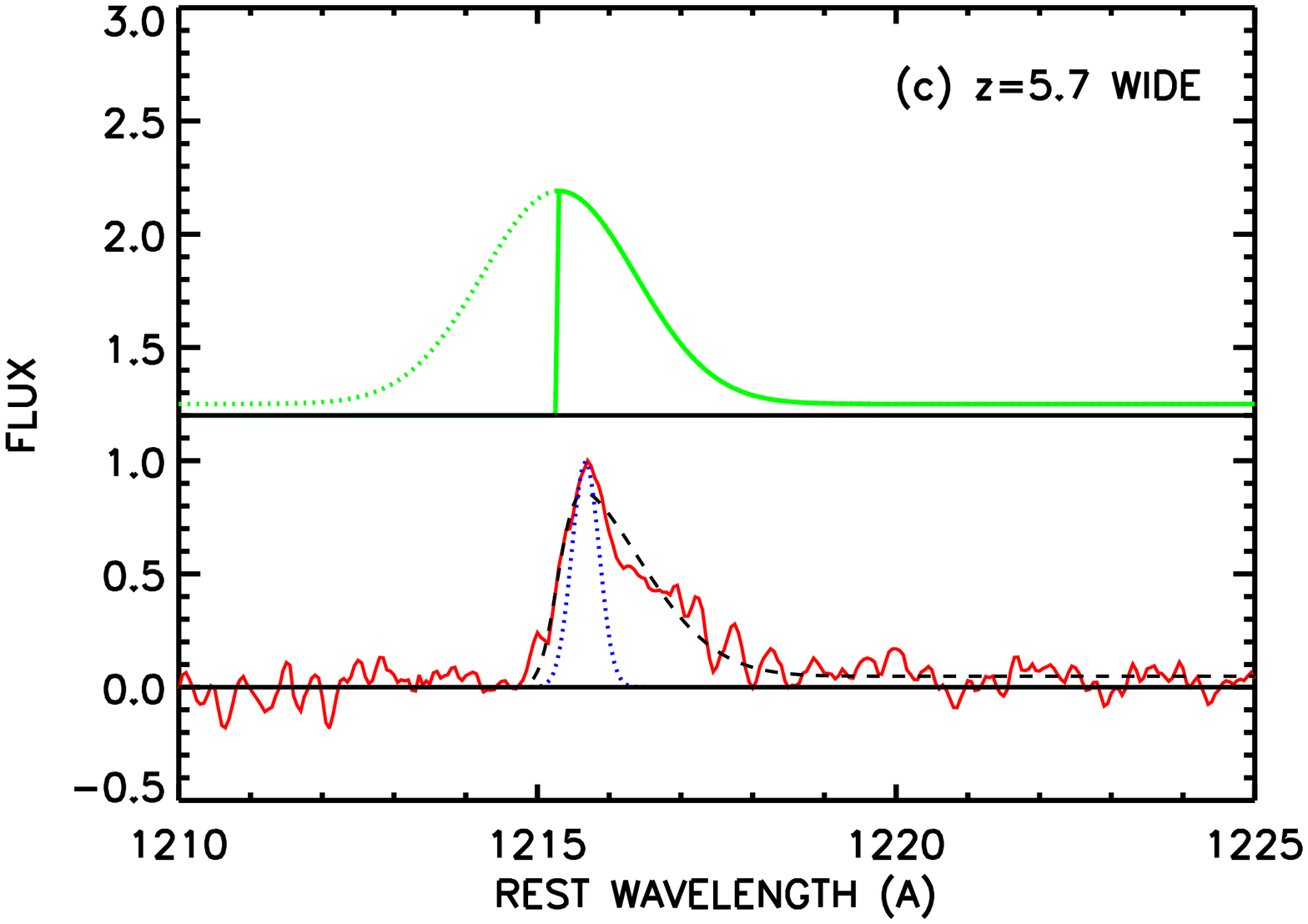}
\includegraphics[width=2.7in,angle=90,scale=0.95]{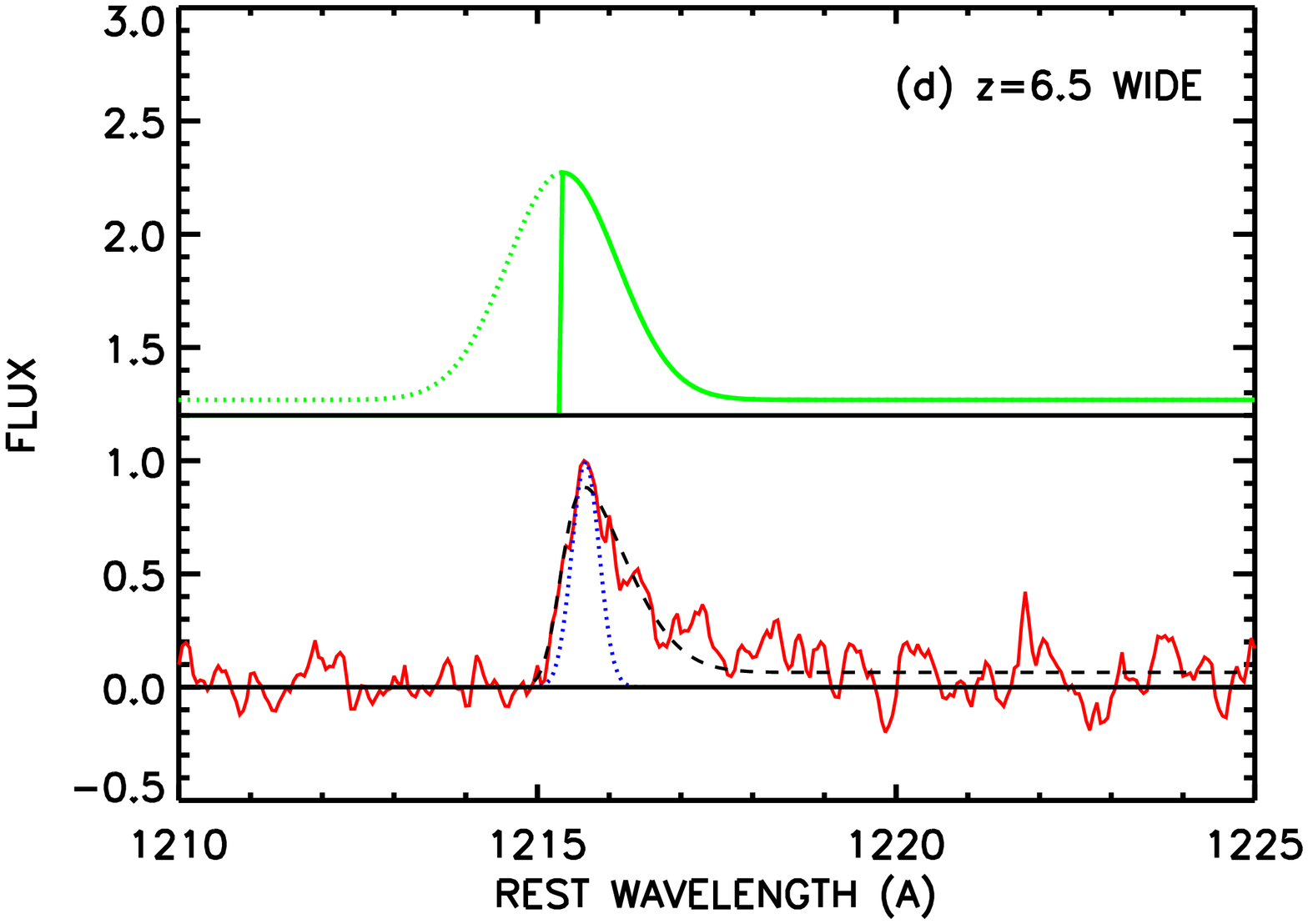}
}
\caption{Adopted fitting procedure. For each spectrum we fitted
a demi-Gaussian consisting of the long-wavelength side of a Gaussian 
profile convolved through the instrument response.
The free parameters are the normalization, the wavelength position,
the width of the Gaussian, and the normalization of the 
long-wavelength continuum. We used the MPFIT programs of 
Markwardt (2009) to make the fit. We show the fits
to the averaged spectra: (a) all $z=5.7$, (b) all $z=6.5$,
(c) wide $z=5.7$, and (d) wide $z=6.5$. In each panel we show
in the upper part the input-truncated Gaussian (green curve) and
in the lower part the fit (black dashed curve) to the observed 
spectrum (red curve).  
The blue dotted line in the lower part shows the instrument 
response.  We use the HWHM of the Gaussian (i.e., the FWHM of 
the demi-Gaussian) to characterize the width of
the lines. The widths for the averaged spectra are 
(a) 0.94~\AA, (b) 0.82~\AA, (c) 1.26~\AA, and (d) 0.90~\AA.
\label{wid_shape}
}
\end{figure*}

The Ly$\alpha$ lines are poorly fit by a Gaussian because of the fairly
sharp cut-off at the short-wavelength side. Thus, in order to provide
a simple fit to the spectra, we used a demi-Gaussian consisting 
only of the long-wavelength side of the Gaussian together
with a constant long-wavelength continuum, as shown in 
Figure~\ref{wid_shape} (green curve).
This parameterization was introduced in H04.
For each spectrum we convolved the demi-Gaussian with the
instrument profile (blue dotted curve) and fitted the result to
the observations using the IDL MPFIT programs of Markwardt (2009).
We show this for
(a) the full averaged spectrum at $z=5.7$;
(b) the full averaged spectrum at $z=6.5$; 
(c) the wide averaged spectrum at $z=5.7$; 
(d) the wide averaged spectrum at $z=6.5$. 
We find  that
this simple model has sufficient freedom with its four free parameters
(the normalization, the cut-off wavelength, the line width, and
the red continuum level) to provide a good fit to all the spectra. 
(This is in agreement with H04's conclusion
for their $z=5.7$ line profile but not with Kashikawa
et al.\ 2006, who found that they could not reproduce the red side
of their $z=6.5$ line profile with this type of model.)
In particular, the shape of the wider spectra are simply reproduced 
by an increase in the width of the Gaussian, and the short-wavelength 
drop in the observed lines is fully consistent with the abrupt cut-off 
in the model.  We define the rest-frame FWHM of the lines as the 
half-width half-maximum (HWHM) of the Gaussian prior to truncation.
We give the HWHM in the tables in the Appendix, together with the 
$1\sigma$ errors, for all the individual spectra.

For the average of all the spectra, the FWHM (as defined in
the previous paragraph) is $0.98\pm0.04$~\AA\ at $z=5.7$ and
$0.81\pm0.08$~\AA\ at $z=6.5$. The corresponding rest-frame 
equivalent widths, which we define as the area
of the line divided by the red continuum level, are
$34\pm2$~\AA\ at $z=5.7$ and $24\pm3$~\AA\ at $z=6.5$.
This suggests that the equivalent widths have dropped
slightly between the two redshifts. However, the
error is primarily in the determination of the red
continuum, and the difficulty of accurately
measuring this quantity and the possibility of systematic errors
should be kept in mind in assessing this result.

\begin{figure}
\centerline{
\includegraphics[width=3.4in,angle=90,scale=0.95]{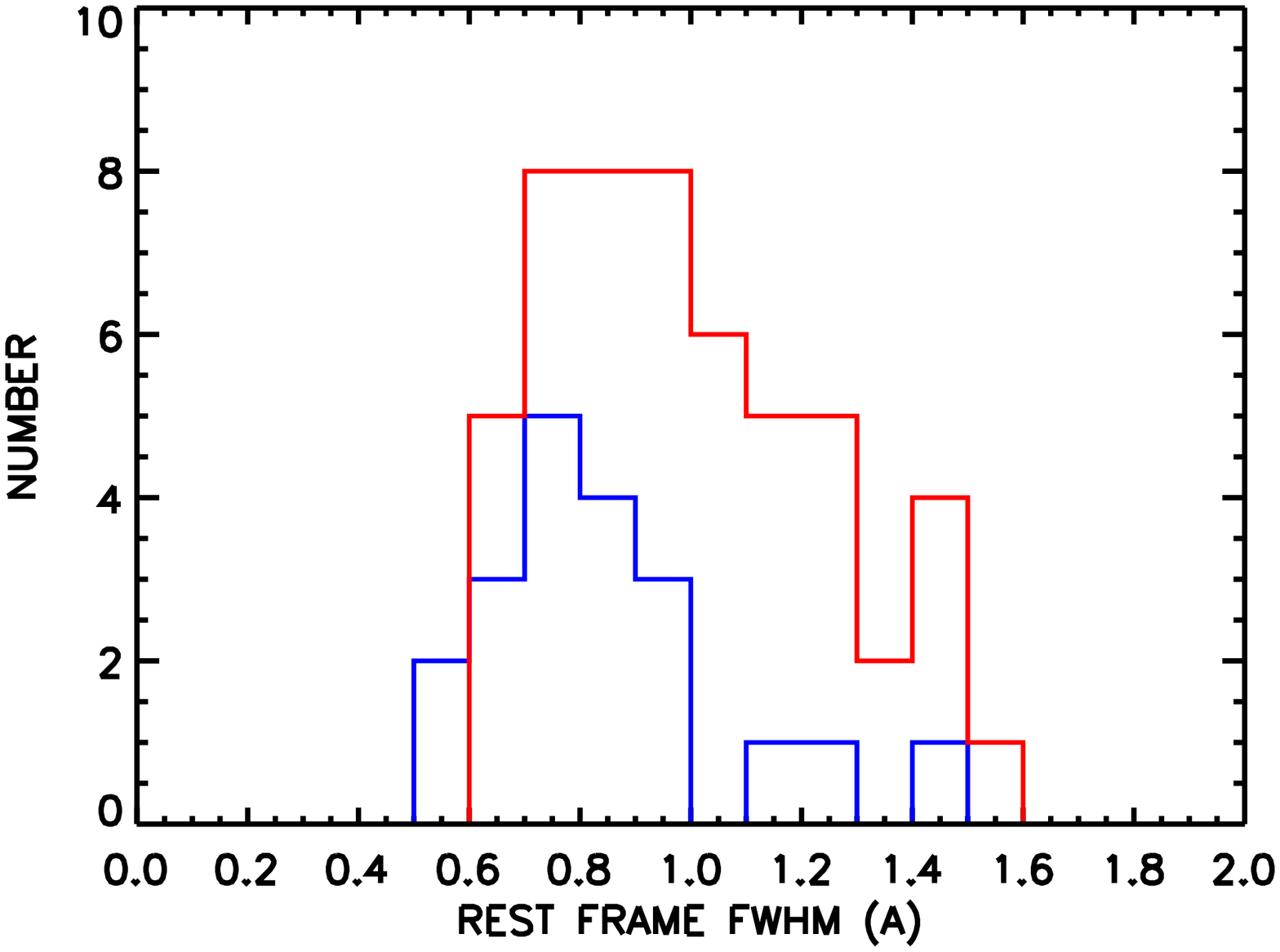}
}
\caption{Distribution of FWHM line widths obtained from
the fitting procedure for the $z=5.7$
sample (red histogram) and for the $z=6.5$ sample (blue histogram).
\label{width_hist}
}
\end{figure}

\begin{figure}
\centerline{
\includegraphics[width=3.4in,angle=90,scale=0.95]{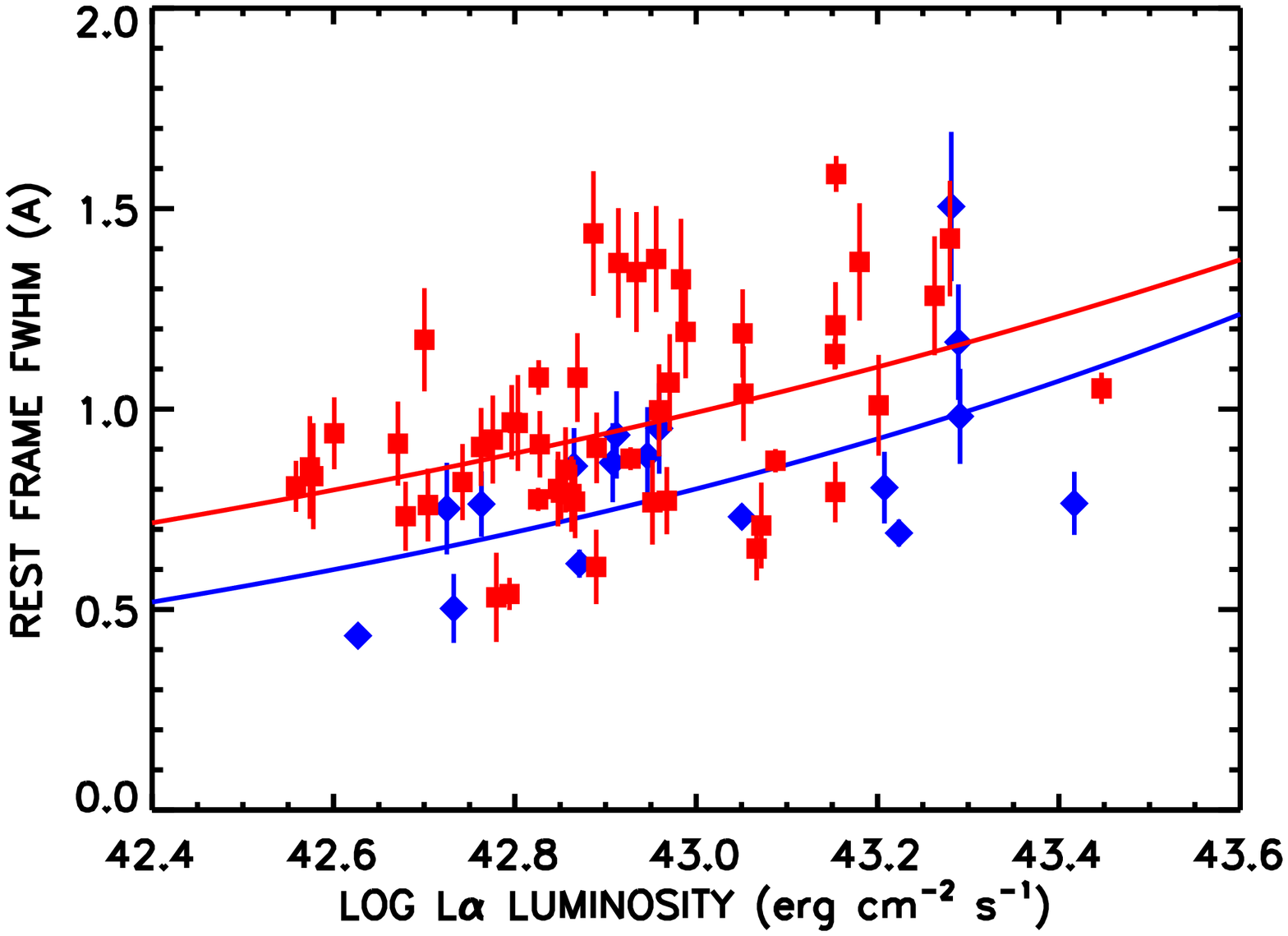}
}
\caption{Distribution of FWHM line widths obtained from
the fitting procedure for the $z=5.7$
sample (red squares) and for the $z=6.5$ sample (blue diamonds).
In both cases we show the $\pm1\sigma$ errors. The blue
and red lines show power-law fits of the form 
FWHM $=A~L_{\alpha}^{a}$.
\label{width_lx}
}
\end{figure}

The line widths are robustly measured and can be
obtained for each of the individual spectra. However, the
long-wavelength continuua are often too weak to be measured in
the individual spectra, so we do not attempt to  measure
equivalent widths in the individual spectra.
We compare the distribution of line widths in the two samples in 
Figure~\ref{width_hist}, where the red histogram shows the $z=5.7$ 
sample and the blue histogram shows the $z=6.5$ sample. 
There is just over a factor of two spread in the  widths, 
which suggests that the spread in galaxy properties combines 
with the transfer effects to produce a fairly uniform
output line with a velocity width in the range 
$150-360$~km~s$^{-1}$.  As with the
averaged spectra, the lines are narrower at the higher
redshift with a median value of $0.77$~\AA\ at $z=6.5$
and $0.92$~\AA\ at $z=5.7$. However, the difference is only
marginally significant. A Mann-Whitney rank sum test rejects
the two samples as being drawn from the same population
at the $5\%$ confidence level.
 
Part of the spread in the line widths appears to be caused by 
a dependence of the FWHM on the Ly$\alpha$ luminosity, $L_\alpha$.
In Figure~\ref{width_lx}
we show the dependence of the deconvolved FWHM measured with the
fitting procedure on  $L_\alpha$. Both the $z=5.7$ and the $z=6.5$
samples appear to show an increase of the width with luminosity.
This is in contrast to Kashikawa et al.\ 2006, who suggested a 
slight increase with decreasing luminosity in their $z=6.5$ sample;
see their Figure~11. The
difference may arise from the higher resolution of the present
observations and the wider range in $L_\alpha$. Ouchi
et al. (2010), using Keck DEIMOS spectra,  reverse
the Kashikawa et al.\ 2006 result and find evidence at the 2.5 sigma
level for a rise in the FWHM with luminosity. The effect is highly
significant in our $z=5.7$ sample: a Mann-Whitney rank sum test
shows that there is only a 0.005 probability that the population
with $\log L_\alpha<42.9$ is drawn from the same distribution
of FWHM as those at brighter luminosities. The Spearman correlation
coefficient is 0.42 at a $3\sigma$ significance for the $z=5.7$
sample and 0.58 at a $2.3\sigma$ significance for the $z=6.5$ sample.

In each case we have fitted a power law of the form
FWHM $=A~L_{\alpha}^{a}$ to the data. For $z=5.7$ we
find $a=0.24\pm0.07$, and for $z=6.5$ we find $a=0.31\pm0.11$.
These fits are shown by the red line ($z=5.7$) and the blue
line ($z=6.5$) in Figure~\ref{width_lx}. There are many
effects which could contribute to there being a relation
between the observed line luminosity and the line width. Possibly
the simplest interpretation is that the higher luminosity
galaxies are more massive and the emerging L$\alpha$ line
is wider. However, detailed modeling, including all of
the line transfer effects, is necessary to fully interpret
this result.

\subsection{Number Counts}
\label{secnum}

The average number of objects detected per SuprimeCam field
is 12.4 at $z=5.7$ and 3.9 at $z=6.5$. The observed ranges
of $6-18$ per field at $z=5.7$ and $1-7$ per field at $z=6.5$ are 
fully consistent with the spread expected from the small number
statistics. We do not require any additional effects
from cosmic variance to understand this, though such
effects may be expected to be present.

\begin{figure*}
\centerline{
\includegraphics[width=3.5in,angle=0,scale=1.]{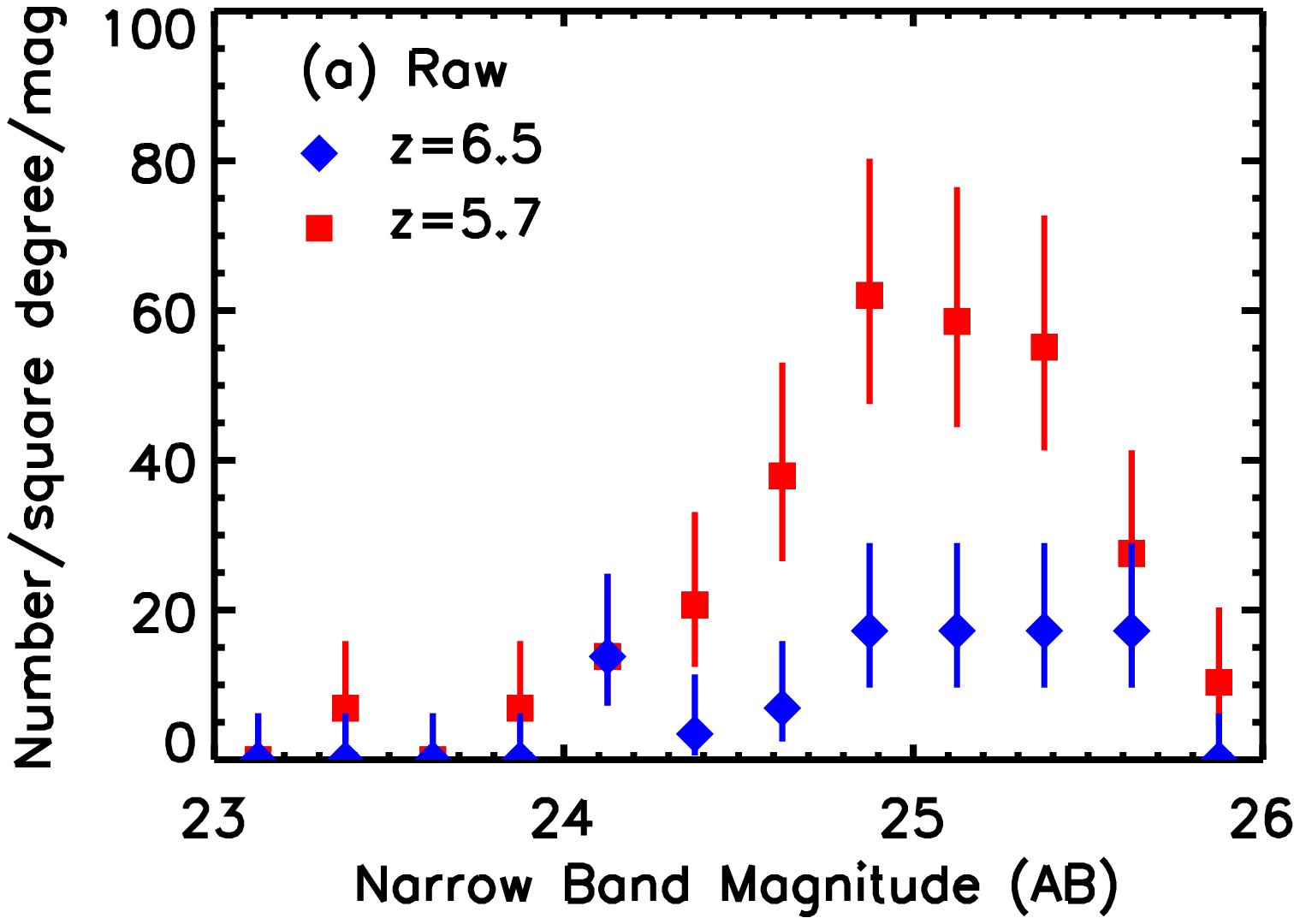}
\includegraphics[width=3.5in,angle=0,scale=1.]{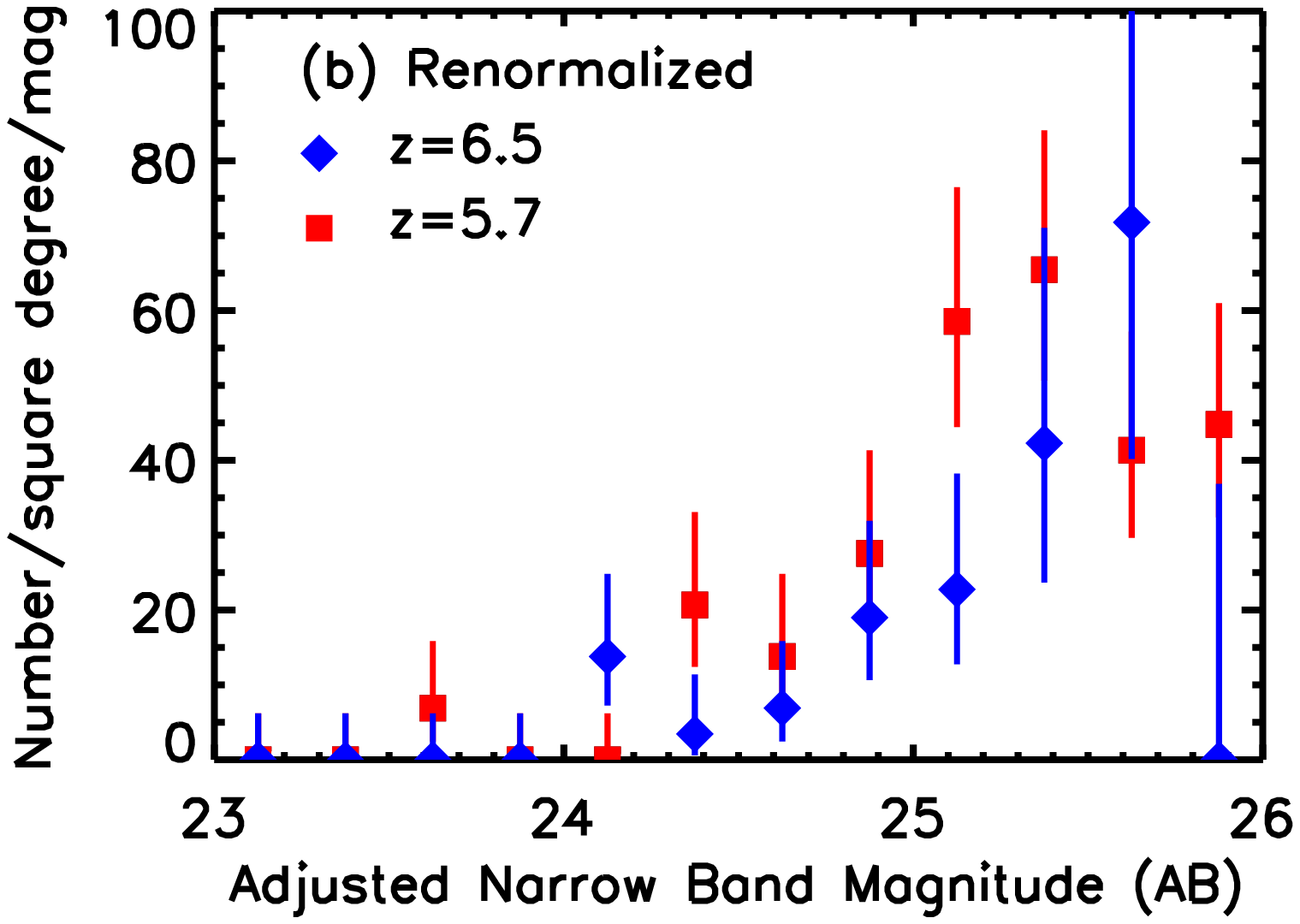}
}
\caption{(a) Observed number counts vs. narrowband
magnitude [red squares---NB816 ($z=5.7$) selected galaxies; 
blue diamonds---NB912 ($z=6.5$) selected galaxies]. In both cases 
the error bars are $\pm1\sigma$ based on the number of objects
in each bin. (b) Comparison of the counts in the NB816 sample (the 
magnitudes were adjusted to equivalent NB912 magnitudes by correcting 
for the relative luminosity distance), corrected for incompleteness in 
the photometric catalog, with the counts in the NB912 sample, also 
corrected for incompleteness.
Between 23 and 25.25, where both sets of counts are near complete, 
the NB816 number counts need to be multiplied by a factor of
$0.47\pm0.13$ to match the NB912 counts.
\label{compare_counts}
}
\end{figure*}

The similarities of the depths of the fields and of the shapes and 
rest-frame widths of the two filters allow us to make a simple estimate 
of the decrease in the number of Ly$\alpha$ emitters with increasing 
redshift directly from the number counts. The number counts in the two 
redshift ranges are shown versus narrowband magnitude in
Figure~\ref{compare_counts}(a). We denote the $z=5.7$ counts by
red squares and the $z=6.5$ counts by blue diamonds. As would be expected 
from the initial selection, the counts rise smoothly to near 
$N_{\rm AB}=25.0-25.5$ and then drop rapidly at fainter magnitudes. 
In this discussion and in the derivation of the LF in
the next subsection, we shall restrict to a sample with
$N_{\rm AB}<25.25$ where we believe the samples in both bands are 
substantially complete both in the initial selection and in the 
spectroscopic followup.

To compare the number counts, we corrected the NB816 magnitudes to 
equivalent NB912 magnitudes by adding the difference in magnitude  
corresponding to the relative luminosity distance.   
In Figure~\ref{compare_counts}(b) we compare 
the $z=5.7$ counts (for the equivalent NB912 magnitudes), corrected 
for the initial photometric catalog incompleteness,
with the $z=6.5$ counts, corrected in the same way.
(The photometric catalog incompleteness at these magnitudes is
small; see Figure~\ref{ncounts}.)
The ratio of the numbers of galaxies above the (equivalent) NB912 magnitude 
of 25.25 at $z=6.5$ and at $z=5.7$ is $0.47\pm0.13$.
The error is $\pm1\sigma$. 
The shapes are fully consistent, and a single multiplicative
renormalization alone is sufficient to match the two sets of counts.

\begin{figure*}
\centerline{
\includegraphics[width=4.5in,angle=0,scale=1.]{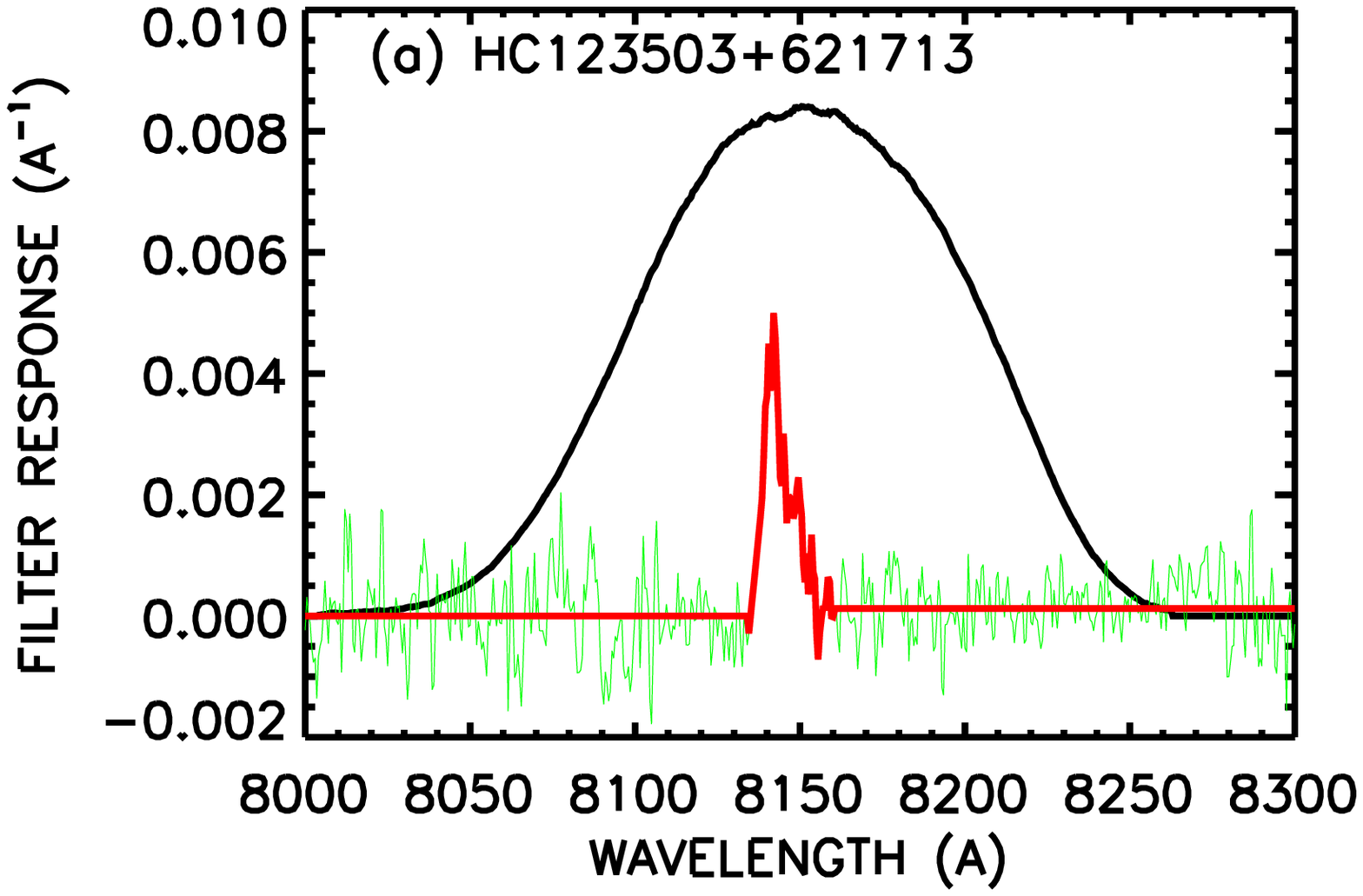}
}
\centerline{
\includegraphics[width=4.5in,angle=0,scale=1.]{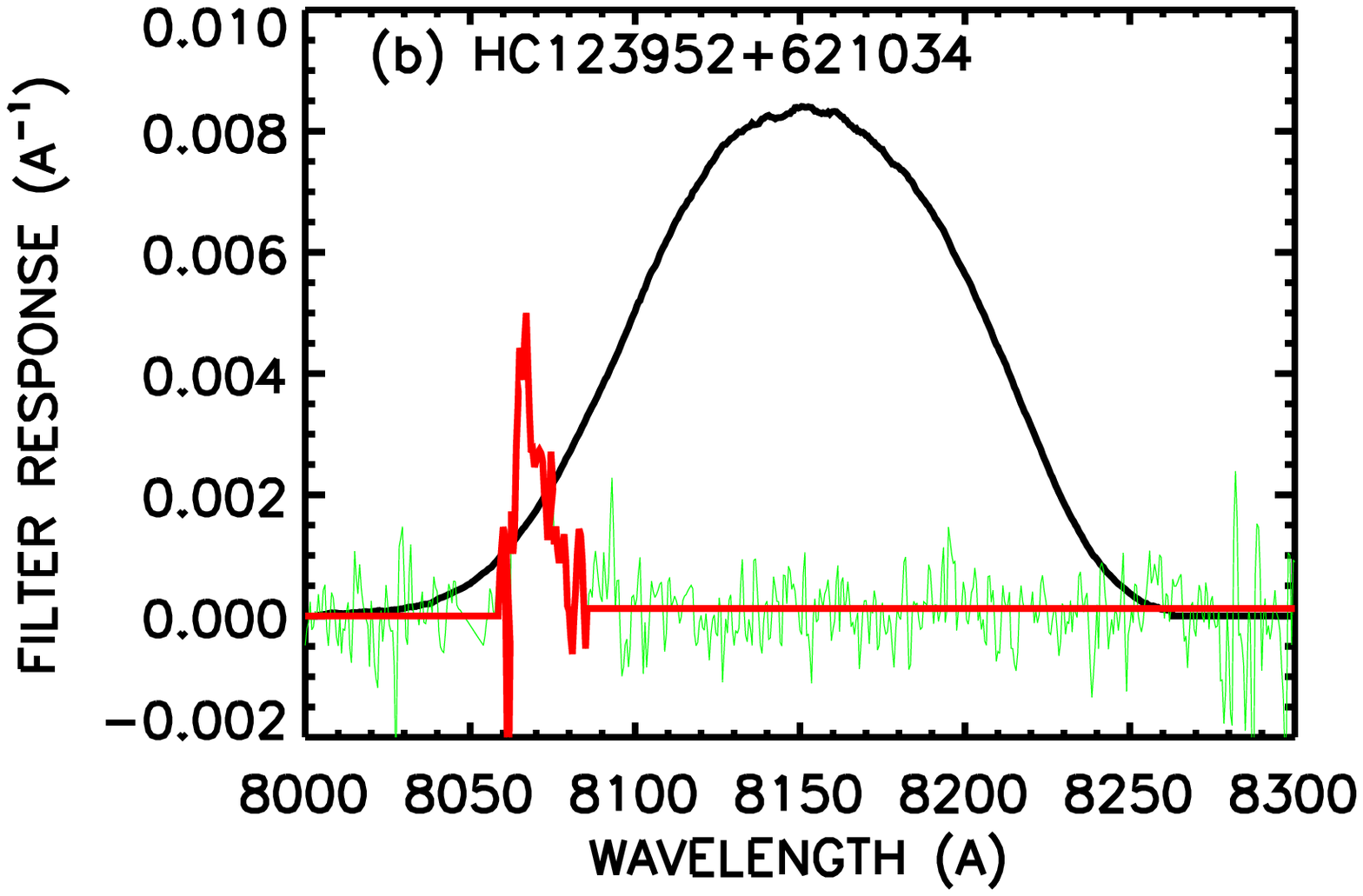}
}
\caption{Illustration of the flux computation for two of the 
$z=5.7$ Ly$\alpha$ emitters. 
The curves show the NB816 filter response (black); the adopted 
spectrum in the observed frame (red), which corresponds to
a constant value of 0.025 of the peak in the spectrum for the red 
continuum, zero for the blue continuum, and the actual spectrum
between the rest-frame wavelengths of 1214.5 to 1218.5~\AA;
and the actual spectrum outside this range (green).
(a) A Ly$\alpha$ emitter that is well centered on the filter.
Nearly 90\% of the contribution to the narrowband magnitude 
comes from the emission line. 
(b) A Ly$\alpha$ emitter lying at the short-wavelength end of the 
filter. Almost 60\% of the contribution to the narrowband 
magnitude comes from the continuum. 
\label{example_z5}
}
\end{figure*}

\subsection{Luminosity Functions}
\label{seclumfun}

In order to compute the Ly$\alpha$ luminosities of the individual
galaxies and the cosmological volumes sampled by the survey,
we must allow for the shapes of the narrowband selection
filters. This is made more complex by the presence of the
continuum, which slightly pulls the selection to the 
short-wavelength side of the filter, as can be seen in 
Figure~\ref{filtshape} and also in Figure~13 of H04 
and in Figure~3 of Kashikawa et al.\ (2006).
The continuum is extremely faint and often undetected in the
spectra and even in the continuum images, so this effect is not
easy to model exactly, but it is necessary to include it
to make a correct conversion from the narrowband magnitude
to a line flux.

To convert the narrowband magnitude to a line flux, we
convolve the observed spectrum through the narrowband filter. 
However, outside the line itself, defined as
the portion of the spectrum between rest-frame wavelengths 
1214.5 and 1218.5~\AA, we use a model continuum. At redder 
wavelengths we use a continuum with a flux of 0.025 times
the peak in the line  
seen in the individual spectra (see Figure~\ref{compare_spectra}). 
The adopted ratio is based on that measured in the averaged
spectrum.
At bluer wavelengths we assume that the continuum is zero. 
We use the narrowband magnitude to flux calibrate the
spectrum and hence to determine the line flux.

We illustrate the procedure in Figure~\ref{example_z5}, where
we show the NB816 filter response (black curve), the adopted 
spectrum (red curve), and the actual spectrum (green curve) 
in the wavelength range where we use the model continuum instead.
In Figure~\ref{example_z5}(a) we show an emitter that is well 
centered on the filter. For this object nearly 90\% of the 
contribution to the narrowband magnitude comes from the emission
line, and the conversion to an emission-line flux should
be robust. In Figure~\ref{example_z5}(b) we show an emitter 
lying at the short-wavelength end of the filter. For this object 
almost 60\% of the contribution to the narrowband magnitude 
comes from the continuum. In order to avoid the uncertainty
associated with the flux conversion in objects like this,
we restrict our subsequent
analysis to galaxies with Ly$\alpha$ wavelengths where the filter
transmission is above 25\% of the peak value. This eliminates
objects where the narrowband flux is continuum dominated.
In column~11 of Tables~\ref{la_table}, \ref{high_la_tab},
and \ref{very_high_la.tab} in the Appendix
we list the derived Ly$\alpha$ luminosities for each object. 

\begin{figure}
\centerline{
\includegraphics[width=3.6in,angle=0,scale=1.]{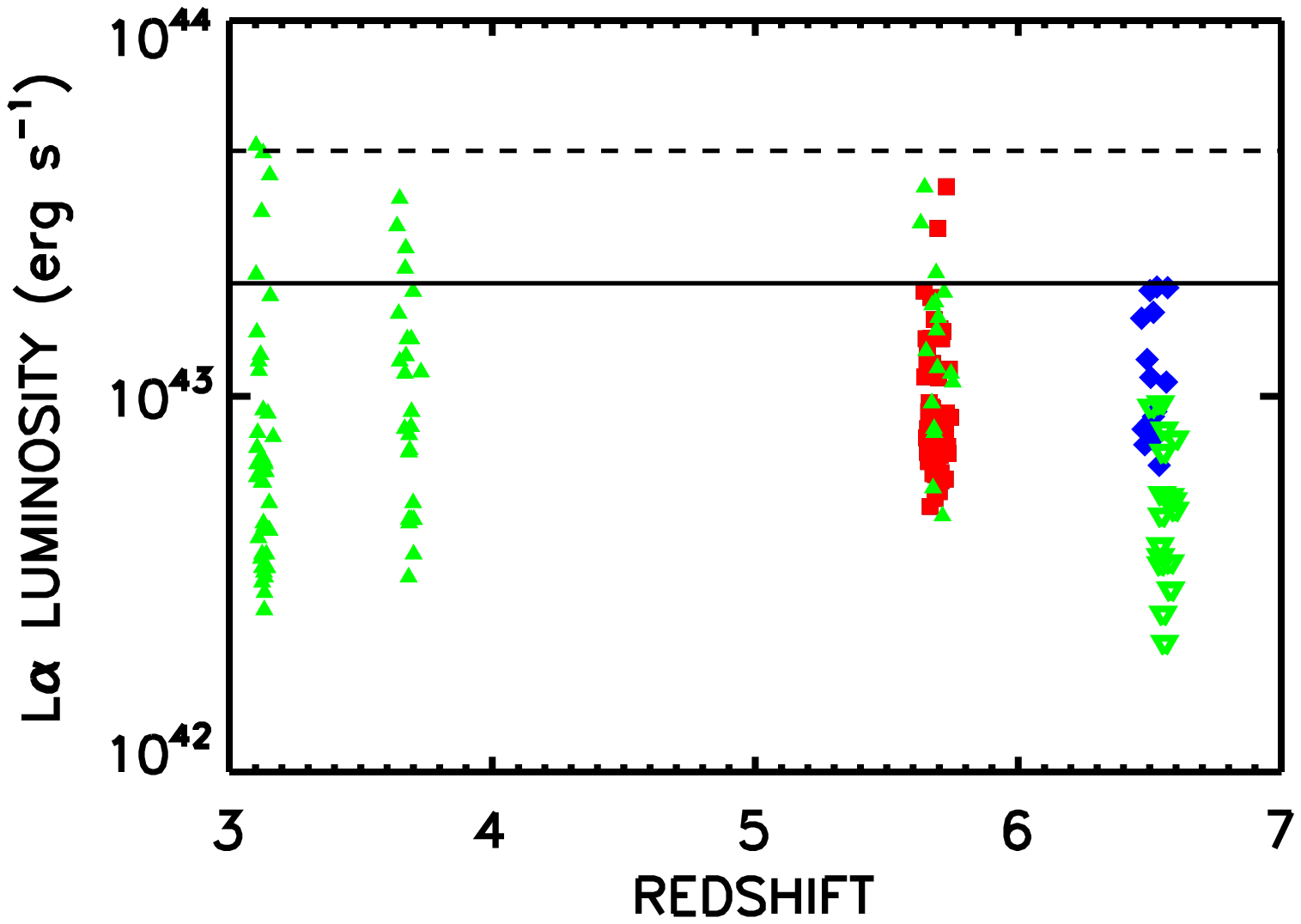}
}
\caption{The observed Ly$\alpha$ luminosity
in each of the redshift intervals (red squares---$z=5.7$ Ly$\alpha$ 
emitters; blue diamonds---$z=6.5$ Ly$\alpha$ emitters).  Only objects 
with narrowband magnitudes less than 25.25 at a wavelength that has 
a filter response above 0.25 are shown.  In order to avoid lensing 
effects, objects within a $10'$ radius of the center of A370 are excluded.  
For comparison, spectroscopically identified objects in 
Ouchi et al.\ (2008) (green solid triangles) and in 
both Taniguchi et al.\ (2005) and Kashikawa et al.\ (2006) 
(green open downward pointing triangles) are also shown. 
The dashed (solid) horizontal line 
shows the maximum luminosity object of $\sim4.5\times10^{43}$~erg~s$^{-1}$ 
($\sim2.0\times10^{43}$~erg~s$^{-1}$) observed at $z\sim3$ ($z\sim6.5$).
\label{lx_z}
}
\end{figure}

\begin{figure}
\centerline{
\includegraphics[width=3.6in,angle=0,scale=1.]{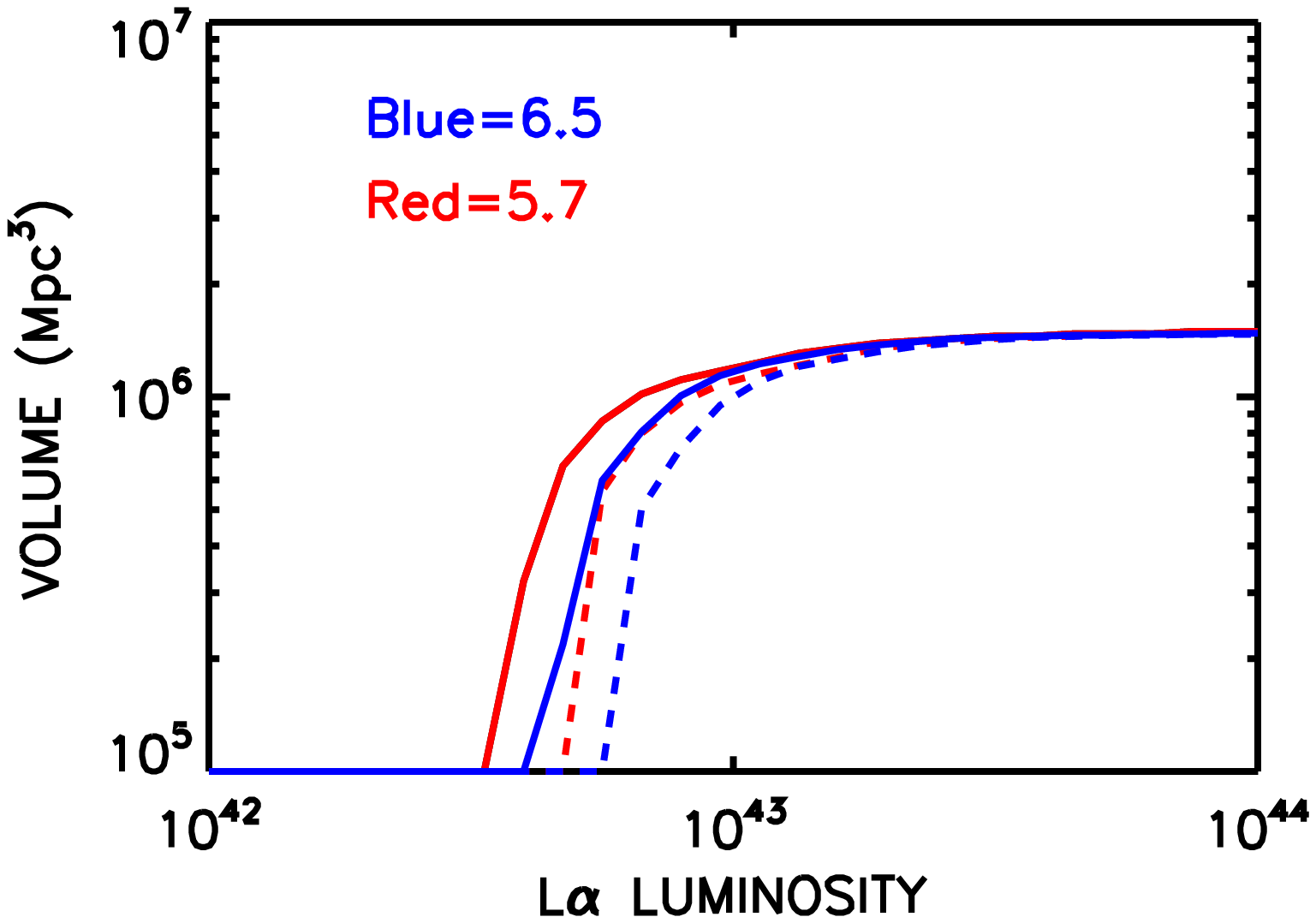}
}
\caption{The observable comoving volume as a function of Ly$\alpha$
luminosity (blue curves---$z=6.5$; red curves---$z=5.7$). 
The solid (dashed) curves are for a limiting narrowband 
magnitude of 25.5 (25.25).
We restrict to wavelengths where the filter
transmission is above 25\% of the peak value.
\label{vol_plot}
}
\end{figure}

We show the distribution of Ly$\alpha$ luminosities for
the samples at $z=5.7$ (red squares) and $z=6.5$ (blue
diamonds) in Figure~\ref{lx_z}.  We only show objects
with $N_{\rm AB}<25.25$ that have a filter transmission above
25\% of the peak. We compare with spectroscopic 
samples at $z=3.1, 3.7$, and $5.7$ from Ouchi et al.\ (2004) 
(green solid triangles) and at $z=6.5$ from both Taniguchi
et al.\ (2005) and Kashikawa et al.\ (2006) (green open downward
pointing triangles). At $z=5.7$ our distribution of luminosities
is very similar to that of Ouchi et al.\ (2008), despite some
methodological differences. (For example, Ouchi et al.\ do not 
account for the filter shape in computing the luminosities but
instead deal with this in subsequent simulations.) However, 
the $z=6.5$ samples of Kashikawa et al.\ and Taniguchi et al.\
are systematically lower than those in the present work.
It appears from their description that their luminosities
are based on uncorrected $2''$ diameter aperture magnitudes.
(The Ouchi et al.\ 2008 luminosities are based on corrected
$2''$ diameter aperture magnitudes.)
The correction to total magnitudes would then raise their
luminosities by factors of $1.3-1.4$, which could account
for a substantial part of the difference. They also
assume a rectangular shape for the narrowband filter
in computing the luminosities (Taniguchi et al.\ 2005's
Equations~6 and 7), which could also result in differences.

The peak luminosities seen in our present samples at
$z=5.7$ and $z=6.5$ are slightly more than a factor of
two less than those seen near $z=3$. This can be
fully understood in terms of the intergalactic absorption
correction, and it appears that the intrinsic luminosities
of the brightest emitters are hardly changing from
$z=6.5$ to $z=3.1$. However, this simple analysis is
dependent on the number of objects in each redshift
sample, and the evolution is best treated by looking at
the LFs, which we now do.

In order to compute the LFs, we must
determine the accessible comoving volume as a function
of luminosity. Here again
the shape of the filter transmission makes the calculation
more complicated. For a rectangular filter the volume is
fixed above the detection threshold. However, for more
complex filter shapes, the wavelength range is a function
of the luminosity and the selection magnitude. 
For a given limiting magnitude, more luminous objects will 
be seen over a wider range of redshifts, since they can
still be detected at lower filter transmissions.

In order to compute the observable comoving volume as a 
function of luminosity and limiting narrowband magnitude, 
we used the averaged spectral profiles of the emitters at
$z=5.7$ and $z=6.5$ (Figure~\ref{compare_spectra}). We 
normalized the appropriate spectrum for the filter 
(i.e., the $z=5.7$ spectrum for the NB816 filter and 
the $z=6.5$ spectrum for the NB912 filter) to correspond 
to a given line luminosity
and then stepped this through the filter, calculating
the narrowband magnitude at each redshift. This allowed
us to determine the redshift range over which a line of
this luminosity would produce a narrowband magnitude
above the narrowband magnitude limit. We show the observed
volumes at $z=6.5$ (blue curves) and at $z=5.7$ (red curves) 
as a function of the Ly$\alpha$ luminosity
in Figure~\ref{vol_plot}. The solid lines show the
values for a limiting narrowband magnitude of 25.5,
and the dashed lines for 25.25.
For 25.25, which we use in the
subsequent calculations, the volume drops rapidly below
$\sim5.6\times10^{42}$~erg~s$^{-1}$ at $z=5.7$ and
below $\sim6.7\times10^{42}$~erg~s$^{-1}$ at $z=6.5$.

\begin{figure}
\centerline{
\plotone{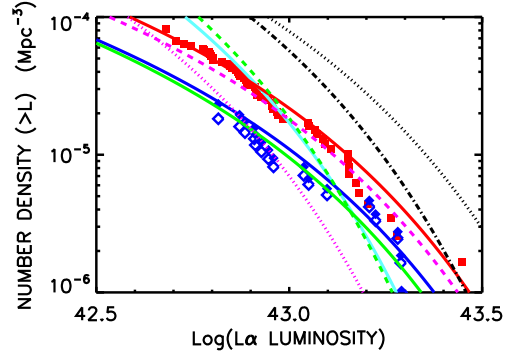}
}
\caption{The cumulative Ly$\alpha$ LFs at 
$z=5.7$ (red squares) and $z=6.5$ (blue diamonds). 
The red and blue solid curves show the maximum likelihood
fits to a Schechter function with an $\alpha=-1.5$ slope. 
The black dash-dotted curve shows the $z=5.7$ 
cumulative LF from Ouchi et al.\ (2008) and the
black dotted curve shows the $z=5.7$ cumulative LF 
of Shimasaku et al.\ (2006), both based on 
their photometric samples.
The purple dashed (dotted) curve shows the 
$z=5.7$ ($z=6.5$) cumulative LF from the spectroscopic 
sample of Malhotra \& Rhoads (2004).
The green solid (dashed) curve shows the $z=6.5$ cumulative 
luminosity function from the spectroscopic (photometric) 
sample of Kashikawa et al.\ (2006) while the solid cyan
curve shows that of Ouchi et al. (2010) which is also
primarily photometric.
\label{cum_lum_fun}
}
\end{figure}

We first computed the cumulative LF, which
is simply the sum of the inverse observable volumes above
a given Ly$\alpha$ luminosity. In order to avoid lensing
effects, we excluded the central $10'$ region in
the A370 field and corrected the volume accordingly.
Only objects with narrowband magnitudes brighter than
25.25 and lying above 25\% of the maximum response in
the filter are included in the sample, and the observable
volumes were computed to correspond to this selection.
We show the results in Figure~\ref{cum_lum_fun}. For
the $z=6.5$ sample we show both the cumulative LF
prior to any incompleteness correction (blue open diamonds) 
and that with the spectroscopic and photometric incompleteness 
correction included (blue solid diamonds). The correction is 
small with this magnitude selection. For the $z=5.7$ sample, 
where the correction is even smaller, we show only the 
incompleteness corrected LF (red solid squares). 

The red and blue solid curves show the maximum likelihood fits 
to the two data samples for a Schechter function with slope 
$\alpha=-1.5$. Because of the limited dynamic range of the data, 
we have not attempted to fit for the slope of the Schechter function 
but rather computed $L_\star$ for fixed values of this quantity.
The normalization $\phi_\star$ is calculated to match the observed 
number of sources over the observed luminosity range.
For $\alpha=-1.5$ and $z=5.7$, the fitted parameters are 
$\log L_\star=43.0^{43.3}_{42.8}$ and 
$\phi_\star=1.1\pm0.2\times10^{-4}$~Mpc$^{-3}$. 
For $\alpha=-1.5$ and $z=6.5$, they are 
$\log L_\star=43.0^{43.6}_{42.7}$ and 
$\phi_\star=0.6\pm0.2\times10^{-4}$~Mpc$^{-3}$. 
Here the luminosities are in units of erg~s$^{-1}$, and the range 
is the 68\% confidence limit. 
Over the fitted range the $z=6.5$ counts are $0.56\pm0.17$
of the $z=5.7$ counts for the $\alpha=-1.5$ case and
the characteristic luminosity is unchanged consistent
with our previous discussion of the number counts. We
summarize the fitted parameter for varying values of
$\alpha$ in table~\ref{tabfits}.

\begin{deluxetable}{lccc}
\tablecolumns{4}
\tablecaption{Luminosity function fits}
\tablewidth{0pt}
\tablehead{
\colhead{Redshift}  &  \colhead{$\alpha$} & \colhead{Log $\L_\star$} &
\colhead{$\phi_\star$}\\ 
\colhead{}  &  \colhead{} & \colhead{erg s$^{-1}$} &
\colhead{$10^{-4}$ Mpc$^{-3}$}} 
\startdata
\noalign{\vskip-2pt}
5.7 &  -1.0 & $42.9^{43.1}_{42.7}$ &  $1.7\pm0.3$\\
5.7 &  -1.5 & $43.0^{43.3}_{42.8}$ &  $1.1\pm0.2$\\
5.7 &  -2.0 & $43.2^{43.7}_{42.9}$ &  $0.5\pm0.1$\\
6.5 &  -1.0 & $42.9^{43.3}_{42.6}$ &  $0.7\pm0.2$\\
6.5 &  -1.5 & $43.0^{43.6}_{42.7}$ &  $0.6\pm0.2$\\
6.5 &  -2.0 & $43.1^{44.3}_{42.7}$ &  $0.3\pm0.1$\\
\enddata
\label{tabfits}
\end{deluxetable}

In Figure~\ref{cum_lum_fun} we also compare the present data 
with measurements from the literature. In all cases
we show the maximum likelihood fits for a Schechter function
with slope $\alpha=-1.5$.  Our $z=5.7$ LF agrees quite well 
with the Malhotra \& Rhoads (2004) LF based on their 
spectroscopic sample (purple dashed curve), 
but it is about a factor of slightly 
more than two lower than either the Ouchi et al.\ (2008)
(black dash-dotted curve) or the Shimasaku et al.\ (2006)
(black dotted curve) LFs, which are based on 
their photometric samples.  Our $z=6.5$ LF is only 
slightly above the Kashikawa et al.\ (2006) LF
based on their spectroscopic sample (green solid curve).
It is similar to the Malhotra \& Rhoads (2004) 
spectroscopically based  sample
(purple dotted curve) at low luminosities,
though the present sample clearly has more objects at high
luminosities than this fit would predict. It is 
about a factor of two lower than the Kashikawa et al.\ (2006)
photometrically based sample (green dashed curve) or
the very similar Ouchi et al.\ (2010) LF (cyan curve) at low 
luminosities, but it is higher at the bright end.

The largest differences appear to be with the photometrically 
based LFs of Ouchi et al.\ (2008) and Kashikawa et al.\ (2006)
and Ouchi et al. (2010).
The presently derived LFs are typically about a factor of two 
lower. This is not a consequence of the magnitude calibrations,
since, as we have discussed previously, both the Ouchi et al.\ 
and the Kashikawa et al.\ magnitude measurements are fainter 
than the present photometry, and this would have the opposite 
effect (i.e., it would raise our LFs relative to theirs rather 
than reduce them). 

The real reason for the reduction in our LFs relative to
theirs seems to come from our having excluded the spectroscopically 
unconfirmed objects. Ouchi et al.\ (2008) observed 29 of
their photometrically selected galaxies and
identified 17 as Ly$\alpha$ emitters. However,
they estimated the contamination based on only the 
objects brighter than a narrowband magnitude of 24.5,
where three-quarters of the objects were identified, 
and concluded that the maximum contamination was 25\%. 
Since this was small, they did not apply a correction.
However, the numbers on which this calculation are based 
are small, and the probability of incorrect selections 
may be expected to increase as we move to fainter magnitudes.
Thus, the bright selection is likely to be inappropriate for 
the full sample.

In the present work we could not spectroscopically confirm 
approximately half of our photometrically selected objects
in our 25.5 magnitude limited sample.
Thus, the factor of two difference between our $z=5.7$ LF 
and the Ouchi et al.\ (2008) $z=5.7$ LF may be entirely due 
to this. It is possible that some of the photometrically 
selected objects that are unconfirmed in the spectroscopy 
are genuine Ly$\alpha$ emitters where the spectroscopy was 
problematic. However, for many, even
with multiple observations we were unable to confirm 
spectroscopically the objects as emitters. We therefore think 
the present spectroscopically based Ly$\alpha$ LFs represent 
the best estimate, and the photometric LFs may be viewed as 
extreme upper limits.

\begin{figure}
\centerline{
\plotone{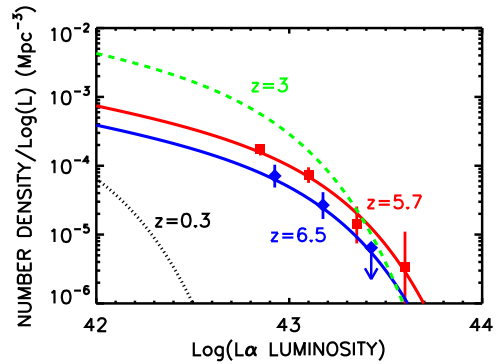}
}
\caption{The Ly$\alpha$ LF at $z=5.7$
is shown with red squares and that at $z=6.5$ is shown
with blue diamonds. The errors are $\pm1\sigma$ based
on the number of objects in each luminosity bin.
The red and blue solid curves show maximum likelihood
fits to a Schechter function with an $\alpha=-1.5$ slope.
The green dashed curve shows
the $z=3.1$ LF measured by Ouchi et al.\ (2008) using
their maximum likelihood fit for $\alpha=-1.5$.
The black dotted curve shows the local Ly$\alpha$ LF at 
$z=0.3$ derived by Cowie et al.\ (2010) from {\em GALEX\/} 
spectroscopy.
\label{lum_fun}
}
\end{figure}

In Figure~\ref{lum_fun} we show the differential LFs at
$z=6.5$ (blue diamonds) and at $z=5.7$ (red squares). 
The error bars are $\pm1\sigma$
based on the Poisson errors corresponding to the number
of objects in the bin. For the $z=6.5$ case we also
show the $1\sigma$ upper limit at the highest luminosity
with the downward pointing arrow. We
show the maximum likelihood fits for $\alpha=-1.5$ with
the blue and red curves. In the figure 
we compare the present LFs with those measured at lower
redshifts. We show the LF at $z=0.3$ derived by Cowie et al.\ (2010)
using {\em GALEX\/} spectroscopy with the black dotted curve.
We show the LF at $z=3.1$ from Ouchi et al.\ (2008) 
with the green dashed curve. Other determinations at this
redshift by van Breukelen et al.\ (2005),
Gronwall et al.\ (2007), and Cowie \& Hu (1998) are 
extremely similar, and we do not plot them separately. 

Clearly the low-redshift evolution of the Ly$\alpha$ LF
between $z=0.3$ and $z=3$ is much more spectacular
than that seen at the higher redshifts. Locally
there are very few luminous Ly$\alpha$ objects
and very little light density in the Ly$\alpha$ emission line.
However, at $z>3$ the maximum luminosity is relatively
invariant, and the number density is falling off
slowly with increasing redshift. For a fixed $\alpha=-1.5$ 
the light density in the Ly$\alpha$ line at $z=5.7$ is about 
22\% of that at $z=3.1$, and the light density in the Ly$\alpha$
line at $z=6.5$ is about 11\% of that at $z=3.1$.
When the values are corrected for the effects
of the intergalactic scattering, which substantially
reduce the observed Ly$\alpha$ luminosities relative
to the intrinsic luminosities, the change between
$z=3.1$ and $z=5.7$ will be less, probably no more
than a decrease of a factor of 2. We shall make a more
detailed comparison to the evolution of the
UV continuum light at these redshifts in a subsequent
paper (L. Cowie et al.\ 2010, in preparation).


\acknowledgements

We are indebted to the staff of the Subaru and Keck
Observatories for their excellent assistance with the
observations. We would also like to thank Yoshi
Taniguchi and Masami Ouchi for useful conversations.
We gratefully acknowledge support from NSF grants
AST 0687850 (E.~M.~H.), AST 0709356 (L.~L.~C.), 
and AST 0708793 (A.~J.~B.), a grant from NASA through an
award issued by JPL 1289080 (E.~M.~H.), the University of
Wisconsin Research Committee with funds granted by the
Wisconsin Alumni Research Foundation, and the David and
Lucile Packard Foundation (A.~J.~B.).

\clearpage


\appendix
\label{appendix}
\newpage

\tablenum{5}
\begin{deluxetable*}{clllccccclc}
\small\addtolength{\tabcolsep}{-4pt}
\tablewidth{0pt}
\tablecaption{$z=5.7$ Ly$\alpha$ Emitters}
\scriptsize
\tablehead{Number & Name & R.A. & Decl. & $N$ & $I$ & Redshift &  Expo & Qual & FWHM & Log(L)\\ & &(J2000) & (J2000) & (AB) & (AB)  & & (hrs) & & \AA & erg/s  \\ (1) & (2) & (3) & (4)  & (5) & (6) & (7) & (8) & (9) & (10) & (11)}
\startdata
       1  &  HC123818+621621  &     189.57700  &     62.27261  &  $ 23.26$  &
$ 25.28$  &   5.7275  &   1  &    2  &  $  1.31\pm0.04$  &    43.55  \cr
       2  &  HC124128+622022  &     190.36998  &     62.33969  &  $ 23.32$  &
$ 26.25$  &   5.6947  &   4  &    1  &  $  1.05\pm0.03$  &    43.44  \cr
       3  &  HC123744+621145  &     189.43600  &     62.19588  &  $ 23.95$  &
$ 25.88$  &   5.6680  &   3  &    1  &  $  1.28\pm 0.14$  &    43.26  \cr
       4  &  HC221832-002844  &     334.63602  &    0.4790830  &  $ 23.96$  &
$ 25.48$  &   5.6800  &   5  &    1  &  $ 0.82\pm0.02$  &    43.20  \cr
       5  &  HC124115+622258  &     190.31601  &     62.38280  &  $ 24.03$  &
$ 25.60$  &   5.7020  &   2  &    1  &  $  1.36\pm 0.14$  &    43.18  \cr
       6  &  HC221811-000500  &     334.54807  &   0.08355599  &  $ 24.06$  &
$ 26.16$  &   5.7086  &   3  &    1  &  $ 0.79\pm0.07$  &    43.15  \cr
       7  &  HC124125+622050  &     190.35699  &     62.34750  &  $ 24.12$  &
$ 26.43$  &   5.7137  &   2  &    1  &  $  1.42\pm 0.13$  &    43.17  \cr
       8  &  HC221654-000538  &     334.22900  &   0.09402800  &  $ 24.12$  &
$ 26.28$  &   5.6758  &   3  &    1  &  $  1.20\pm 0.10$  &    43.15  \cr
       9  &  HC221720-002007  &     334.33701  &    0.3353610  &  $ 24.30$  &
$ 25.84$  &   5.6706  &   3  &    1  &  $ 0.65\pm0.07$  &    43.06  \cr
      10  &  HC221720-001737  &     334.33701  &    0.2938329  &  $ 24.30$  &
$ 25.84$  &   5.6672  &   2  &    1  &  $  1.58\pm0.04$  &    43.15  \cr
      11  &  HC123503+621713  &     188.76300  &     62.28719  &  $ 24.31$  &
$ 27.67$  &   5.6975  &   1  &    1  &  $  1.17\pm 0.12$  &    43.04  \cr
      12  &  HC221802-001431  &     334.50903  &    0.2421390  &  $ 24.31$  &
$ 26.19$  &   5.6738  &   4  &    1  &  $ 0.87\pm0.02$  &    43.08  \cr
      13  &  HC221843-004439  &     334.67996  &    0.7441939  &  $ 24.44$  &
$ 26.22$  &   5.6550  &   2  &    1  &  $ 0.83\pm0.09$  &    43.10  \cr
      14  &  HC221728-001918  &     334.37000  &    0.3217220  &  $ 24.49$  &
$ 25.49$  &   5.6426  &   4  &    1  &  $  1.42\pm 0.14$  &    43.27  \cr
      15  &  HC221733-002216  &     334.38800  &    0.3713330  &  $ 24.56$  &
$ 26.15$  &   5.6531  &   3  &    1  &  $ 0.77\pm0.08$  &    43.09  \cr
      16  &  HC123651+621936  &     189.21498  &     62.32680  &  $ 24.58$  &
$ 26.33$  &   5.6750  &   3  &    1  &  $ 0.77\pm0.08$  &    42.96  \cr
      17  &  HC123652+622152  &     189.21700  &     62.36460  &  $ 24.60$  &
$ 27.18$  &   5.6861  &   6  &    1  &  $ 0.87\pm0.02$  &    42.92  \cr
      18  &  HC221705-001300  &     334.27301  &    0.2169169  &  $ 24.61$  &
$ 26.33$  &   5.6700  &   3  &    1  &  $  1.06\pm 0.12$  &    42.97  \cr
      19  &  HC124033+621838  &     190.14000  &     62.31069  &  $ 24.61$  &
$ 26.48$  &   5.6502  &   1  &    1  &  $  1.13\pm0.03$  &    43.15  \cr
      20  &  HC123626+620346  &     189.11000  &     62.06300  &  $ 24.65$  &
$ 27.35$  &   5.6998  &   2  &    1  &  $  1.04\pm0.04$  &    42.91  \cr
      21  &  HC024015-012946  &     40.065208  &    -1.496333  &  $ 24.70$  &
$ 27.36$  &   5.7105  &   1  &    1  &  $ 0.95\pm 0.10$  &    42.90  \cr
      22  &  HC123613+620748  &     189.05600  &     62.13000  &  $ 24.70$  &
$ 26.12$  &   5.6345  &   1  &    1  &  $ 0.76\pm 0.10$  &    42.95  \cr
      23  &  HC023953-013627  &     39.972916  &    -1.607750  &  $ 24.70$  &
$ 25.79$  &   5.6928  &   2  &    1  &  $ 0.84\pm 0.10$  &    42.85  \cr
      24  &  HC221656-001446  &     334.23499  &    0.2463060  &  $ 24.71$  &
$ 26.53$  &   5.6621  &   2  &    1  &  $  1.32\pm 0.15$  &    42.98  \cr
      25  &  HC123903+621444  &     189.76300  &     62.24569  &  $ 24.71$  &
$ 26.58$  &   5.7374  &   1  &    1  &  $ 0.70\pm 0.10$  &    43.07  \cr
      26  &  HC170647+434520  &     256.69598  &     43.75569  &  $ 24.78$  &
$ 26.67$  &   5.7084  &   6  &    1  &  $  1.43\pm 0.15$  &    42.88  \cr
      27  &  HC170648+435813  &     256.70099  &     43.97050  &  $ 24.78$  &
$ 26.81$  &   5.7265  &   2  &    1  &  $  1.37\pm 0.13$  &    42.95  \cr
      28  &  HC221731-000937  &     334.38202  &    0.1602780  &  $ 24.79$  &
$ 26.74$  &   5.6843  &   5  &    1  &  $ 0.78\pm0.09$  &    42.86  \cr
      29  &  HC024035-013626  &     40.146709  &    -1.607500  &  $ 24.79$  &
$ 27.24$  &   5.6796  &   4  &    2  &  $  1.50\pm 0.18$  &    42.87  \cr
      30  &  HC123835+620643  &     189.64702  &     62.11200  &  $ 24.83$  &
$ 26.87$  &   5.6938  &   1  &    2  &  $ 0.84\pm 0.12$  &    42.80  \cr
      31  &  HC221739-002545  &     334.41299  &    0.4292219  &  $ 24.84$  &
$ 25.90$  &   5.6599  &   2  &    1  &  $ 0.99\pm 0.11$  &    42.95  \cr
      32  &  HC221710-001240  &     334.29199  &    0.2112780  &  $ 24.85$  &
$ 25.88$  &   5.6221  &   2  &    1  &  $  1.42\pm0.07$  &    43.27  \cr
      33  &  HC124043+621534  &     190.18199  &     62.25969  &  $ 24.86$  &
$ 27.85$  &   5.7033  &   3  &    1  &  $  1.24\pm 0.13$  &    42.84  \cr
      34  &  HC221707-002744  &     334.28000  &    0.4624719  &  $ 24.86$  &
$ 26.81$  &   5.7241  &   1  &    1  &  $  1.36\pm 0.13$  &    42.91  \cr
      35  &  HC024059-012737  &     40.246414  &    -1.460472  &  $ 24.87$  &
$ 26.18$  &   5.7224  &   5  &    1  &  $ 0.90\pm0.08$  &    42.89  \cr
      36  &  HC221716-001325  &     334.31799  &    0.2238609  &  $ 24.89$  &
$ 26.09$  &   5.6444  &   3  &    1  &  $  1.03\pm 0.11$  &    43.05  \cr
      37  &  HC123717+621759  &     189.32401  &     62.29988  &  $ 24.90$  &
$ 26.55$  &   5.6610  &   2  &    3  &  $  1.88\pm 0.16$  &    42.91  \cr
      38  &  HC024138-013616  &     40.411419  &    -1.604611  &  $ 24.92$  &
$ 26.34$  &   5.6841  &   3  &    1  &  $ 0.96\pm 0.11$  &    42.80  \cr
      39  &  HC024104-012750  &     40.267582  &    -1.464000  &  $ 24.93$  &
$ 26.50$  &   5.7237  &   5  &    1  &  $  1.22\pm 0.12$  &    42.88  \cr
      40  &  HC170704+435135  &     256.76898  &     43.85988  &  $ 24.93$  &
$ 25.86$  &   5.6157  &   3  &    2  &  $ 0.74\pm 0.16$  &    42.98  \cr
      41  &  HC221706-001222  &     334.27798  &    0.2063060  &  $ 24.95$  &
$ 26.58$  &   5.6510  &   2  &    1  &  $ 0.60\pm0.09$  &    42.88  \cr
      42  &  HC170704+435811  &     256.76801  &     43.96980  &  $ 24.97$  &
$ 26.08$  &   5.6992  &   1  &    1  &  $ 0.73\pm0.09$  &    42.74  \cr
      43  &  HC024056-012845  &     40.235203  &    -1.479333  &  $ 24.97$  &
$ 27.56$  &   5.7063  &   6  &    1  &  $ 0.96\pm0.09$  &    42.79  \cr
      44  &  HC024121-013220  &     40.340084  &    -1.539055  &  $ 25.00$  &
$ 26.70$  &   5.6759  &   5  &    1  &  $ 0.53\pm0.04$  &    42.79  \cr
\enddata
\label{la_table}
\end{deluxetable*}

\tablenum{5}
\begin{deluxetable*}{clllccccclc}
\small\addtolength{\tabcolsep}{-4pt}
\renewcommand\baselinestretch{1.0}
\tablewidth{0pt}
\tablecaption{$z=5.7$ Ly$\alpha$ Emitters}
\scriptsize
\tablehead{Number & Name & R.A. & Decl. & $N$ & $I$ & Redshift &  Expo & Qual & FWHM & Log(L)\\ & &(J2000) & (J2000) & (AB) & (AB)  & & (hrs) & & \AA & erg/s  \\ (1) & (2) & (3) & (4)  & (5) & (6) & (7) & (8) & (9) & (10) & (11)}
\startdata
      45  &  HC024038-013500  &     40.161293  &    -1.583389  &  $ 25.01$  &
$ 26.81$  &   5.7063  &   3  &    1  &  $ 0.92\pm 0.10$  &    42.77  \cr
      46  &  HC221733-004202  &     334.39001  &    0.7005829  &  $ 25.01$  &
$ 26.81$  &   5.6180  &   2  &    1  &  $  1.00\pm 0.12$  &    43.20  \cr
      47  &  HC170657+434626  &     256.73898  &     43.77400  &  $ 25.02$  &
$ 25.80$  &   5.6599  &   5  &    1  &  $  1.07\pm 0.11$  &    42.86  \cr
      48  &  HC221816-003647  &     334.56903  &    0.6131939  &  $ 25.06$  &
$ 26.42$  &   5.7314  &   1  &    1  &  $ 0.76\pm0.09$  &    42.86  \cr
      49  &  HC124052+621119  &     190.21700  &     62.18880  &  $ 25.07$  &
$ 26.79$  &   5.6835  &   1  &    1  &  $ 0.62\pm0.03$  &    42.72  \cr
      50  &  HC123516+620508  &     188.81900  &     62.08561  &  $ 25.08$  &
$ 26.43$  &   5.7220  &   1  &    1  &  $ 0.53\pm 0.11$  &    42.77  \cr
      51  &  HC124130+622621  &     190.37903  &     62.43938  &  $ 25.10$  &
$ 26.47$  &   5.6560  &   1  &    3  &  $ 0.73\pm0.04$  &    42.91  \cr
      52  &  HC221752-003615  &     334.47000  &    0.6043059  &  $ 25.12$  &
$ 28.08$  &   5.7330  &   1  &    1  &  $ 0.80\pm0.09$  &    42.84  \cr
      53  &  HC221740-002414  &     334.42004  &    0.4040830  &  $ 25.13$  &
$ 25.95$  &   5.6350  &   4  &    1  &  $  1.18\pm 0.10$  &    43.05  \cr
      54  &  HC024029-013919  &     40.120998  &    -1.655500  &  $ 25.15$  &
$ 26.15$  &   5.6427  &   2  &    1  &  $  1.02\pm 0.12$  &    42.95  \cr
      55  &  HC124107+622316  &     190.28101  &     62.38780  &  $ 25.15$  &
$ 27.03$  &   5.6858  &   2  &    1  &  $ 0.70\pm0.09$  &    42.66  \cr
      56  &  HC124103+622152  &     190.26601  &     62.36460  &  $ 25.16$  &
$ 26.50$  &   5.7430  &   2  &    1  &  $  1.04\pm 0.10$  &    42.94  \cr
      57  &  HC024042-013316  &     40.176788  &    -1.554556  &  $ 25.18$  &
$ 29.33$  &   5.6543  &   6  &    1  &  $ 0.79\pm0.06$  &    42.85  \cr
      58  &  HC221845-003405  &     334.69000  &    0.5683060  &  $ 25.18$  &
$ 27.62$  &   5.6909  &   4  &    1  &  $ 0.73\pm0.08$  &    42.67  \cr
      59  &  HC123607+620838  &     189.03300  &     62.14411  &  $ 25.20$  &
$ 27.10$  &   5.6400  &   2  &    1  &  $ 0.85\pm 0.10$  &    42.95  \cr
      60  &  HC221652-001639  &     334.22000  &    0.2777499  &  $ 25.20$  &
$ 28.89$  &   5.6589  &   4  &    1  &  $  1.07\pm0.04$  &    42.82  \cr
      61  &  HC170614+434815  &     256.56201  &     43.80438  &  $ 25.23$  &
$ 102.2$  &   5.6655  &   1  &    2  &  $ 0.90\pm 0.13$  &    42.70  \cr
      62  &  HC123952+621034  &     189.97000  &     62.17630  &  $ 25.25$  &
$ 27.43$  &   5.6356  &   1  &    1  &  $  1.19\pm 0.11$  &    42.98  \cr
      63  &  HC170707+435530  &     256.78299  &     43.92511  &  $ 25.25$  &
$ 26.07$  &   5.6992  &   4  &    1  &  $ 0.91\pm 0.10$  &    42.67  \cr
      64  &  HC024111-012855  &     40.299500  &    -1.482222  &  $ 25.25$  &
$ 26.40$  &   5.6954  &   4  &    1  &  $ 0.72\pm0.03$  &    42.67  \cr
      65  &  HC123609+620244  &     189.03897  &     62.04569  &  $ 25.26$  &
$ 30.28$  &   5.7183  &   1  &    1  &  $  1.29\pm 0.14$  &    42.72  \cr
      66  &  HC123612+620420  &     189.05000  &     62.07230  &  $ 25.32$  &
$ 27.52$  &   5.7410  &   1  &    1  &  $ 0.77\pm0.02$  &    42.82  \cr
      67  &  HC221810-003622  &     334.54199  &    0.6061940  &  $ 25.35$  &
$ 27.46$  &   5.6885  &   2  &    3  &  $ 0.94\pm 0.13$  &    42.57  \cr
      68  &  HC123821+621046  &     189.59100  &     62.17961  &  $ 25.39$  &
$ 26.45$  &   5.6490  &   2  &    2  &  $  1.17\pm 0.10$  &    42.83  \cr
      69  &  HC170641+440756  &     256.67099  &     44.13230  &  $ 25.40$  &
$ 102.2$  &   5.6923  &   1  &    1  &  $ 0.93\pm0.08$  &    42.60  \cr
      70  &  HC221658-000836  &     334.24402  &    0.1433890  &  $ 25.41$  &
$ 26.88$  &   5.7353  &   2  &    1  &  $ 0.90\pm0.09$  &    42.76  \cr
      71  &  HC221725-000752  &     334.35501  &    0.1312779  &  $ 25.45$  &
$ 26.88$  &   5.7509  &   1  &    1  &  $ 0.98\pm 0.10$  &    42.93  \cr
      72  &  HC221904-002520  &     334.76797  &    0.4224439  &  $ 25.46$  &
$ 101.9$  &   5.6242  &   2  &    1  &  $  1.03\pm 0.12$  &    42.96  \cr
      73  &  HC123532+621445  &     188.88399  &     62.24589  &  $ 25.46$  &
$ 29.18$  &   5.7026  &   1  &    3  &  $ 0.83\pm 0.13$  &    42.57  \cr
      74  &  HC221658-001849  &     334.24500  &    0.3137220  &  $ 25.48$  &
$ 26.67$  &   5.6281  &   2  &    1  &  $  1.34\pm 0.14$  &    42.93  \cr
      75  &  HC221909-004121  &     334.79001  &    0.6893330  &  $ 25.49$  &
$ 27.54$  &   5.6882  &   2  &    2  &  $  1.17\pm 0.13$  &    42.59  \cr
      76  &  HC124015+621739  &     190.06601  &     62.29419  &  $ 25.49$  &
$ 29.16$  &   5.6913  &   2  &    1  &  $ 0.85\pm 0.12$  &    42.57  \cr
      77  &  HC124154+622439  &     190.47501  &     62.41088  &  $ 25.49$  &
$ 102.2$  &   5.7448  &   2  &    1  &  $ 0.91\pm0.08$  &    42.82  \cr
      78  &  HC123558+621017  &     188.99498  &     62.17150  &  $ 25.51$  &
$ 29.42$  &   5.6718  &   2  &    3  &  $  2.35\pm 0.20$  &    42.58  \cr
      79  &  HC124007+621245  &     190.03101  &     62.21269  &  $ 25.51$  &
$ 27.29$  &   5.7388  &   1  &    1  &  $ 0.81\pm0.09$  &    42.74  \cr
      80  &  HC221750-002907  &     334.45898  &    0.4854440  &  $ 25.53$  &
$ 28.21$  &   5.6945  &   1  &    3  &  $ 0.20\pm0.04$  &    42.50  \cr
      81  &  HC024031-013230  &     40.132500  &    -1.541778  &  $ 25.57$  &
$-27.44$  &   5.7104  &   2  &    1  &  $  1.15\pm0.04$  &    42.57  \cr
      82  &  HC221744-004024  &     334.43600  &    0.6734719  &  $ 25.58$  &
$ 28.61$  &   5.6860  &   2  &    2  &  $  1.11\pm0.03$  &    42.54  \cr
      83  &  HC221757-004118  &     334.48898  &    0.6885830  &  $ 25.68$  &
$ 101.9$  &   5.6845  &   1  &    2  &  $ 0.85\pm 0.11$  &    42.47  \cr
      84  &  HC221846-003156  &     334.69299  &    0.5324170  &  $ 25.68$  &
$ 26.67$  &   5.6409  &   1  &    1  &  $ 0.76\pm0.09$  &    42.70  \cr
      85  &  HC123756+622415  &     189.48500  &     62.40438  &  $ 25.74$  &
$ 28.84$  &   5.6499  &   1  &    1  &  $ 0.90\pm 0.12$  &    42.60  \cr
      86  &  HC221812-004330  &     334.55197  &    0.7251110  &  $ 25.77$  &
$ 27.86$  &   5.6440  &   1  &    1  &  $  1.17\pm 0.12$  &    42.70  \cr
      87  &  HC221840-003720  &     334.67004  &    0.6223610  &  $ 25.79$  &
$ 27.76$  &   5.6590  &   2  &    2  &  $ 0.57\pm0.02$  &    42.55  \cr
      88  &  HC221802-003007  &     334.51001  &    0.5020279  &  $ 25.80$  &
$ 27.02$  &   5.6598  &   2  &    3  &  $  1.18\pm0.05$  &    42.53  \cr
\enddata
\end{deluxetable*}

\begin{figure*}[h]
\figurenum{A1}
\includegraphics[width=8.5in,angle=0,scale=0.95]{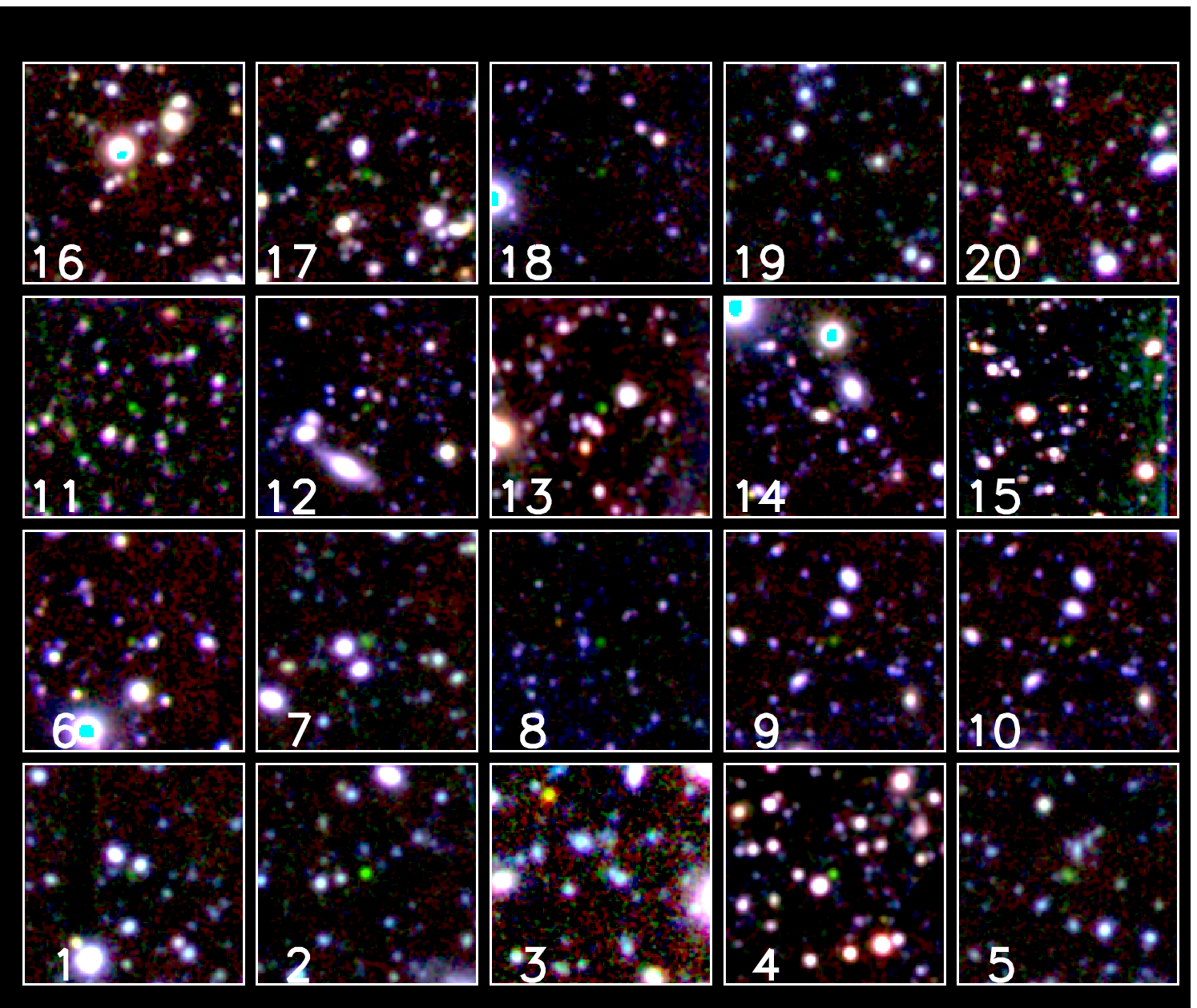}
\caption{Images of the spectroscopically confirmed $z=5.7$ Ly$\alpha$ 
emitter sample.  For each object we show a $40''$ thumbnail around 
the emitter with blue=$R$-band, green=F816 narrowband, and red=$z$-band. 
The emitter appears as a green object at the center of the thumbnail.
The numerical label corresponds to the number in Table~A2
and Figure~A2.
\label{figA1:z5_images}
}
\end{figure*}

\begin{figure*}[h]
\figurenum{A1}
\includegraphics[width=8.5in,angle=0,scale=0.95]{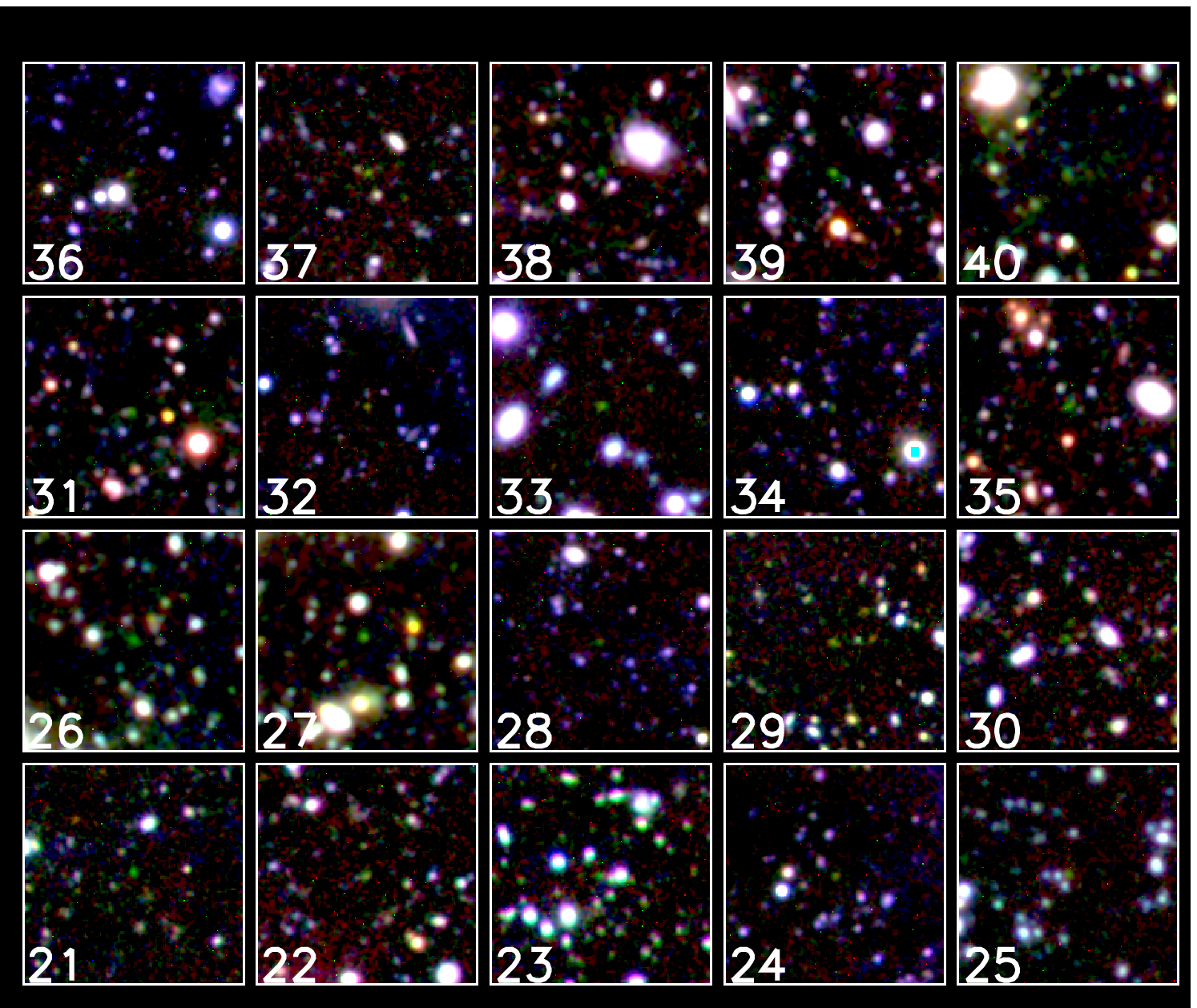}
\caption{Images of the spectroscopically
confirmed $z=5.7$ Ly$\alpha$ emitter sample (continued).
}
\end{figure*}

\begin{figure*}[h]
\figurenum{A1}
\includegraphics[width=8.5in,angle=0,scale=0.95]{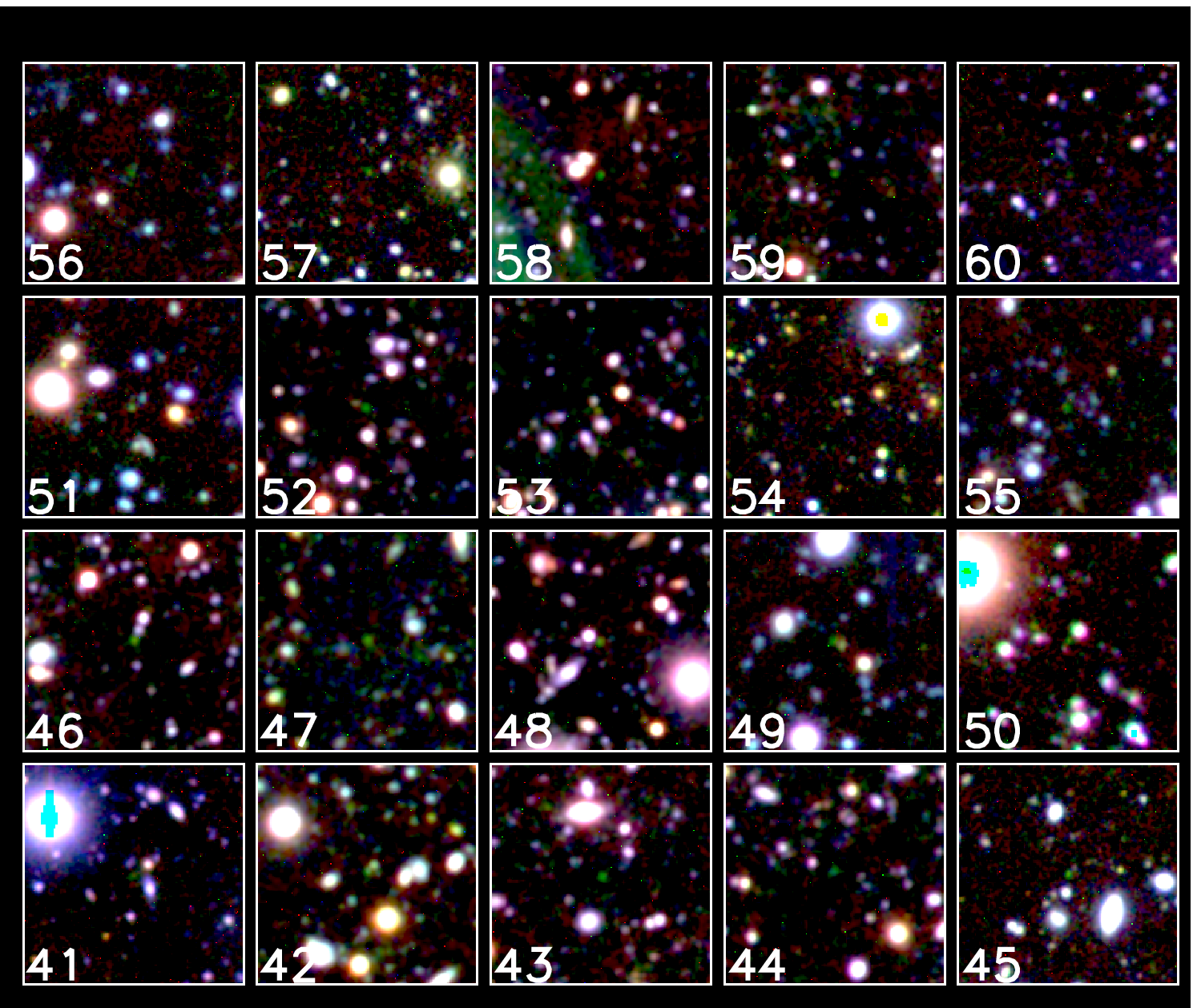}
\caption{Images of the spectroscopically 
confirmed $z=5.7$ Ly$\alpha$ emitter sample (continued).
}
\end{figure*}

\begin{figure*}[h]
\figurenum{A1}
\includegraphics[width=8.5in,angle=0,scale=0.95]{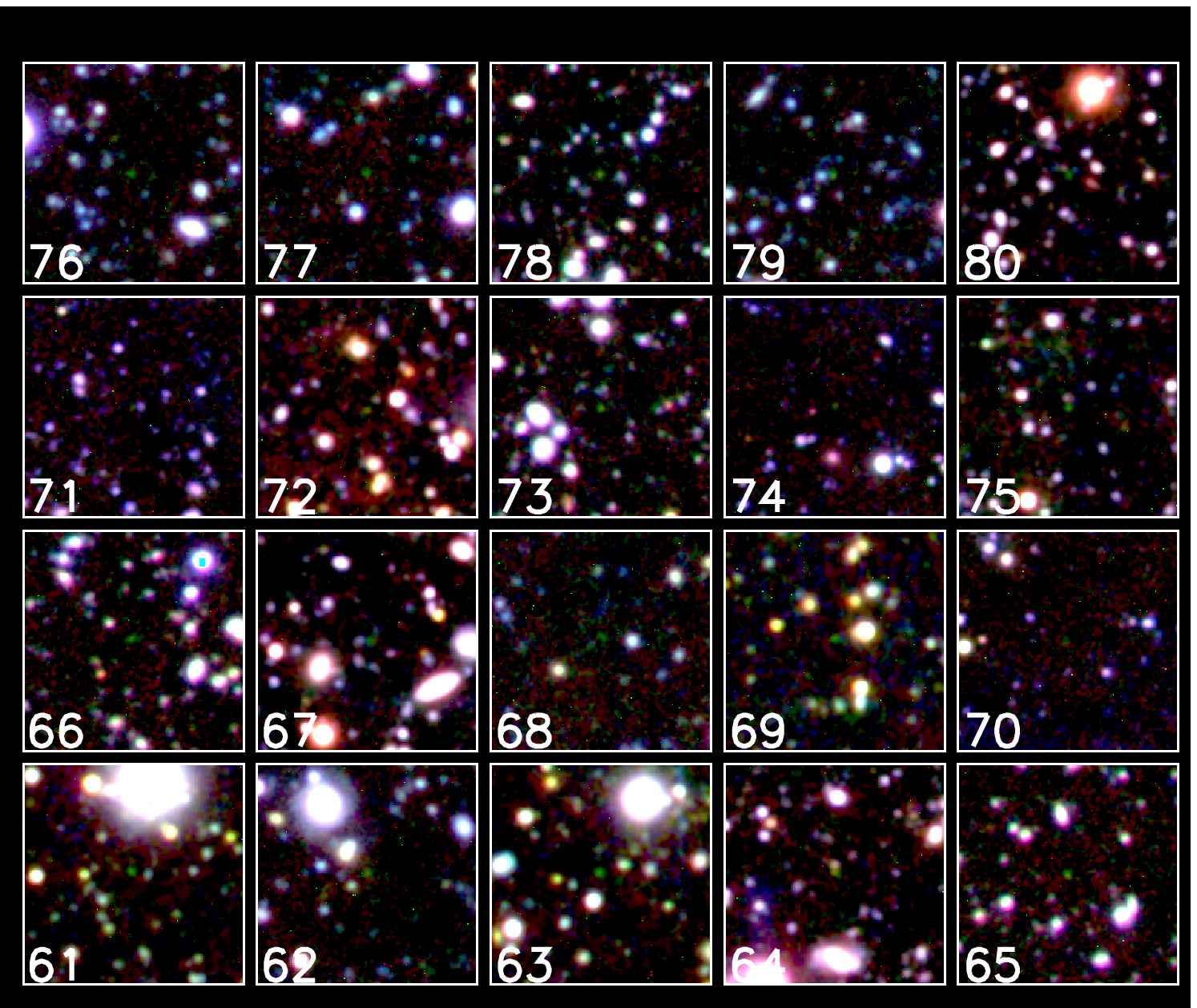}
\caption{Images of the spectroscopically
confirmed $z=5.7$ Ly$\alpha$ emitter sample (continued).
}
\end{figure*}

\begin{figure*}[h]
\figurenum{A1}
\includegraphics[width=8.5in,angle=0,scale=0.95]{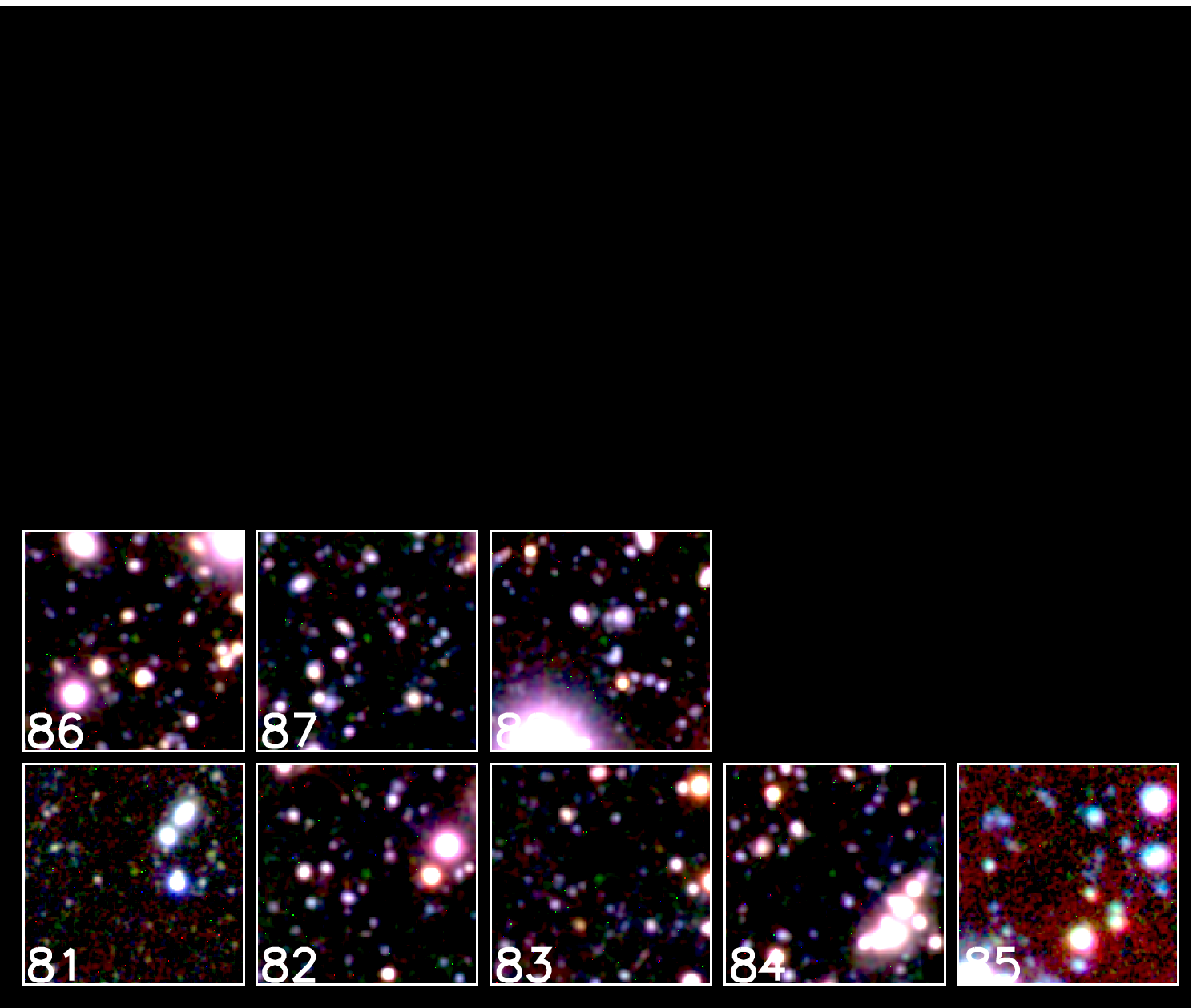}
\caption{Images of the spectroscopically
confirmed $z=5.7$ Ly$\alpha$ emitter sample (continued).
}
\end{figure*}

\begin{figure*}[h]
\figurenum{A2}
\includegraphics[width=4.6in,angle=90,scale=0.95]{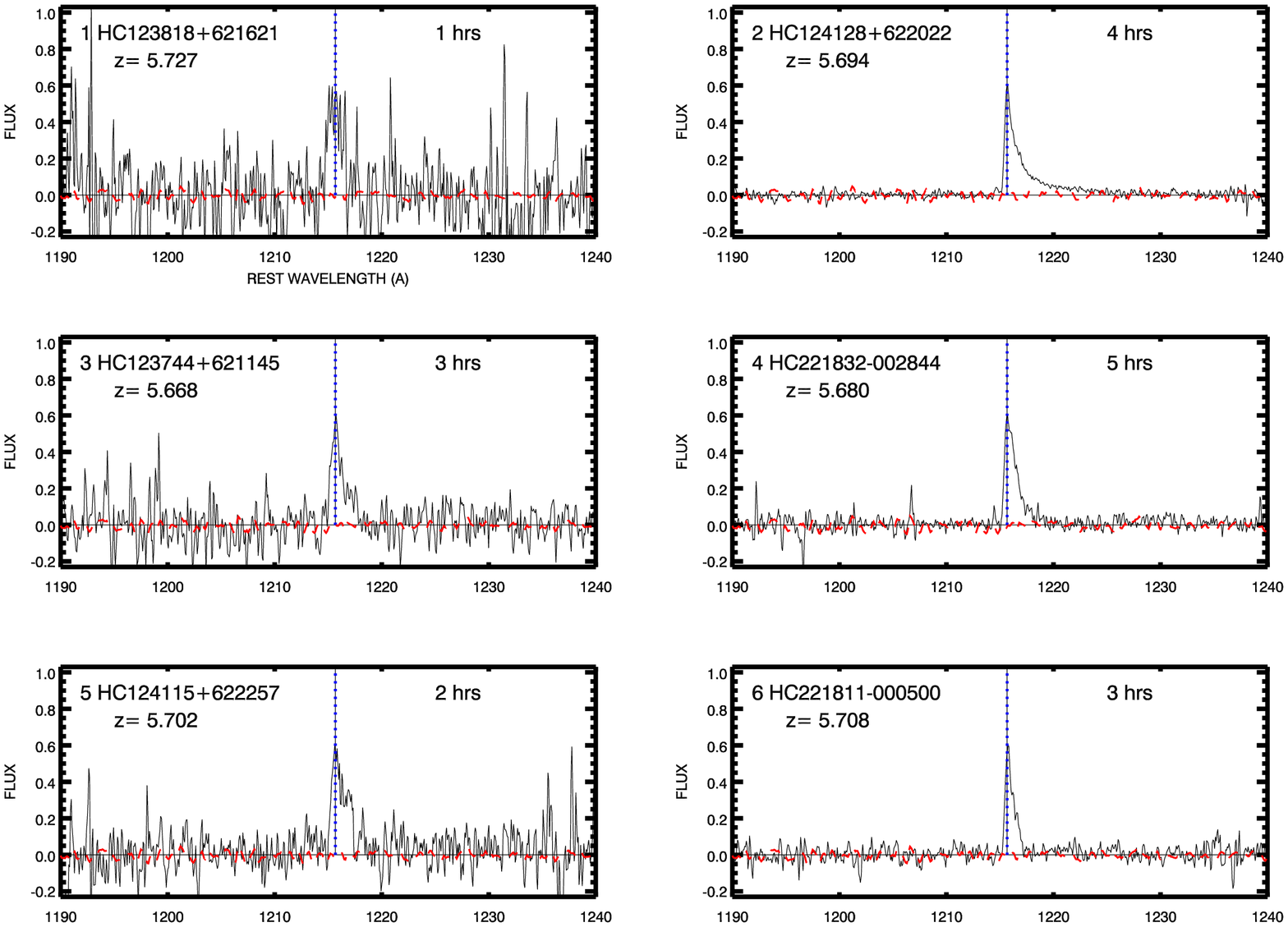}
\includegraphics[width=4.6in,angle=90,scale=0.95]{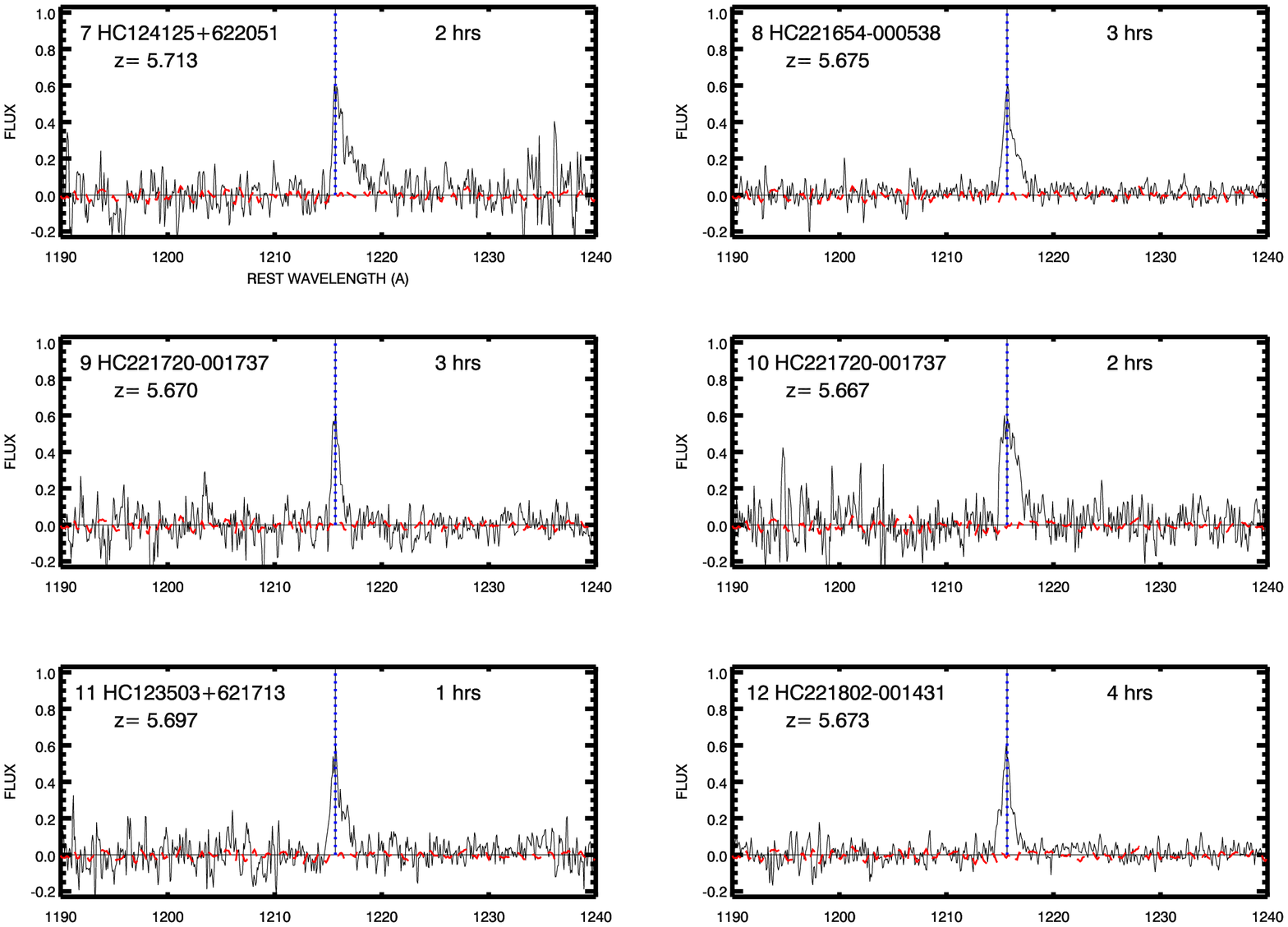}
\caption{Spectra of the spectroscopically
confirmed $z=5.7$ Ly$\alpha$ emitter sample.
For each object we show the spectrum (black) compared with the
average spectral shape of the entire sample (red dashed).
The blue dotted line shows the position of the spectrum peak,
which we use to define the redshift. The name of the object
and its redshift is given in the upper left, and the exposure
time in hours is given in the upper right. 
\label{figA2:z5_spectra}
}
\end{figure*}

\begin{figure*}[h]
\figurenum{A2}
\includegraphics[width=4.6in,angle=90,scale=0.95]{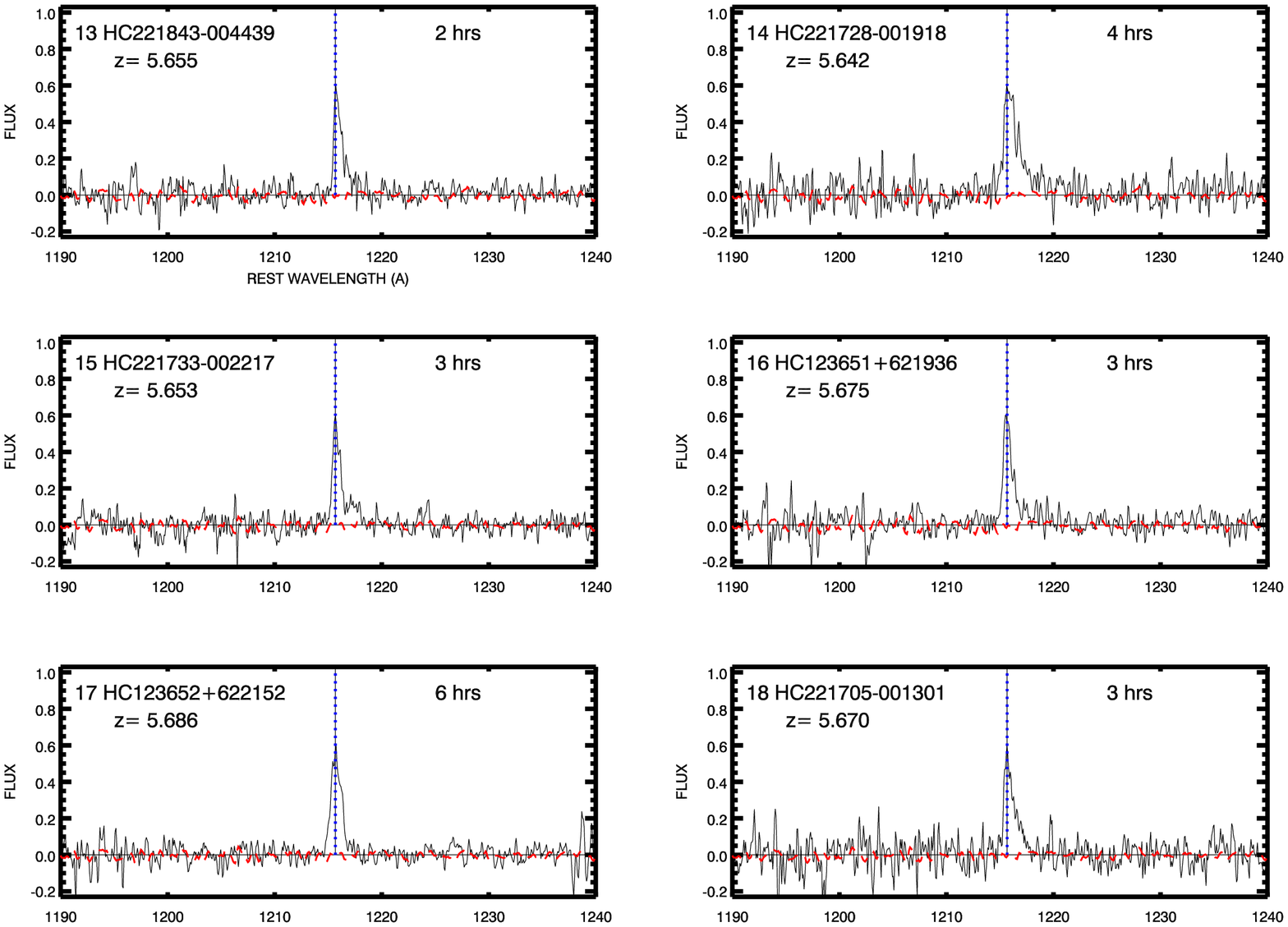}
\includegraphics[width=4.6in,angle=90,scale=0.95]{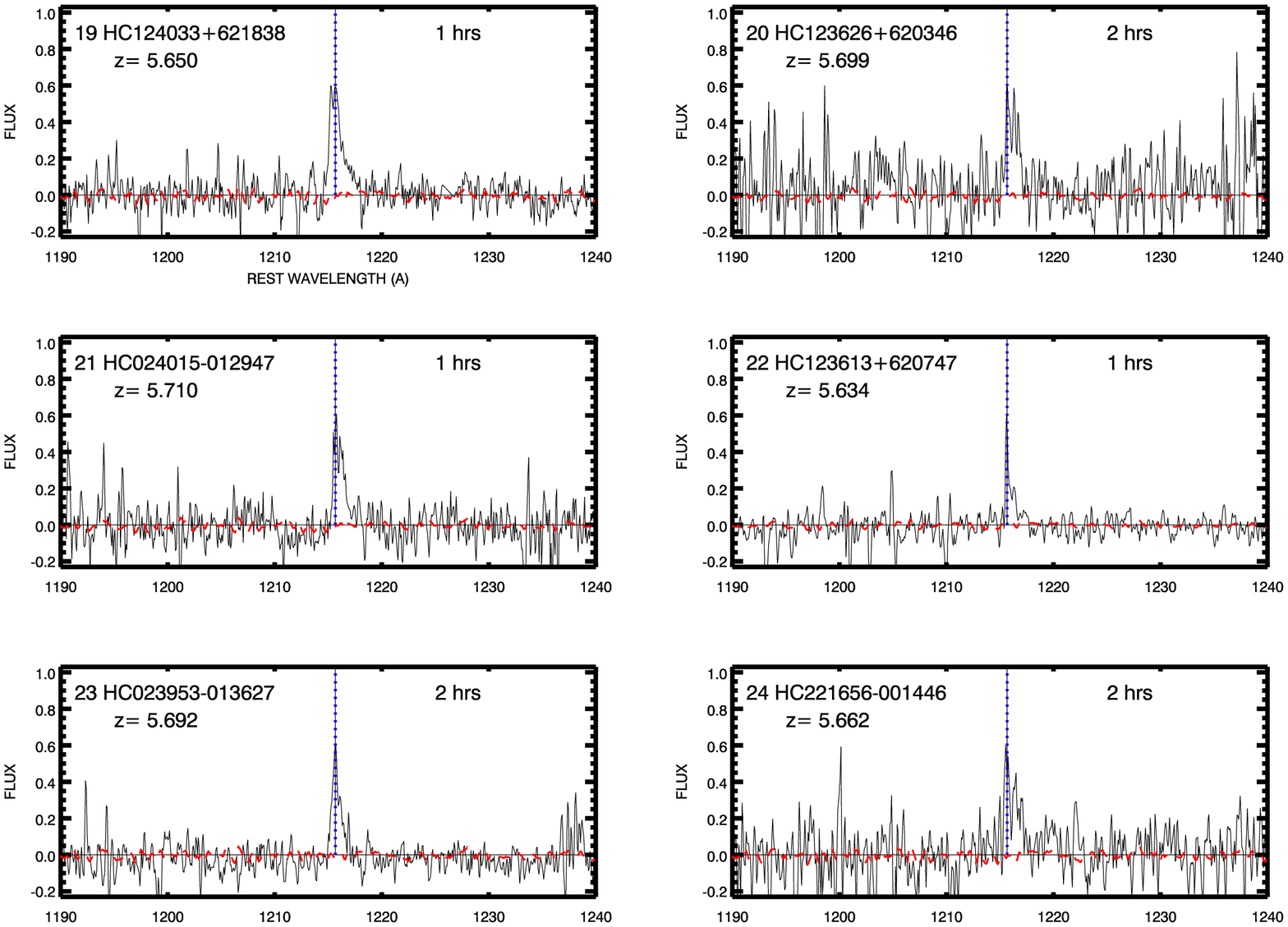}
\caption{Spectra of the spectroscopically 
confirmed $z=5.7$ Ly$\alpha$ emitter sample (continued).
}
\end{figure*}

\begin{figure*}[h]
\figurenum{A2}
\includegraphics[width=4.6in,angle=90,scale=0.95]{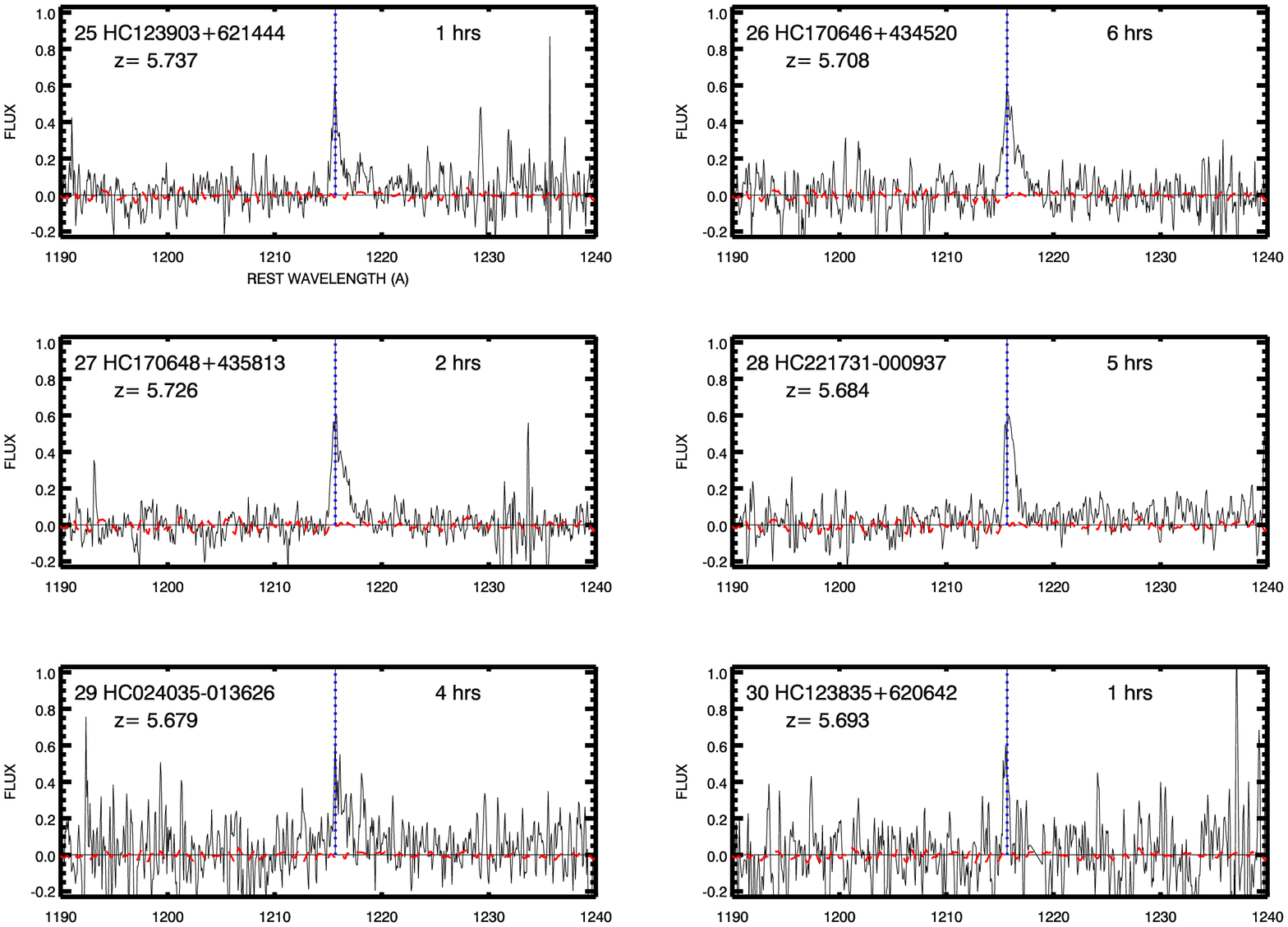}
\includegraphics[width=4.6in,angle=90,scale=0.95]{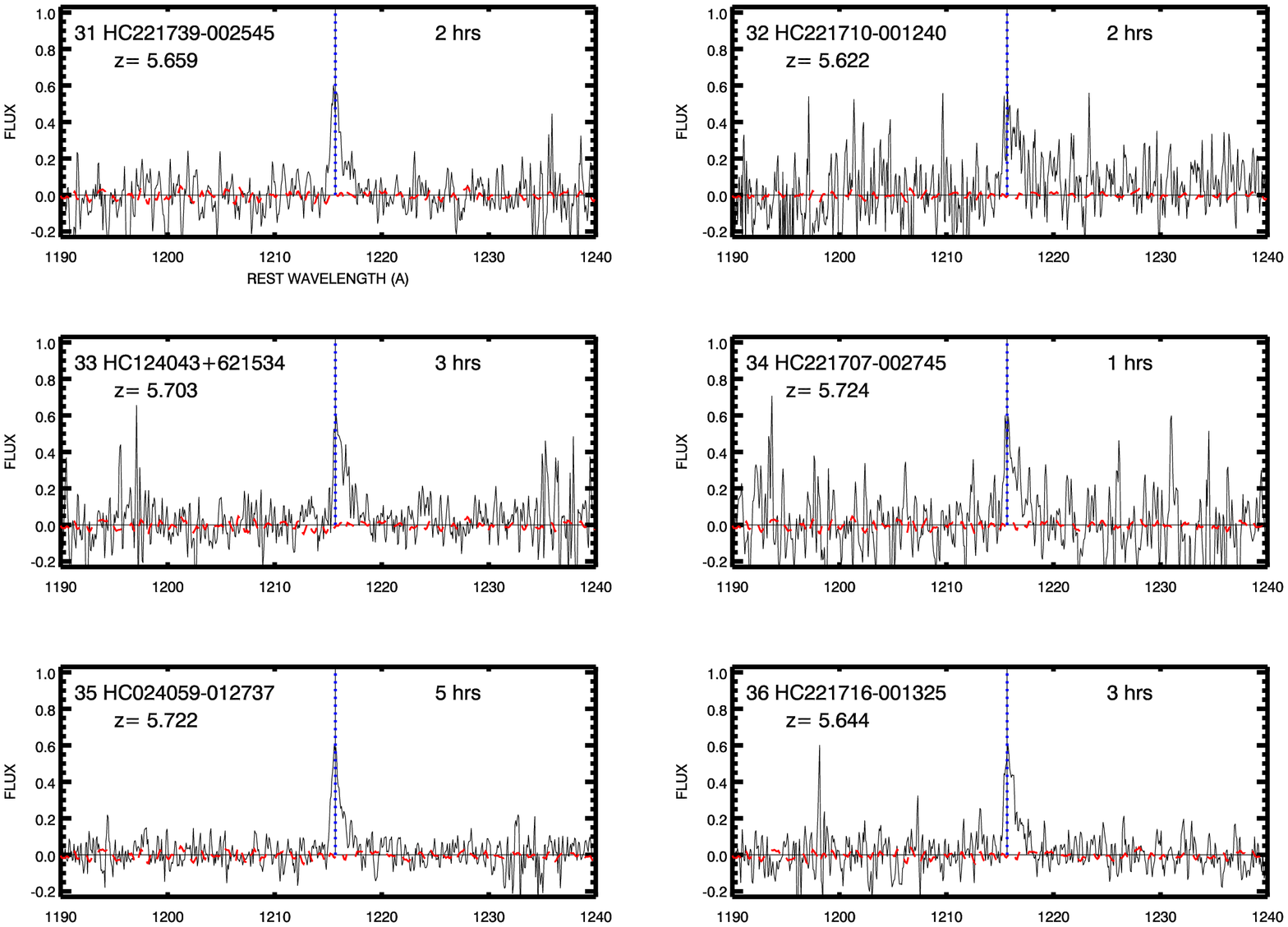}
\caption{Spectra of the spectroscopically 
confirmed $z=5.7$ Ly$\alpha$ emitter sample (continued).
}
\end{figure*}

\begin{figure*}[h]
\figurenum{A2}
\includegraphics[width=4.6in,angle=90,scale=0.95]{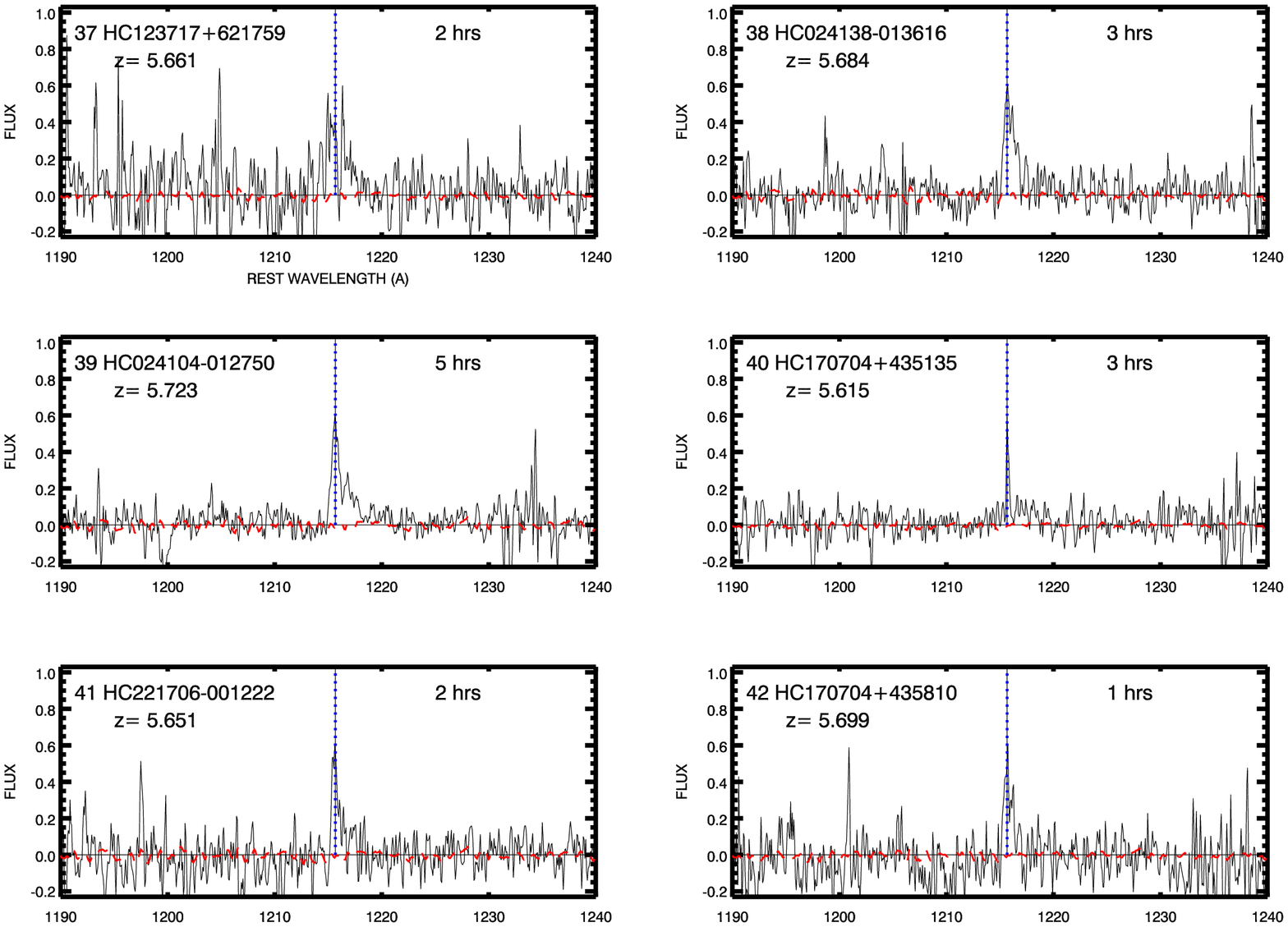}
\includegraphics[width=4.6in,angle=90,scale=0.95]{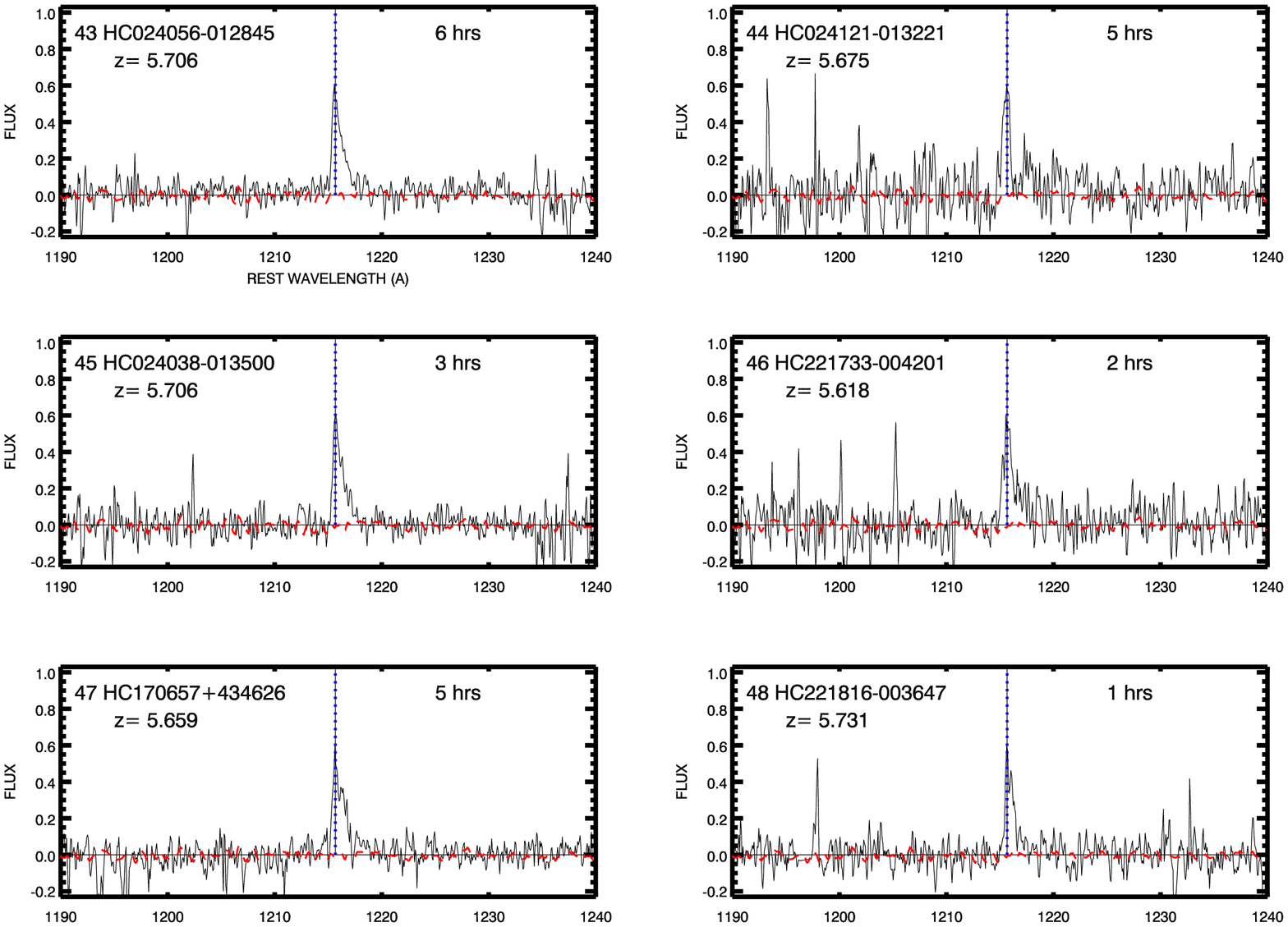}
\caption{Spectra of the spectroscopically 
confirmed $z=5.7$ Ly$\alpha$ emitter sample (continued).
}
\end{figure*}

\begin{figure*}[h]
\figurenum{A2}
\includegraphics[width=4.6in,angle=90,scale=0.95]{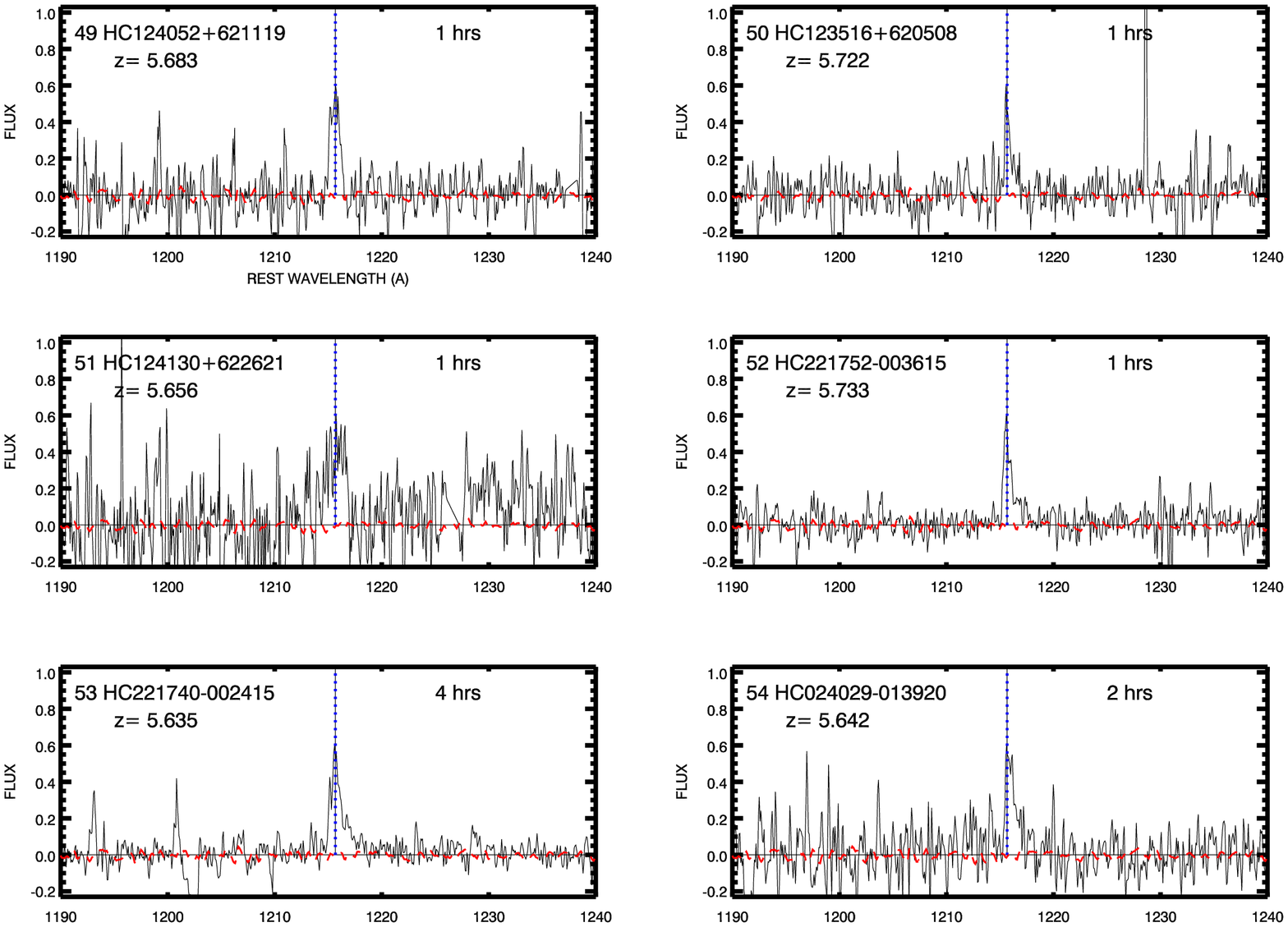}
\includegraphics[width=4.6in,angle=90,scale=0.95]{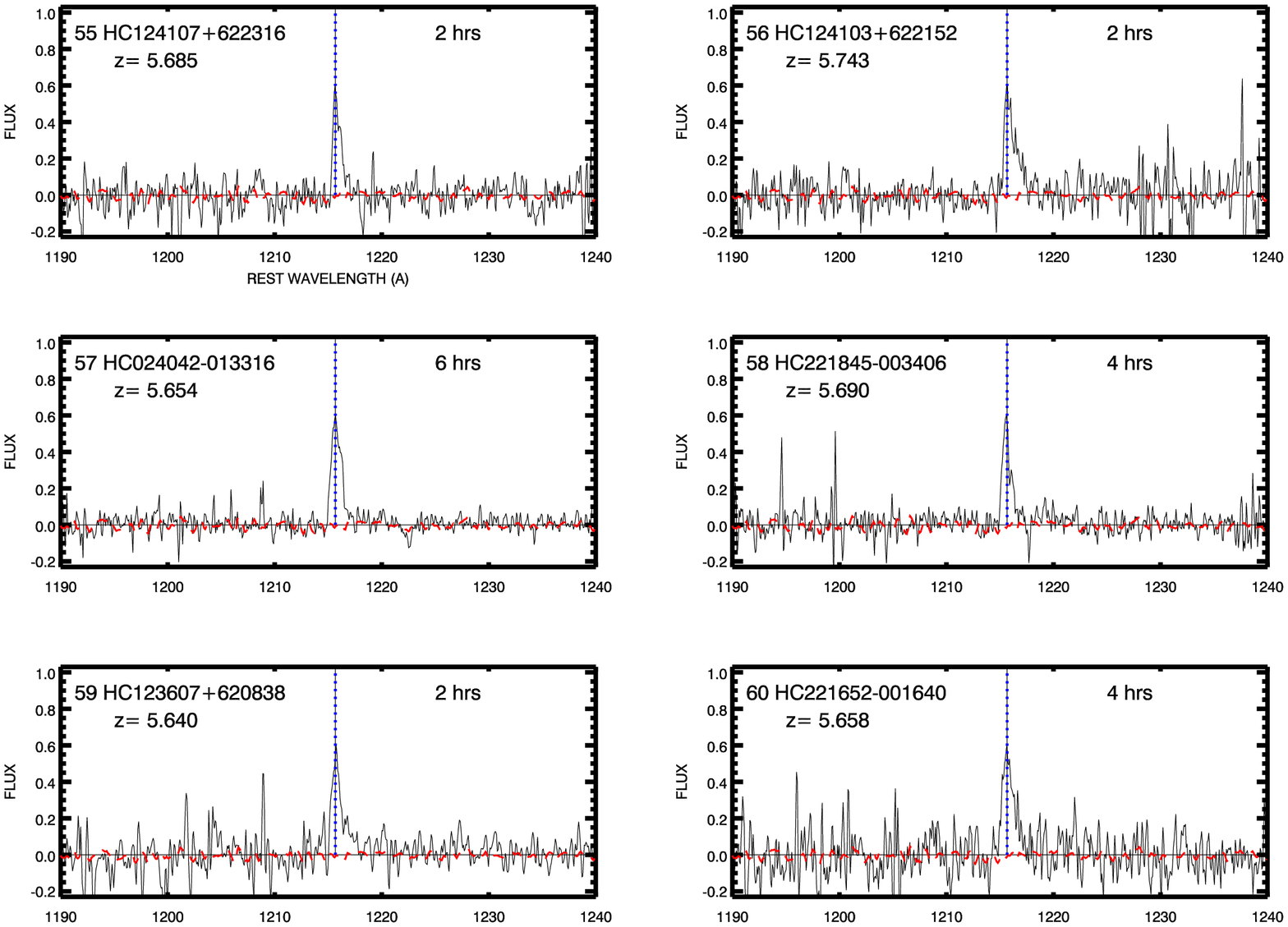}
\caption{Spectra of the spectroscopically 
confirmed $z=5.7$ Ly$\alpha$ emitter sample (continued).
}
\end{figure*}

\begin{figure*}[h]
\figurenum{A2}
\includegraphics[width=4.6in,angle=90,scale=0.95]{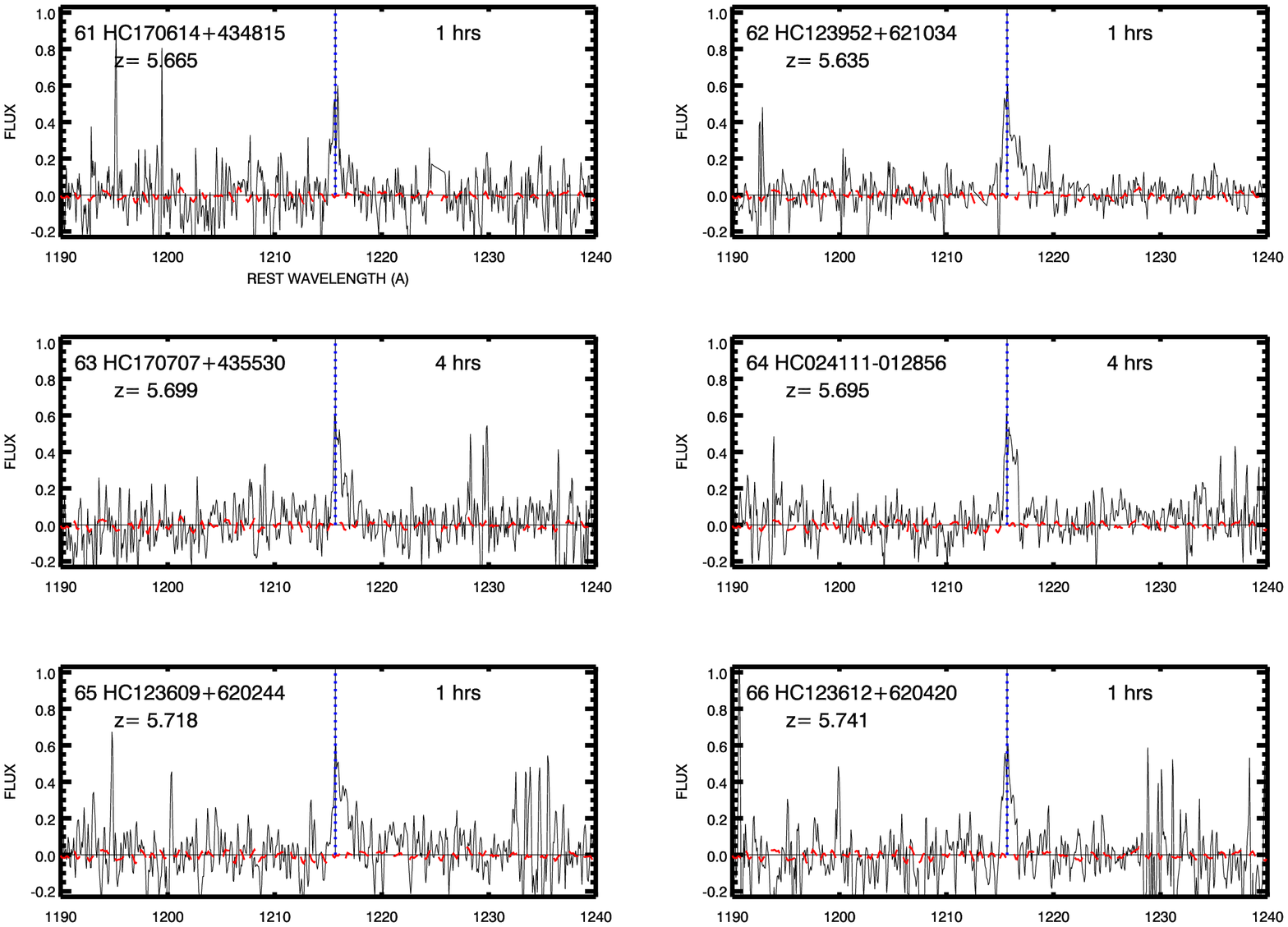}
\includegraphics[width=4.6in,angle=90,scale=0.95]{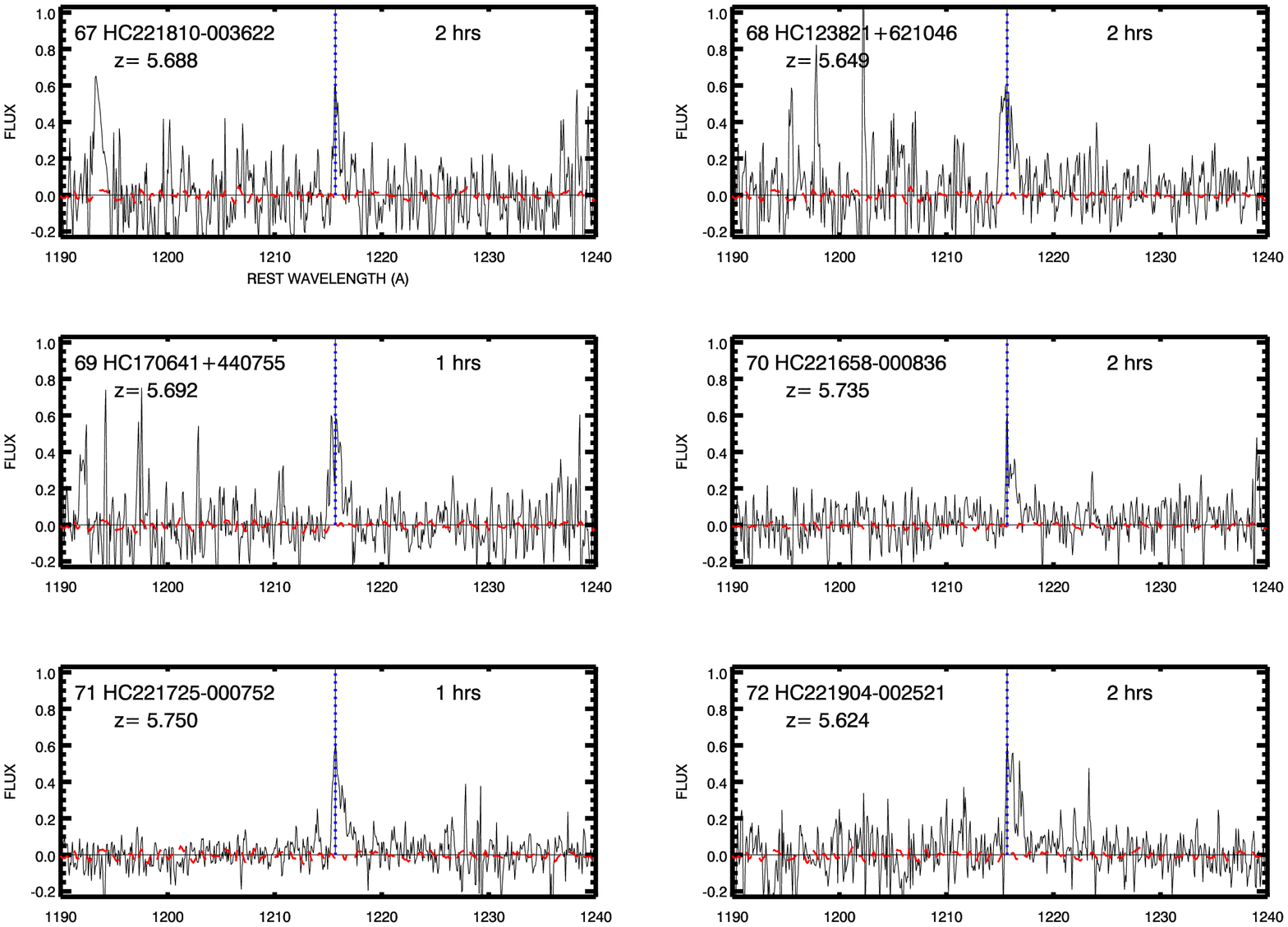}
\caption{Spectra of the spectroscopically 
confirmed $z=5.7$ Ly$\alpha$ emitter sample (continued).
}
\end{figure*}

\begin{figure*}[h]
\figurenum{A2}
\includegraphics[width=4.6in,angle=90,scale=0.95]{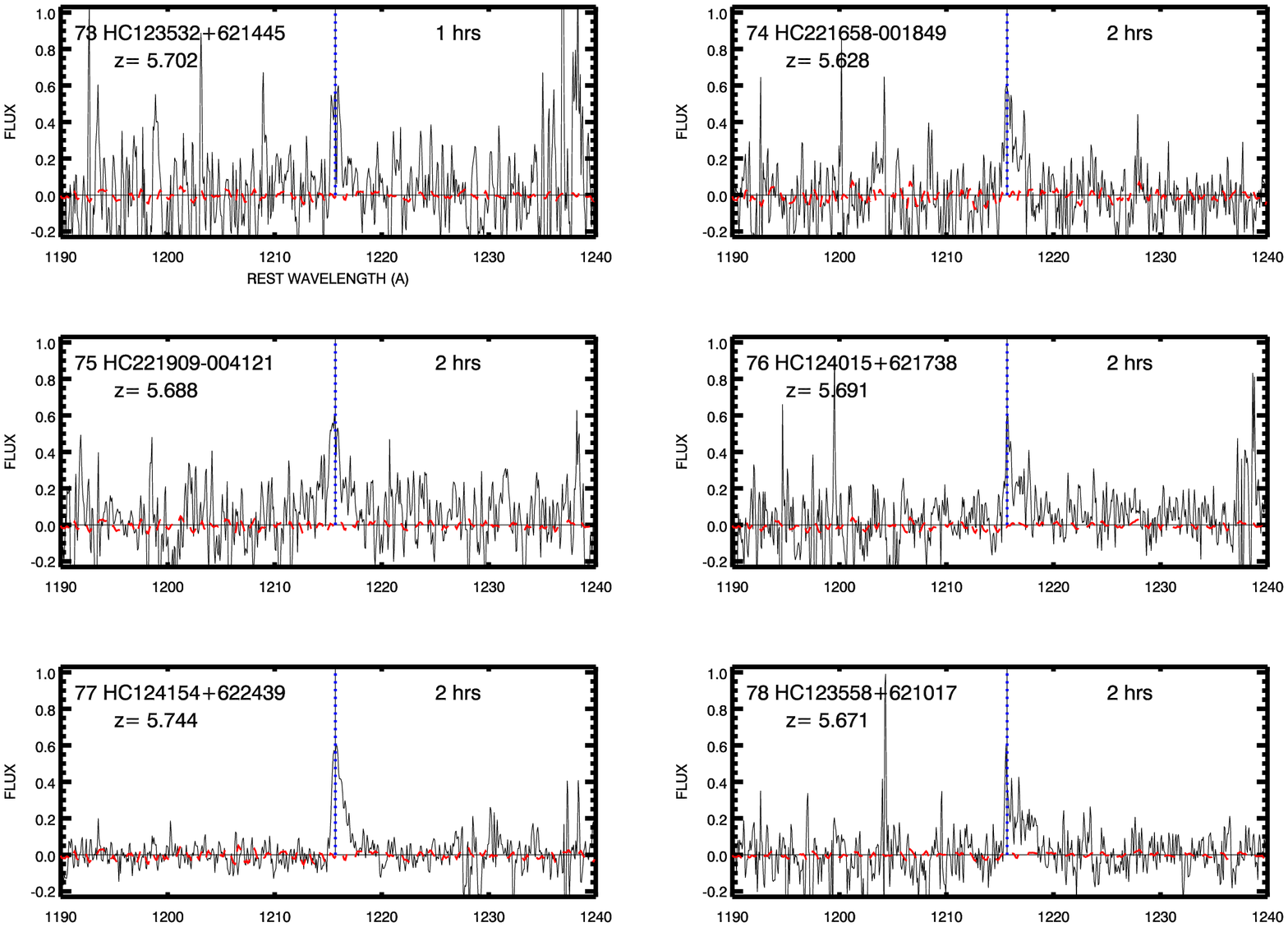}
\includegraphics[width=4.6in,angle=90,scale=0.95]{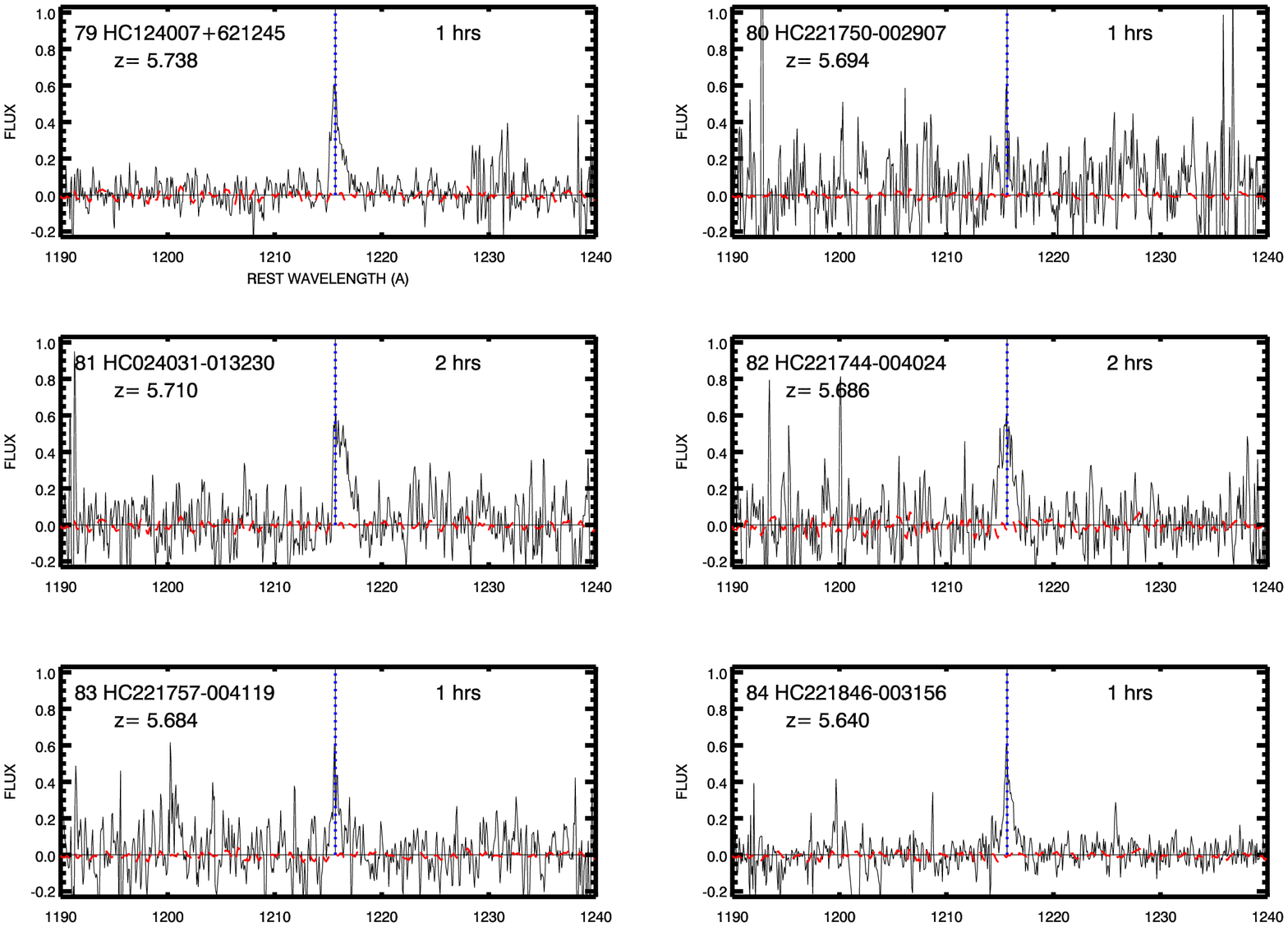}
\caption{Spectra of the spectroscopically 
confirmed $z=5.7$ Ly$\alpha$ emitter sample (continued).
}
\end{figure*}

\begin{figure*}[h]
\figurenum{A2}
\includegraphics[width=4.6in,angle=90,scale=0.95]{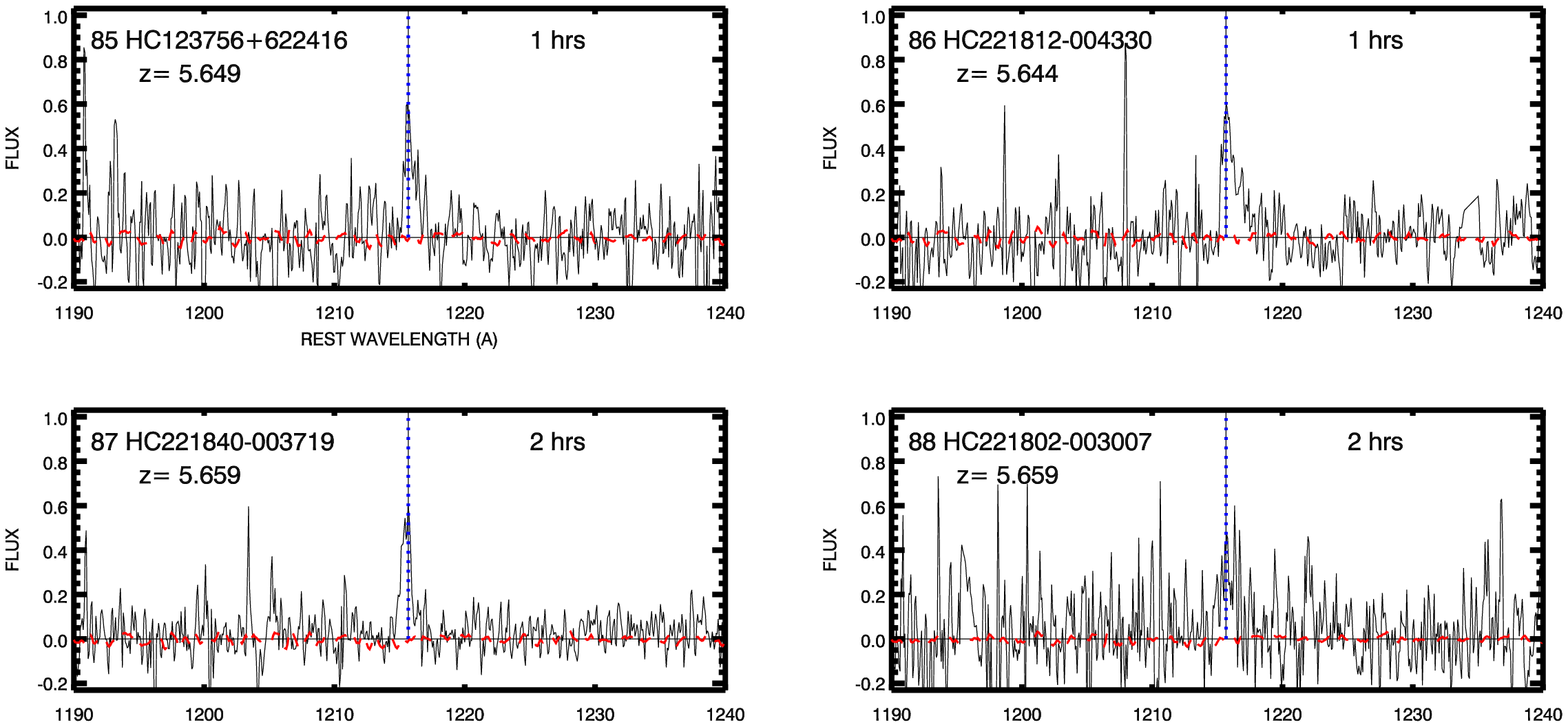}
\caption{Spectra of the spectroscopically 
confirmed $z=5.7$ Ly$\alpha$ emitter sample (continued).
}
\end{figure*}

\clearpage
\setcounter{figure}{1}

\tablenum{6}
\begin{deluxetable*}{clllccccclc}
\small\addtolength{\tabcolsep}{-4pt}
\renewcommand\baselinestretch{1.0}
\tablewidth{0pt}
\tablecaption{$z=6.5$ Ly$\alpha$ Emitters}
\scriptsize
\tablehead{Number & Name & R.A. & Decl. & $N$ & $z$ & Redshift &  Expo & Qual & FWHM & Log(L)\\ & &(J2000) & (J2000) & (AB) & (AB)  & & (hrs) & & \AA & erg/s  \\ (1) & (2) & (3) & (4)  & (5) & (6) & (7) & (8) & (9) & (10) & (11)}
\startdata
       1  &  HC221741-003134  &     334.42105  &    0.5262219  &  $ 24.01$  &
$ 26.60$  &   6.5008  &   2  &    1  &  $  1.50\pm 0.18$  &    43.28  \cr
       2  &  HC170716+435039  &     256.81799  &     43.84438  &  $ 24.06$  &
$ 26.27$  &   6.5272  &   1  &    1  &  $ 0.98\pm 0.11$  &    43.29  \cr
       3  &  HC221725-001119  &     334.35703  &    0.1886940  &  $ 24.12$  &
$ 25.20$  &   6.5142  &   2  &    1  &  $ 0.69\pm0.03$  &    43.22  \cr
       4  &  HC023954-013332  &     39.978001  &    -1.559111  &  $ 24.24$  &
$ 25.18$  &   6.5587  &   6  &    1  &  $ 0.76\pm0.07$  &    43.41  \cr
       5  &  HC221742-001808  &     334.42896  &    0.3023610  &  $ 24.36$  &
$ 26.15$  &   6.4692  &   3  &    1  &  $ 0.80\pm0.08$  &    43.20  \cr
       6  &  HC221749-004825  &     334.45700  &    0.8070280  &  $ 24.50$  &
$ 27.33$  &   6.4903  &   1  &    1  &  $ 0.83\pm 0.10$  &    43.09  \cr
       7  &  HC221827-004727  &     334.61499  &    0.7909169  &  $ 24.53$  &
$ 27.81$  &   6.5036  &   1  &    1  &  $ 0.73\pm0.03$  &    43.05  \cr
       8  &  HC221831-004012  &     334.63306  &    0.6700829  &  $ 24.80$  &
$ 26.53$  &   6.5683  &   2  &    1  &  $  1.16\pm 0.14$  &    43.28  \cr
       9  &  HC221848-004353  &     334.70294  &    0.7315830  &  $ 24.81$  &
$ 27.87$  &   6.5232  &   2  &    1  &  $ 0.95\pm 0.11$  &    42.95  \cr
      10  &  HC124215+621729  &     190.56500  &     62.29150  &  $ 24.84$  &
$ 27.21$  &   6.5140  &   1  &    1  &  $ 0.88\pm 0.12$  &    42.94  \cr
      11  &  HC123512+621911  &     188.80232  &     62.31977  &  $ 24.87$  &
$ 27.04$  &   6.4975  &   2  &    3  &  $ 0.55\pm0.04$  &    42.92  \cr
      12  &  HC124001+621946  &     190.00801  &     62.32960  &  $ 24.95$  &
$-26.28$  &   6.5121  &   1  &    2  &  $ 0.85\pm 0.12$  &    42.88  \cr
      13  &  HC221823-004631  &     334.59802  &    0.7754439  &  $ 25.01$  &
$ 26.40$  &   6.4836  &   1  &    2  &  $ 0.78\pm 0.10$  &    42.90  \cr
      14  &  HC221738-000909  &     334.41000  &    0.1526670  &  $ 25.01$  &
$ 27.41$  &   6.4811  &   2  &    1  &  $ 0.61\pm0.03$  &    42.87  \cr
      15  &  HC023939-013451  &     39.914791  &    -1.581028  &  $ 25.04$  &
$ 28.21$  &   6.5309  &   3  &    1  &  $ 0.86\pm0.09$  &    42.90  \cr
      16  &  HC024055-014315  &     40.229206  &    -1.721000  &  $ 25.07$  &
$-27.52$  &   6.4749  &   2  &    1  &  $ 0.93\pm 0.10$  &    42.91  \cr
      17  &  HC221801-002220  &     334.50806  &    0.3722780  &  $ 25.24$  &
$ 99.20$  &   6.5360  &   2  &    3  &  $ 0.49\pm0.07$  &    42.81  \cr
      18  &  HC024121-012300  &     40.341000  &    -1.383417  &  $ 25.26$  &
$ 26.45$  &   6.5051  &   4  &    1  &  $ 0.76\pm0.08$  &    42.76  \cr
      19  &  HC024134-013642  &     40.394211  &    -1.611778  &  $ 25.26$  &
$ 26.70$  &   6.4680  &   3  &    1  &  $ 0.85\pm0.09$  &    42.86  \cr
      20  &  HC221733-004304  &     334.39102  &    0.7177780  &  $ 25.27$  &
$ 101.3$  &   6.5726  &   1  &    3  &  $ 0.28\pm0.09$  &    43.09  \cr
      21  &  HC023949-013121  &     39.956711  &    -1.522667  &  $ 25.42$  &
$ 25.35$  &   6.5640  &   2  &    1  &  $ 0.76\pm0.08$  &    42.99  \cr
      22  &  HC024004-012252  &     40.017708  &    -1.381278  &  $ 25.47$  &
$ 26.37$  &   6.5024  &   1  &    1  &  $ 0.43\pm0.02$  &    42.62  \cr
      23  &  HC023939-013432  &     39.914417  &    -1.575667  &  $ 25.61$  &
$-27.12$  &   6.5485  &   2  &    2  &  $ 0.44\pm0.03$  &    42.76  \cr
      24  &  HC024014-012414  &     40.060917  &    -1.403972  &  $ 25.65$  &
$ 26.33$  &   6.5454  &   1  &    2  &  $ 0.36\pm0.03$  &    42.71  \cr
      25  &  HC221858-004553  &     334.74194  &    0.7649999  &  $ 25.67$  &
$ 101.3$  &   6.5556  &   2  &    1  &  $ 0.87\pm 0.12$  &    42.81  \cr
      26  &  HC024001-014100  &     40.007500  &    -1.683388  &  $ 25.68$  &
$-25.91$  &   6.5444  &   1  &    1  &  $ 0.75\pm 0.11$  &    42.72  \cr
      27  &  HC023927-013523  &     39.863293  &    -1.589833  &  $ 25.70$  &
$ 27.47$  &   6.4497  &   3  &    1  &  $ 0.50\pm0.08$  &    42.73  \cr
\enddata
\label{high_la_tab}
\end{deluxetable*}

\begin{figure*}[h]
\figurenum{A3}
\includegraphics[width=8.5in,angle=0,scale=0.95]{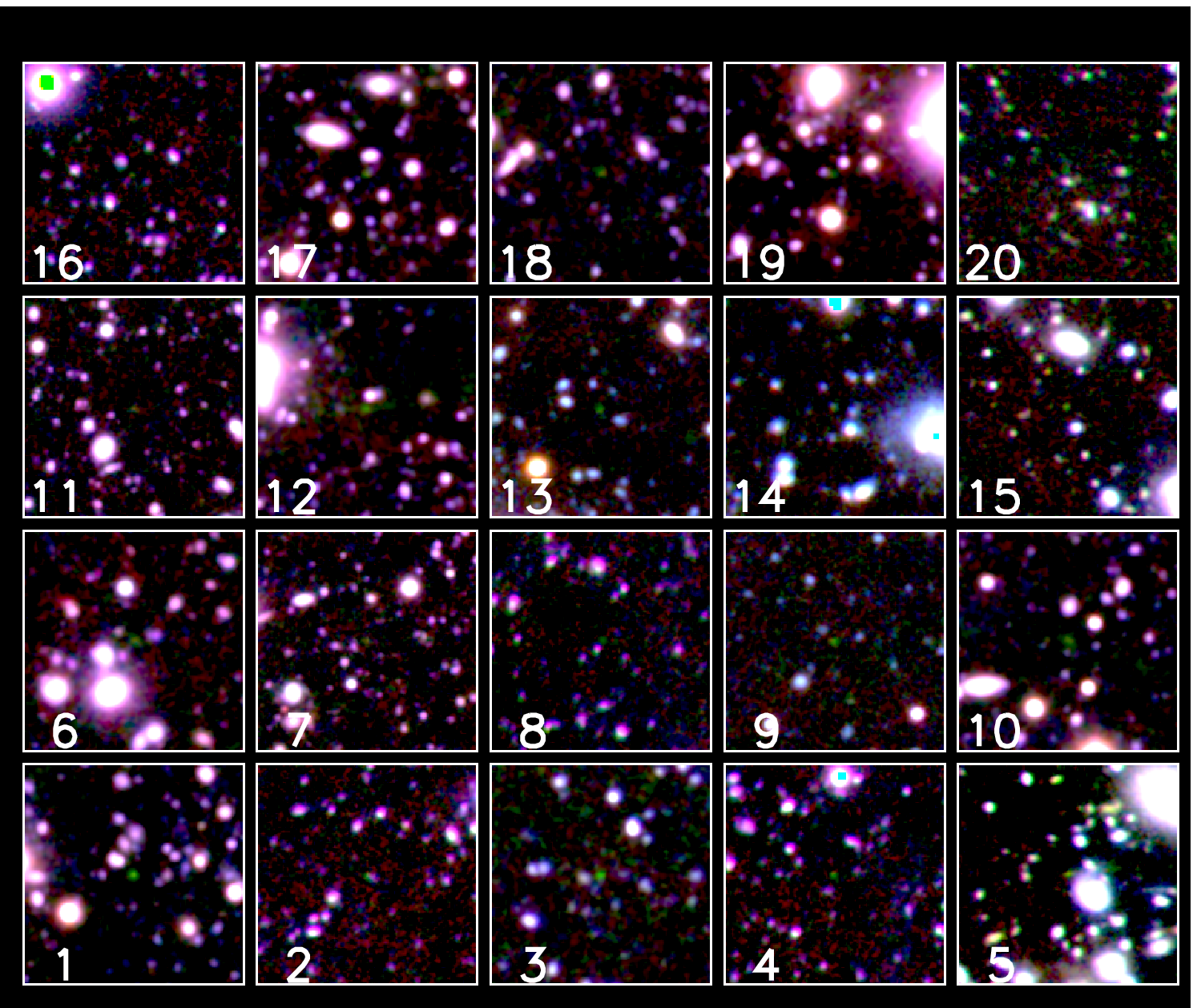}
\caption{Images of the spectroscopically
confirmed $z=6.5$ Ly$\alpha$ emitter sample.
For each object we show a $40''$ thumbnail around the emitter with
blue=$I$-band, green=F912 narrowband, and red=$z$-band. The emitter
appears as a green object at the center of the thumbnail.
The numerical label corresponds to the number in Table~A3
and Figure~A3.
\label{figA3:z6_images}
}
\end{figure*}

\setcounter{figure}{2}

\begin{figure*}[h]
\figurenum{A3}
\includegraphics[width=8.5in,angle=0,scale=0.95]{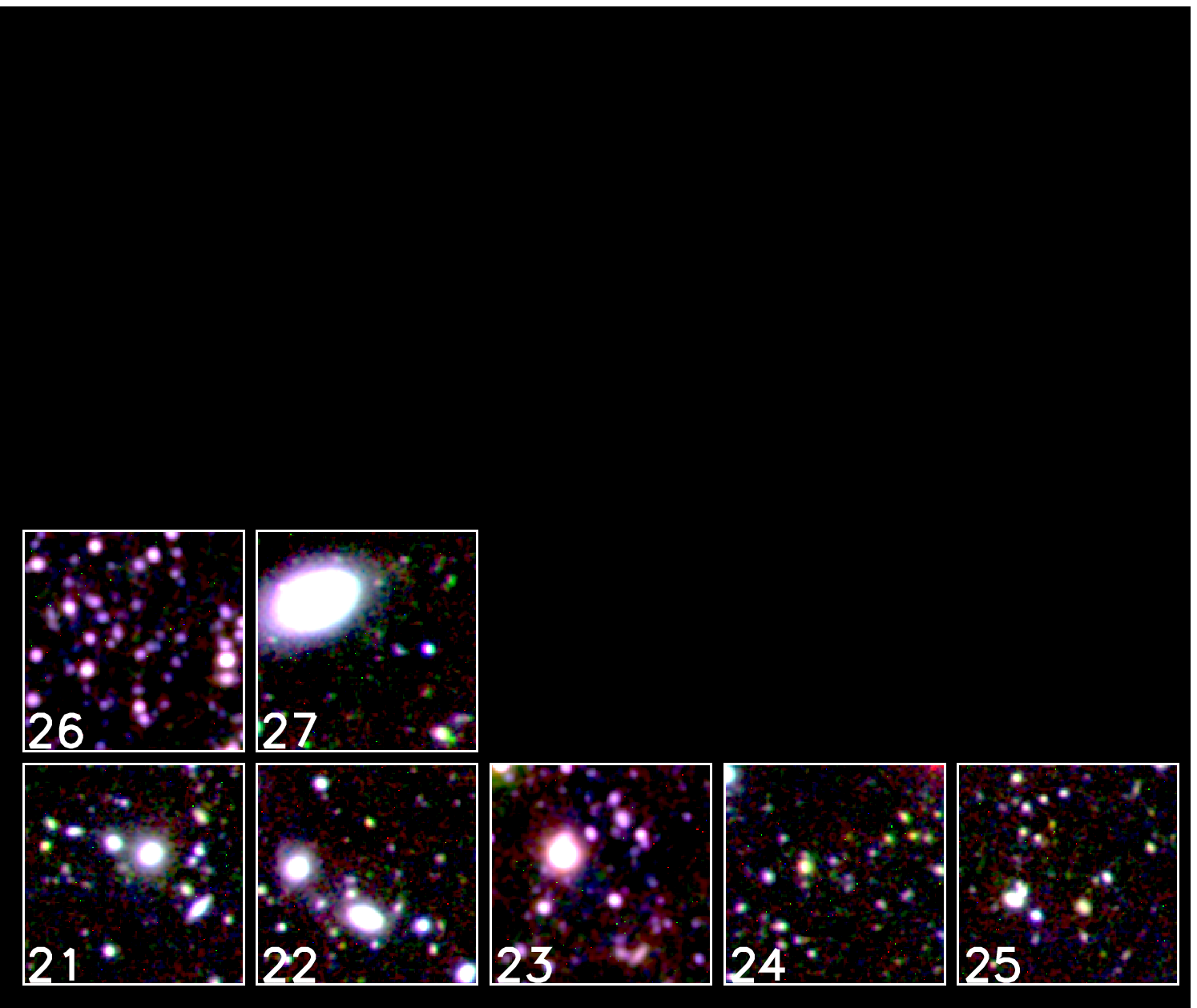}
\caption{Images of the spectroscopically
confirmed $z=6.5$ Ly$\alpha$ emitter sample (continued).
}
\end{figure*}

\begin{figure*}[h]
\figurenum{A4}
\includegraphics[width=4.6in,angle=90,scale=0.95]{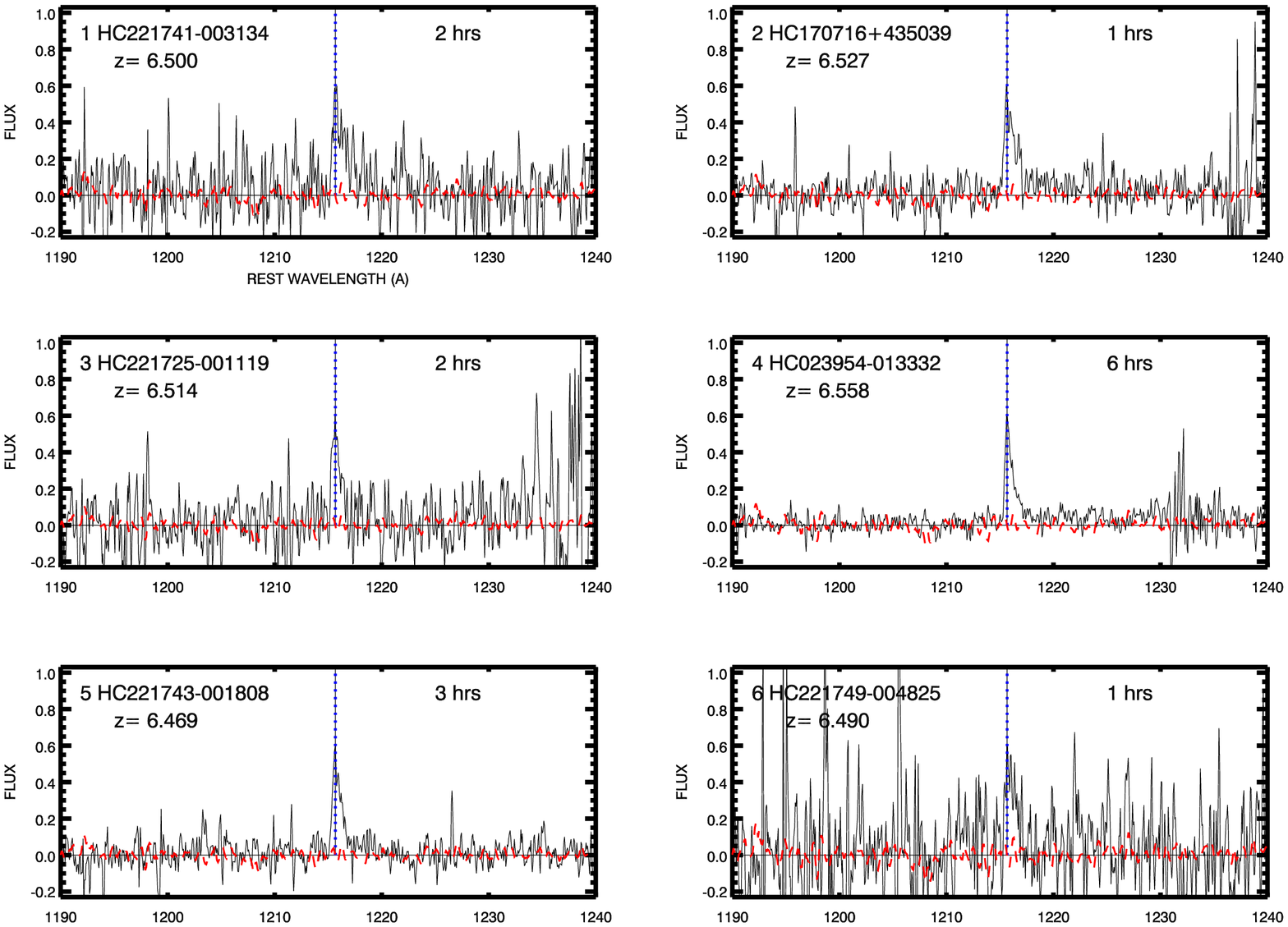}
\includegraphics[width=4.6in,angle=90,scale=0.95]{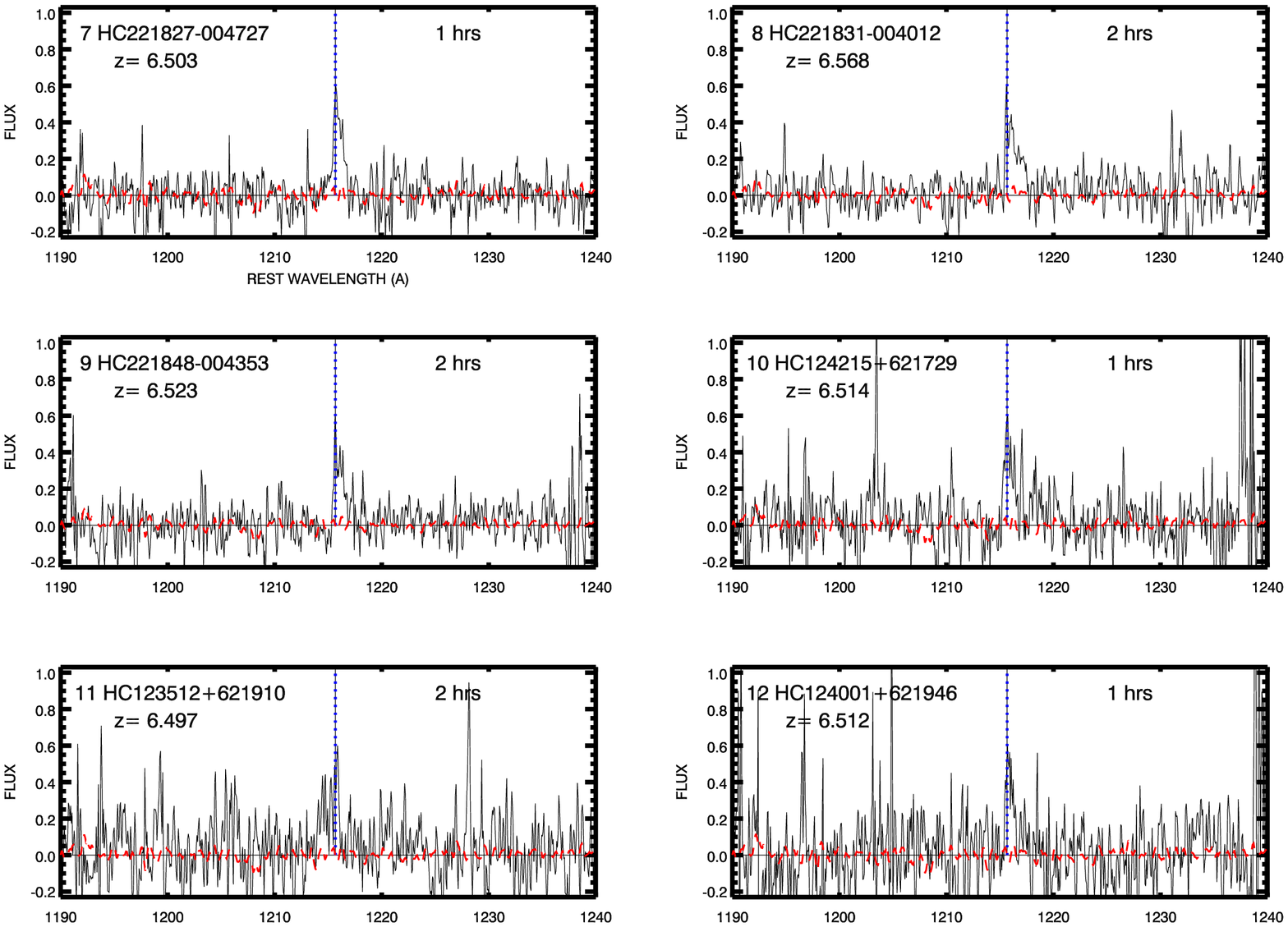}
\caption{Spectra of the spectroscopically
confirmed $z=6.5$ Ly$\alpha$ emitter sample.
For each object we show the spectrum (black) compared with the
average spectral shape of the entire sample (red dashed).
The blue dotted line shows the position of the spectrum peak,
which we use to define the redshift. The name of the object
and its redshift is given in the upper left, and the exposure
time is given in hours in the upper right. 
\label{figA4:z6_spectra}
}
\end{figure*}

\setcounter{figure}{3}

\begin{figure*}[h]
\figurenum{A4}
\includegraphics[width=4.6in,angle=90,scale=0.95]{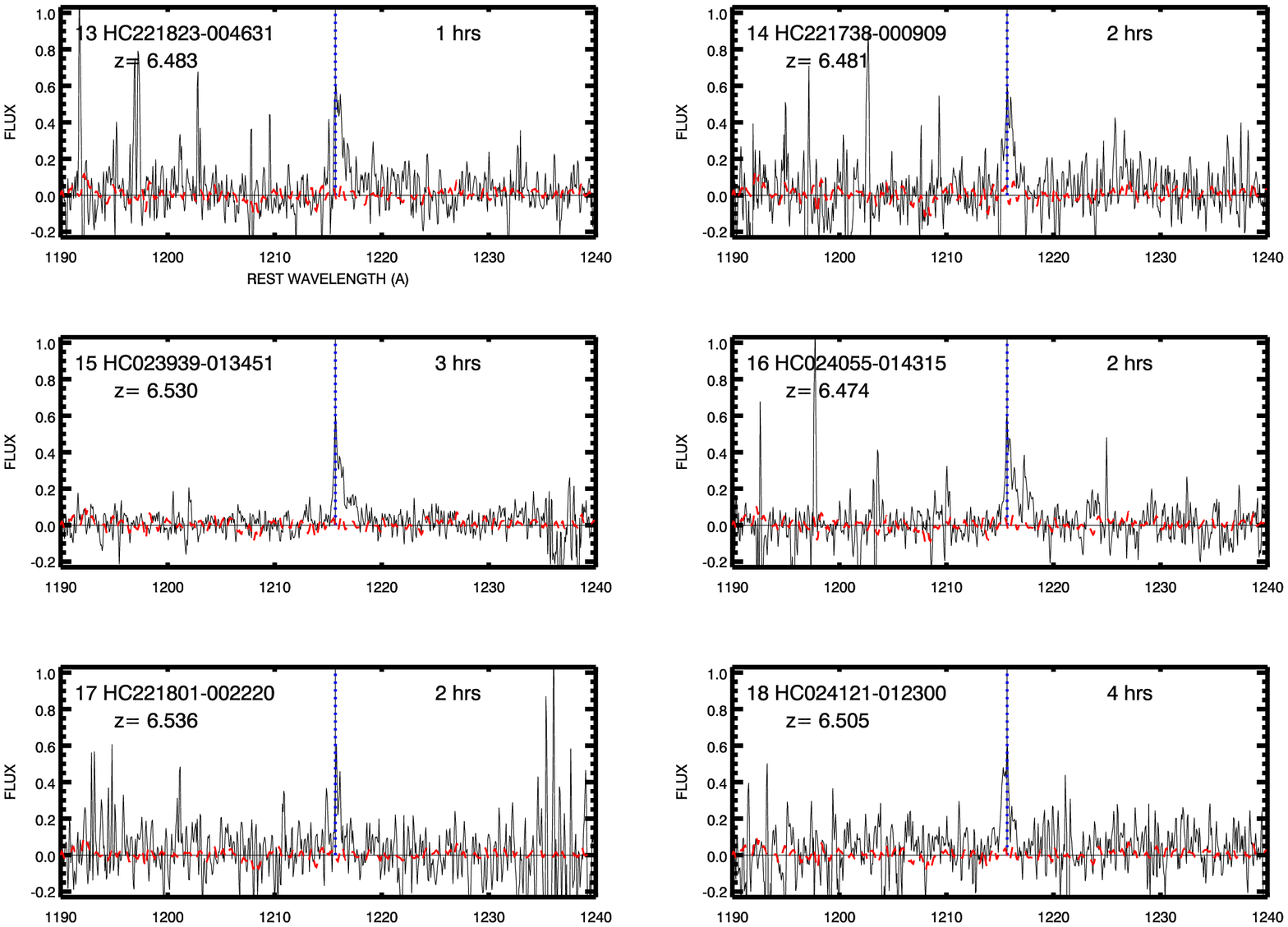}
\includegraphics[width=4.6in,angle=90,scale=0.95]{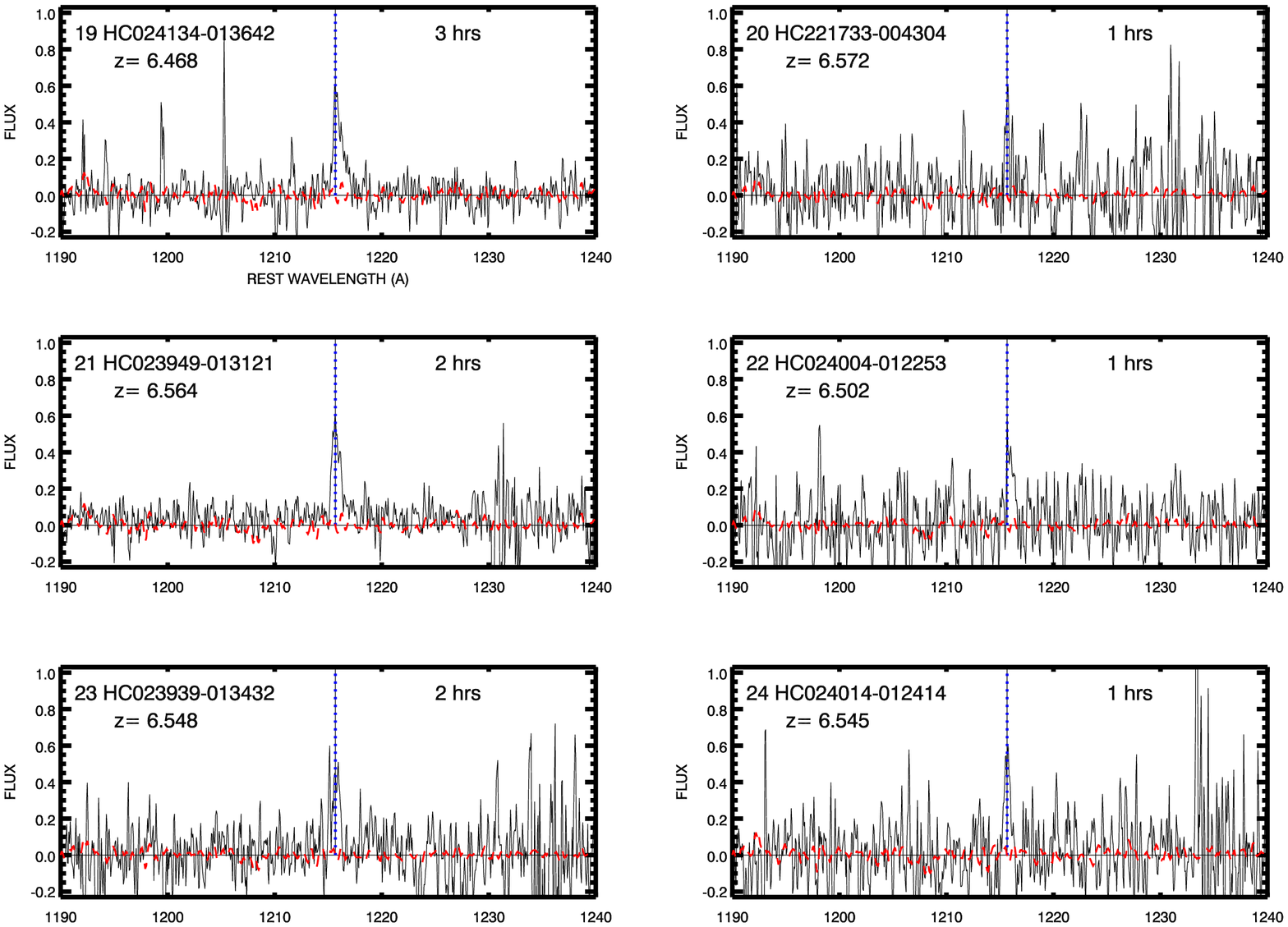}
\caption{Spectra of the spectroscopically 
confirmed $z=6.5$ Ly$\alpha$ emitter sample (continued).
}
\end{figure*}

\clearpage

\begin{figure*}[h]
\figurenum{A4}
\includegraphics[width=4.6in,angle=90,scale=0.95]{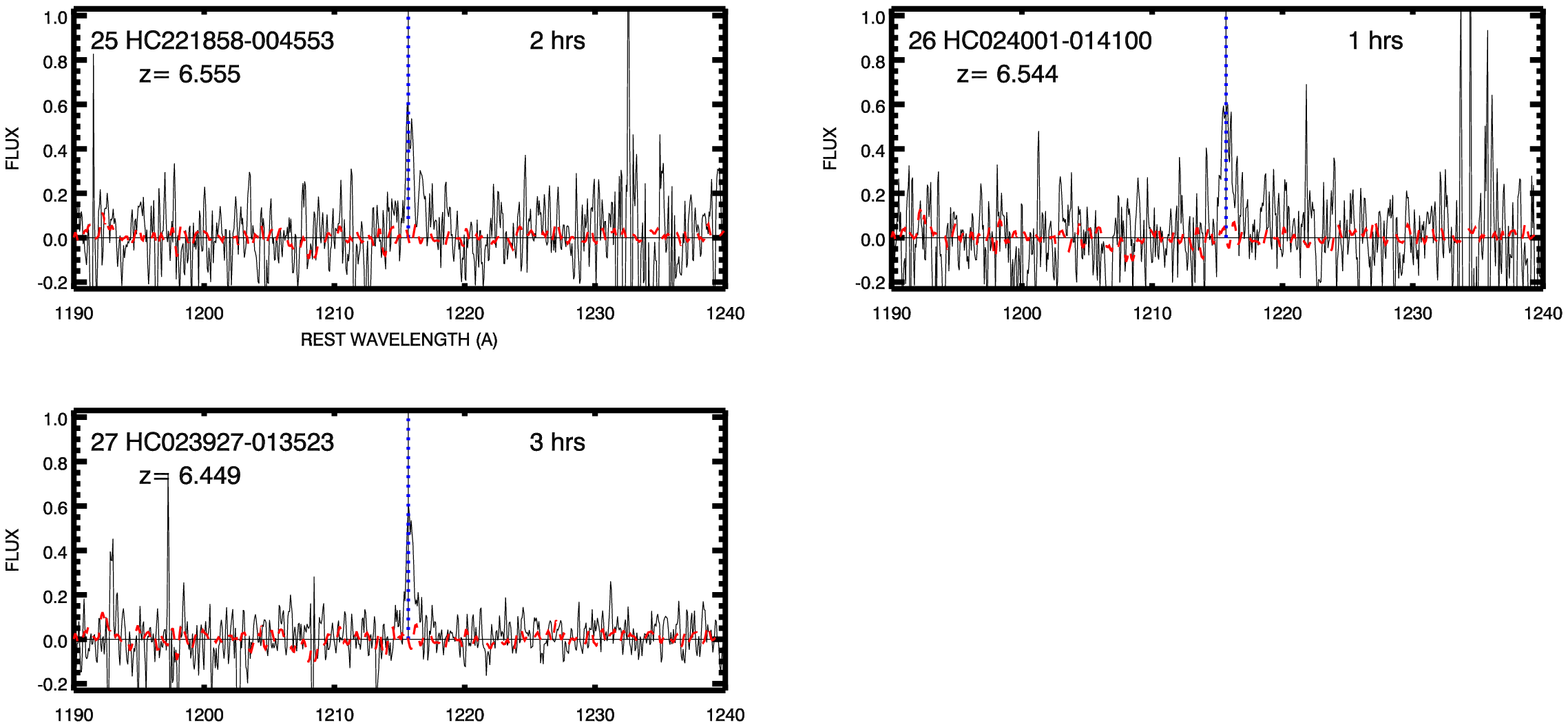}
\caption{Spectra of the spectroscopically 
confirmed $z=6.5$ Ly$\alpha$ emitter sample (continued).
}
\end{figure*}

\newpage
\tablenum{7}
\begin{deluxetable*}{clllcccccl}
\small\addtolength{\tabcolsep}{-4pt}
\renewcommand\baselinestretch{1.0}
\tablewidth{0pt}
\tablecaption{$z=6.5$ Ly$\alpha$ Emitters (GOODS-N)}
\scriptsize
\tablehead{Number & Name & R.A. & Decl. & $N$ & $z$ & Redshift &  Expo & Qual & FWHM\\ & &(J2000) & (J2000) & (AB) & (AB)  & & (hrs) & & (\AA)  \\ (1) & (2) & (3) & (4)  & (5) & (6) & (7) & (8) & (9) & (10)}
\startdata
       1  &  HC123725+621227  &     189.35800  &     62.20769  &  $ 23.83$  &
$-28.80$  &   6.5593  &   5  &    1  &  $ 0.77\pm0.06$  \cr
       2  &  HC123602+621404  &     189.00943  &     62.23466  &  $ 24.85$  &
$-26.84$  &   6.5610  &   2  &    1  &  $ 0.70\pm 0.10$  \cr
       3  &  HC123637+621022  &     189.15700  &     62.17283  &  $ 25.37$  &
$-26.27$  &   6.5428  &   2  &    3  &  $ 0.77\pm 0.11$  \cr
\enddata
\label{very_high_la.tab}
\end{deluxetable*}

\begin{figure*}
\includegraphics[width=8.5in,angle=0,scale=0.95]{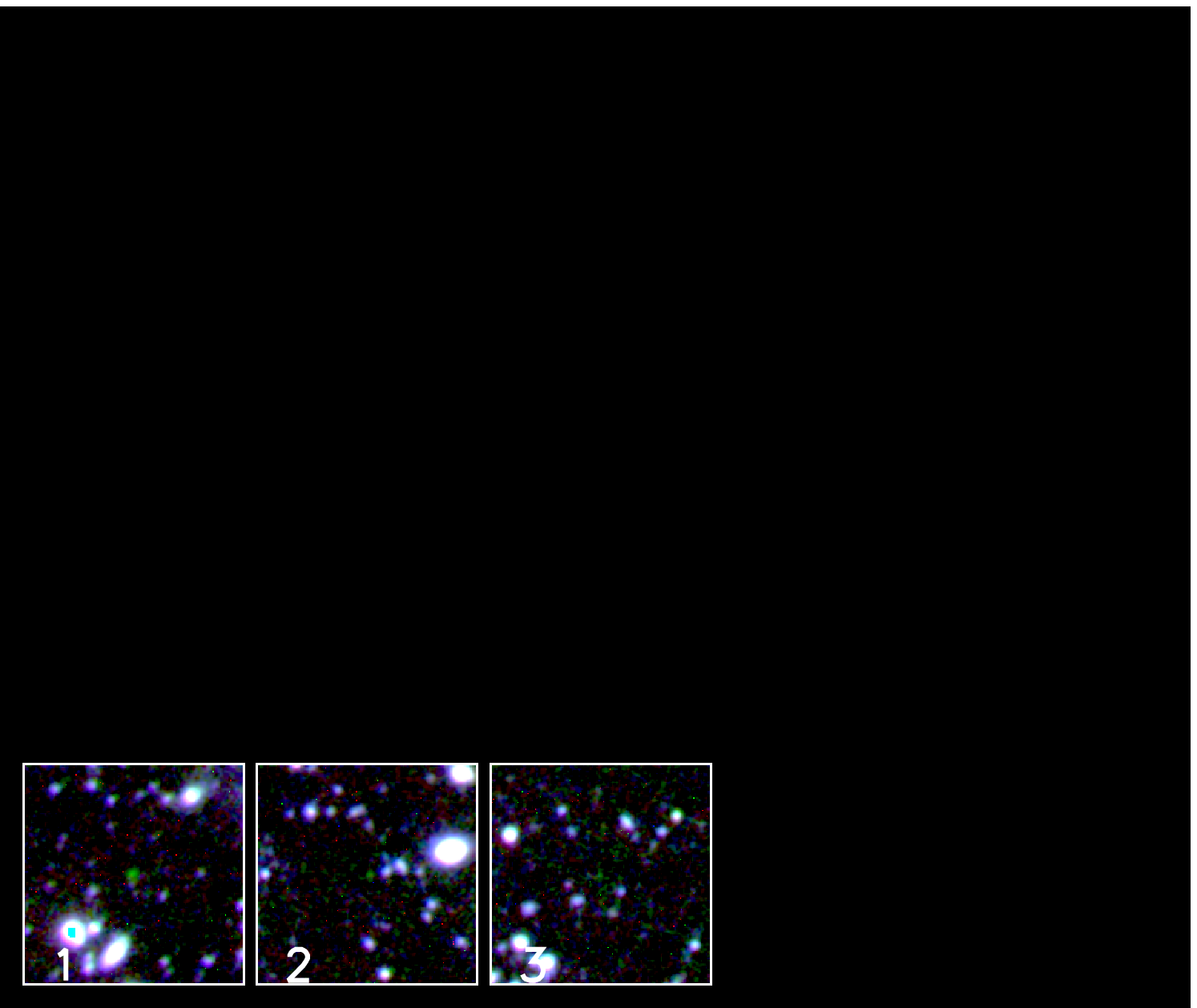}
\figurenum{A5}
\caption{Images of the spectroscopically
confirmed $z=6.5$ Ly$\alpha$ emitter sample in the GOODS-N.
For each object we show a $40''$ thumbnail around the emitter with
blue=$I$-band, green=F921 narrowband, and red=$z$-band. The emitter
appears as a green object at the center of the thumbnail.
The numerical label corresponds to the number in Table~A4 and Figure~A6.
\label{figA5:z6_images}
}
\end{figure*}

\begin{figure*}
\figurenum{A6}
\includegraphics[width=4.6in,angle=90,scale=0.95]{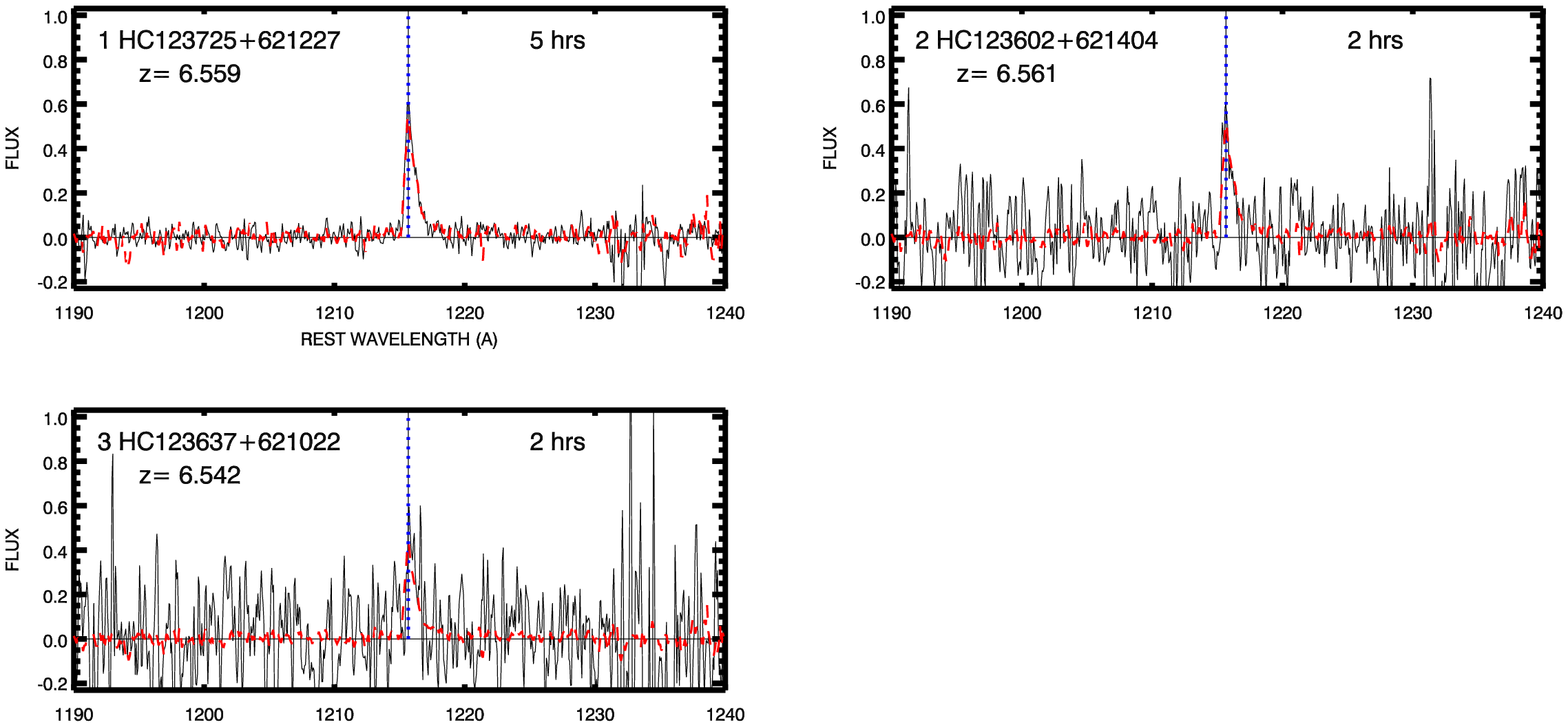}
\caption{Spectra of the spectroscopically
confirmed $z=6.5$ Ly$\alpha$ emitter sample.
For each object we show the spectrum (black) compared with the
average spectral shape of the entire sample (red dashed).
The blue dotted line shows the position of the spectrum peak,
which we use to define the redshift. The name of the object
and its redshift is given in the upper left, and the exposure
time in hours is given in the upper right. 
\label{figA6:z6_spectra_plus}
}
\end{figure*}

\end{document}